\documentclass[12pt,a4paper]{article}
\usepackage{graphicx}
\usepackage[T1]{fontenc}
\usepackage[utf8]{inputenc}
\usepackage{textcomp}
\usepackage[sc,osf]{mathpazo}
\usepackage{a4wide}  
\usepackage{latexsym,amsthm,amsfonts,amsmath,mathrsfs,amssymb}
\usepackage{dsfont}
\usepackage{accents}
\usepackage[nosort]{cite}
\usepackage{booktabs} 
\usepackage[unicode,implicit]{hyperref}
\hypersetup{%
  pdftitle    = {Higher-form symmetries in supergravity,
      scalar charges and black-hole thermodynamics}
  pdfkeywords = {gravity,  supergravity, Kaluza-Klein, compactification,
    higher-form, black hole, thermodynamics, scalar charges},
  pdfauthor   = {Gabriele Barbagallo, Jos\'e Luis V. Cerdeira,
    Carmen G\'omez-Fayr\'en, Patrick Meessen and Tom\'as Ort\'{\i}n},
  plainpages  = true,
  colorlinks  = true,
  citecolor   = blue,
  urlcolor    = red,
  linkcolor   = black
}
\newcommand{\hepth}[1]{{\tt
\href{http://www.arXiv.org/abs/hep-th/#1}{hep-th/#1}}}
\newcommand{\grqc}[1]{{\tt
\href{http://www.arXiv.org/abs/gr-qc/#1}{gr-qc/#1}}}

\newcommand{\arxiv}[1]{{\tt arXiv:\href{http://www.arXiv.org/abs/#1}{#1}}}
\newcommand{\doi}[1]{{\tt DOI:\href{http://dx.doi.org/#1}{#1}}}

\allowdisplaybreaks

\makeatletter
\@addtoreset{equation}{section}
\makeatother

\begin{document}

\begin{flushright}
\small
IFT-UAM/CSIC-25-159\\
December 27\textsuperscript{rd}, 2025\\
\normalsize
\end{flushright}

\begin{center}

  {\Large {\bf Higher-form symmetries in supergravity,\\[.3cm]
      scalar charges and black-hole thermodynamics}}
 
\vspace{.5cm}

\renewcommand{\thefootnote}{\alph{footnote}}

{\sl G.~Barbagallo,$^{1,}$}\footnote{Email: {\tt gabriele.barbagallo[at]estudiante.uam.es}}
{\sl J.L.V.~Cerdeira}$^{1,}$\footnote{Email: {\tt jose.verez-fraguela[at]estudiante.uam.es}}
{\sl C.~G\'omez-Fayr\'en,$^{1,}$}\footnote{Email: {\tt carmen.gomez-fayren[at]estudiante.uam.es}}
{\sl P.~Meessen$^{2,3}$}\footnote{Email: {\tt meessenpatrick[at]uniovi.es}}
{\sl and T.~Ort\'{\i}n}$^{1,}$\footnote{Email: {\tt Tomas.Ortin[at]csic.es}}

\setcounter{footnote}{0}
\renewcommand{\thefootnote}{\arabic{footnote}}
\vspace{.5cm}

${}^{1}${\it\small Instituto de F\'{\i}sica Te\'orica UAM/CSIC\\
C/ Nicol\'as Cabrera, 13--15,  C.U.~Cantoblanco, E-28049 Madrid, Spain}

\vspace{0.1cm}

${}^{2}${\it\small HEP Theory Group, Departamento de F\'{\i}sica, Universidad de Oviedo\\
  Calle Leopoldo Calvo Sotelo 18, E-33007 Oviedo, Spain}\\

\vspace{0.1cm}

${}^{3}${\it\small Instituto Universitario de Ciencias y Tecnolog\'{\i}as Espaciales
  de Asturias (ICTEA)\\ Calle de la Independencia, 13, E-33004 Oviedo, Spain}

\vspace{.5cm}

{\bf Abstract}
\end{center}
\begin{quotation}
  {\small Minimal 5-dimensional supergravity compactified on a circle gives
    the T$^{3}$ model of $\mathcal{N}=2,d=4$ supergravity, whose duality group
    is SL$(2,\mathbb{R})$. We study exhaustively the relations between all the
    local and global symmetries of both theories and between the corresponding
    conserved currents and charges, including the on-shell closed generalized
    Komar charges associated to isometries.  We find that the 2-dimensional
    subgroup of SL$(2,\mathbb{R})$ that does not include electric-magnetic
    transformations is realized as a higher-form symmetry group that acts on
    the 5-dimensional metric and vector field. Using the generalized Komar
    charges we compute the Smarr formulas for black holes, showing that they
    are identical once the relations between all the 5- and 4-dimensional
    thermodynamical quantities are taken into account, which is only possible
    if certain constraints on the fields are satisfied. We notice that
    on-shell closed 5-dimensional 3-form charges give, upon dimensional
    reduction, on-shell closed 3-form currents and 2-form charges. The
    dimensional reduction of the 5-dimensional generalized Komar 3-form charge
    associated to a Killing vector gives a new 4-dimensional on-shell closed
    3-form current which must be associated to a new global symmetry of the
    theory when it admits that Killing vector. Some of the results that we
    have derived are valid for theories of Einstein--Maxwell-like theories of
    $(p+1)$-forms with Chern--Simons terms, which includes 11-dimensional
    supergravity as a particular example.}
\end{quotation}

\newpage
\pagestyle{plain}

\tableofcontents


\section{Introduction}

Minimal 5-dimensional supergravity \cite{Cremmer:1980gs} is, in spite of its
apparent simplicity, a very rich theory that admits many interesting solutions
describing a wide range of objects, supersymmetric \cite{Gauntlett:2002nw} and
non-supersymmetric, including black holes and black rings, Kaluza-Klein
monopoles, pp-waves and fuzzballs. For this reason and due to its similarities
with 11-dimensional supergravity \cite{Cremmer:1978km} it has been the subject
of many studies.  In particular, the Smarr formula \cite{Smarr:1972kt} of the
black holes of this 5d theory was first derived in Ref.~\cite{Gauntlett:1998fz}
and, in Ref.~\cite{Gibbons:2013tqa} (see also Ref.~\cite{Kunduri:2018qqt})
Gibbons and Warner showed that the contribution of an additional 4-form term
associated to non-trivial topology allows for horizonless, regular, massive
solutions (fuzzballs).

The derivation of the Smarr formula in those references was based on the
techniques explained, for instance, in \cite{Townsend:1997ku}, but, as
originally done in Refs.~\cite{Bardeen:1973gs,Carter:1973rla} and more
recently explained in Refs.~\cite{Kastor:2008xb,Kastor:2010gq}, there is a
more transparent derivation based on the construction of a generalization of
the Komar charge of General Relativity \cite{Komar:1958wp}, \textit{i.e.}~a
$(d-2)$-form associated to a Killing vector which is closed on-shell (which
means that its form depends on the theory). 

Let $\mathbf{K}[l]$ denote the generalized Komar charge associated with a 
Killing vector $l$, choose $l$ to generate the horizon and take $\Sigma^{d-1}$
to be a spacelike hypersurface whose only boundaries are spatial infinity S$^{d-2}_{\infty}$
and the bifurcation surface $\mathcal{BH}$, {\em i.e.\/}
\begin{equation}
  \partial \Sigma^{d-1}
  =
  \mathrm{S}^{d-2}_{\infty}\cup \mathcal{BH}
  \, .
\end{equation}
Using the on-shell closedness\footnote{We denote with $\doteq$ equations which may only
  be satisfied on shell.} of $\mathbf{K}[l]$
\begin{equation}
  d\mathbf{K}[l]
  \doteq
  0\,,
\end{equation}
we can use Stokes' theorem to obtain
\begin{equation}
  0
  \doteq
  \int_{\Sigma^{d-1}}d\mathbf{K}[l]
  =
  \int_{\mathrm{S}^{d-2}_{\infty}}\mathbf{K}[l]
-
  \int_{\mathcal{BH}}\mathbf{K}[l]\, .
\end{equation}
This identity relates asymptotic gravitational charges (such as mass and angular momentum)
to quantities defined on the horizon, such as the entropy:
it is the Smarr formula for the black holes (or rings) of the theory,
and is a generalization of the Gauss law of electromagnetism.

It is important to realize that we could equally have chosen a hypersurface
$\Sigma^{d-1}$ extending from spatial infinity to any other surface different
from the bifurcation surface, placed in the interior or the exterior of the
horizon. The choice of the bifurcation surface as the interior boundary is just
very convenient because it allows us to calculate the integrals very easily in
terms of known and meaningful quantities in black-hole physics. Stokes' theorem,
nevertheless, guarantees that the same result (equal to the value of the
integral at infinity) would be obtained by integrating over any other interior
boundary. If the integral at infinity does not vanish, the interior boundary
can only be a singularity or the spatial infinity of another asymptotically
flat region if we allow for non-trivial topology. If the integral at infinity
vanishes, the positive mass theorem
\cite{Schon:1979rg,Schon:1981vd,Witten:1981mf} tells us that we are dealing
with flat spacetime, ruling out the existence of globally regular, static,
asymptotically flat solutions.\footnote{See Ref.~\cite{Gibbons:2013tqa} for an
  interesting historical review with many references on the problem of
  existence of such solitonic solutions, and Ref.~\cite{Ballesteros:2024prz}
  for a more recent exposition that emphasizes the use of the generalized
  Komar charges.}

Minimal 5-dimensional supergravity, though, does admit globally regular,
massive, asymptotically-flat solutions called \textit{fuzzballs}
\cite{Giusto:2004kj,Bena:2005va,Berglund:2005vb}, whose existence and
properties are the basis on which Mathur's fuzzball paradigm is
based.\footnote{Two recent reviews on this topic with references are
  \cite{Bena:2022rna,Mathur:2024ify}.} As we mentioned at the beginning of the
introduction, the existence of fuzzball solutions can be understood by the
contribution of an additional 4-form term to the Smarr formula
\cite{Gibbons:2013tqa}. Since the standard Smarr formula follows from the
on-shell closure of the generalized Komar 3-form charge, the existence of
fuzzballs suggests that this 3-form charge is no longer closed in these
solutions and that the obstruction to its closure is, precisely, the 4-form
identified by Gibbons and Warner. 

This discussion shows that generalized Komar charges and, actually, any other
kind of on-shell closed charges whose integrals satisfy generalized Gauss laws
or fail to do so due to certain topological obstructions, are very powerful
tools to study black holes and solitons. For instance, the on-shell closed
$(d-2)$-form scalar charges defined in
\cite{Pacilio:2018gom,Ballesteros:2023iqb} have been used to derive a no-hair
theorem that we will later use to prove that regular black-hole scalar charges
cannot be independent quantities. One of the goals of this paper is to
construct on-shell (and off-shell) closed charges of minimal 5-dimensional
supergravity, including its generalized Komar charge, identifying the possible
obstructions to its closure.

Generalized Komar charges have been constructed long ago for simple extensions
of General Relativity in vacuum: General Relativity with a cosmological
constant in Ref.~\cite{Magnon:1985sc} and for the cosmological
Einstein--Maxwell theory in Ref.~\cite{Bazanski:1990qd}.  During the last few
years, our group has been studying the construction of generalized Komar
charges of different, much more general, theories
\cite{Ortin:2021ade,Mitsios:2021zrn,Meessen:2022hcg,Ortin:2022uxa,Ballesteros:2023iqb,Gomez-Fayren:2023wxk,Bandos:2023zbs,Zatti:2023oiq,Gomez-Fayren:2024cpl,Ballesteros:2024prz,Bandos:2024pns,Ortin:2024emt,Ortin:2024mmg,Bandos:2024rit,Cerdeira:2025elp,Barbagallo:2025fkg,Cerdeira:2025agq}
building on the observations made in Ref.~\cite{Liberati:2015xcp} which
suggest a more systematic approach based on the symmetries of the theory and
on the assumed symmetries of the solutions
\cite{Barnich:2001jy,Barnich:2003xg}.

In particular, as part of our long-term program of deriving all the
thermodynamic relations of supergravity and string-theory black holes directly
in higher dimensions,\footnote{Often, calculating directly in higher
  dimensions is simpler, but a good understanding of the relations between the
  higher- and lower-dimensional thermodynamical functions and variables
  (entropy, temperature, conserved charges, dual thermodynamical potentials
  etc.) is necessary.} we have started exploring the relation between the
charges of theories related by Kaluza--Klein (KK) compactification. In
\cite{Gomez-Fayren:2023wxk} we studied the simplest case: pure gravity in 5
dimensions compactified on a circle, that gives a 4-dimensional
Einstein--Maxwell-Dilaton (EMD) theory with $a=\sqrt{3}$. The 4-dimensional
theory has a global symmetry that leads to an on-shell closed 3-form current
and, using the trick of Refs.~\cite{Pacilio:2018gom,Ballesteros:2023iqb}, to
an on-shell closed 2-form charge whose 5-dimensional origin was unknown since
pure 5-dimensional Einstein gravity does not seem to have any global
symmetries. In Ref.~\cite{Gomez-Fayren:2024cpl}, though, we showed that with
the topologically non-trivial KK boundary conditions, it does have a global
higher-form-type symmetry whose closed 4-form current and 3-form charge give
rise to the already known 4-dimensional 3-form current and 2-form charge. The
5-dimensional 3-form charge can be combined with the standard Komar charge so
that its integral at infinity gives directly the 4-dimensional mass, which is
the only meaningful definition of mass in asymptotically KK spacetimes,
instead of a linear combination of the 4-dimensional mass and the scalar
charge \cite{Barbagallo:2025fkg}.

We would like to extend this work in pure 5-dimensional gravity with KK
boundary conditions to include matter. Thus our second goal in this paper will
be to study the relation between the symmetries and charges of minimal
5-dimensional supergravity and those of the so-called T$^{3}$ model of
$\mathcal{N}=2,d=4$ supergravity, which is obtained when the former is
compactified on a circle.

Minimal 5-dimensional supergravity with trivial boundary conditions does not
have any global symmetry but the T$^{3}$ model has a global SL$(2,\mathbb{R})$
symmetry group and, based on our experience in the pure gravity case, we
expect that KK boundary conditions give rise to higher-form symmetries that
must act on the matter fields as well. Finding these symmetries and
relating them to the 4-dimensional ones will also be one of our goals. Relating
the 5- and 4-dimensional currents and charges will be another goal and we will
find some surprises when we study the dimensional reduction of the
5-dimensional generalized Komar 3-form charge. To start with, using KK
boundary conditions one can show that, in general, there is an obstruction for
it to be closed on-shell: the obstruction is related to one of the scalar charges. This
allows us to give a full characterization of the possible soliton solutions
with the given boundary conditions. Furthermore, its dimensional reduction
gives a new 4-dimensional on-shell closed 3-form current, which should be
related to an as yet unidentified global symmetry of the 4-dimensional theory
when we assume the existence of a 4-dimensional Killing vector \cite{kn:BCMO}.

This paper is organized as follows: in Section~\ref{sec-higherCS} we study the
symmetries of theories of $(p+1)$-form potentials with Chern--Simons terms
coupled to gravity because they can all be treated simultaneously. Minimal
5-dimensional supergravity is the case of interest but the case of
11-dimensional supergravity is also automatically covered. In
Section~\ref{sec-thetheory} we study minimal 5-dimensional supergravity with
asymptotically-flat boundary conditions: its symmetries (local and global),
conserved charges and the thermodynamics of its black objects. We do the same
of the T$^{3}$ model in Section~\ref{sec-thet3model} and, in
Section~\ref{sec-relation5d4d} we relate all the results of the two preceding
sections. Section~\ref{sec-discussion} contains a discussion of our results,
of questions that remain unanswered and of possible future
directions. Appendix~\ref{app-generalized} review the definition of the
electric momentum map (related to the electrostatic potential of the
black-hole horizon) and a generalization related to the generalized symmetric
ansatz. Appendix~\ref{app:magneticmomentummap} does the same for the magnetic
momentum map.\footnote{The magnetic momentum map equation for minimal
  5-dimensional supergravity was found in Ref.~\cite{Gibbons:2013tqa}. The
  obstruction to the closure of the generalized Komar charge originates in
  this equation.} Appendix~\ref{app-4-5fields} contains a summary of the
relations between the 4- and 5-dimensional fields of the two theories we are
considering and appendix~\ref{app-4-5chargesandpotentials} contains a summary
of the relations between the 4- and 5-dimensional black-hole charges and
potentials. Finally, Appendix~\ref{app-KKBHs} contains some simple
solutions of the two theories related by dimensional reduction that provide
examples of the relations between masses, charges etc.

\section{The symmetries in higher-rank-form theories
  with Chern--Simons  terms}
\label{sec-higherCS}

In this section we want to study the general case of $d$-dimensional actions
of $(p+1)$-form potentials $V$ coupled to gravity, which we will describe with
the Vielbein $e^{a}$.\footnote{Our conventions are those of
  Ref.~\cite{Ortin:2015hya} and we use differential-form language
  throughout. In particular, our metric has mostly-minus signature and our
  Levi-Civita spin connection and curvature satisfy
  \begin{subequations}
    \begin{align}
      \mathcal{D}e^{a}
      & =
        de^{a}-\omega^{a}{}_{b}\wedge e^{b}
        =
        0\,,
      \\
      & \nonumber \\
      R_{ab}
      & =
        d \omega_{ab} - \omega_{ac} \wedge \omega^{c}{}_{b}\, .      
    \end{align}
  \end{subequations}
} We are interested in actions with general Chern--Simons (CS) terms of the
form

\begin{equation}
  \label{eq:CStheoryaction}
  \begin{aligned}
  S[e,V]
  & =
    \frac{1}{16\pi G_{N}^{(d)}}\int \left\{
    (-1)^{d-1}\star (e^{a}\wedge e^{b}) \wedge R_{ab}
+\frac{(-1)^{(p+1)d}}{2} G\wedge \star G
    +\gamma \underbrace{G\wedge \cdots \wedge G}_\text{N times}\wedge V\right\}
    \\
    & \\
    & \equiv
     \int \mathbf{L} \,,
  \end{aligned}
\end{equation}

\noindent
where $\gamma$ is a constant with dimensions $[L]^{N-2}$ and $G=dV$ is the
$(p+2)$-form field strength.\footnote{An alternative, friendlier but less
  compact, expression for $\star (e^{a}\wedge e^{b}) \wedge R_{ab}$ is
  \begin{equation}
      \frac{1}{(d-2)!}\varepsilon_{a_{1}\cdots a_{d-2}bc}
    e^{a_{1}}\wedge \cdots \wedge e^{a_{d-2}} \wedge R^{bc}\,.
  \end{equation}
}

For odd $p$, the CS term would vanish identically for $N>1$ and would be a
total derivative for $N=1$. Consequently, we will consider the case of $p$ being even.
The CS term can
only be introduced consistently when $d+1$ is an integer multiple of $p+2$,
which must be even. Then, $d$ must be odd and
$(-1)^{pd} = (-1)^{d+(p+1)}=(-1)^{Np}=+1$ etc.

Two particularly interesting instances of the action
Eq.~(\ref{eq:CStheoryaction}) are (the bosonic sectors of) minimal
($\mathcal{N}=1$) 5-dimensional supergravity
\cite{Cremmer:1980gs,Chamseddine:1980mpx}, studied in
Section~\ref{sec-thetheory}, with

\begin{equation}
  d=5\,,
  \hspace{.5cm}
  p=0\,,
  \hspace{.5cm}
  N=2\,,
  \hspace{.5cm}
 \gamma=\tfrac{1}{3^{3/2}}\,, 
\end{equation}

\noindent
and $\mathcal{N}=1,d=11$ supergravity \cite{Cremmer:1978km}, with

\begin{equation}
  d=11\,,
  \hspace{.5cm}
  p=2\,,
  \hspace{.5cm}
  N=2\,,
  \hspace{.5cm}
 \gamma=-\tfrac{1}{12^{4}}\,, 
\end{equation}

\noindent
which we will study elsewhere \cite{kn:BCMO}.

Under a general variation of the fields,

\begin{equation}
  \label{eq:generalvariationCStheory}
  \delta S
  =
  \int_{\mathcal{M}} \left\{\mathbf{E}_{a}\wedge \delta e^{a}
    +\mathbf{E}\wedge \delta V
    +d\mathbf{\Theta}(e,V,\delta e,\delta V) \right\}\,,
\end{equation}

\noindent
where the equations of motion $\mathbf{E}_{a},\mathbf{E}$ and the
pre-symplectic potential $\mathbf{\Theta}(e,V,\delta e,\delta V)$ are given,
respectively, by\footnote{
An alterative, friendlier but less
  compact, expression for $\imath_{a}\star (e^{b}\wedge  e^{c})\wedge R_{bc}$ is
  \begin{equation}
      \frac{1}{(d-3)!}\varepsilon_{a b_{1}\cdots b_{d-3}cd}
    e^{b_{1}}\wedge \cdots \wedge e^{b_{d-3}} \wedge R^{cd}\,.
  \end{equation}
  }

\begin{subequations}
  \begin{align}
    \mathbf{E}_{a}
    & =
      \imath_{a}\star (e^{b}\wedge  e^{c})\wedge R_{bc}
      +\tfrac{1}{2}\left[\imath_{a}G\wedge \star G
      -G\wedge \imath_{a}\star G\right]\,,
    \\
    & \nonumber \\
    \label{eq:EVCS}
    \mathbf{E}
    & =
      -d\star G
      +(N+1)\gamma \underbrace{G\wedge \cdots \wedge G}_\text{N times}\,,
    \\
    & \nonumber \\
    \label{eq:ThetaCS}
    \mathbf{\Theta}(e,V,\delta e,\delta V)
    & =
      -\star (e^{a}\wedge e^{b})\wedge \delta \omega_{ab}
      -\delta V\wedge\left[\star G -\gamma N \underbrace{G\wedge \cdots
      \wedge G}_\text{N-1 times}\wedge V\right]\,.
  \end{align}
\end{subequations}

\subsection{Symmetries, Noether identities and conserved charges}
\label{sec-symmetryandNoetehridentitiesforms}

The theories we are considering always have two local symmetries: a $p$-form gauge symmetry and
invariance under general coordinate transformations (GCTs). The global
symmetries of the theory depend on several factors but, most importantly, on
the topology of the manifolds on which we define the action. Since we are
interested in minimal 5-dimensional supergravity and its relation to the
$T^{3}$ model of $\mathcal{N}=2,d=4$ supergravity, we will consider generic
manifolds with no particular topology and manifolds with a compact dimension
that can be used for the dimensional reduction of the theory.

\subsubsection{$p$-form gauge symmetries }
\label{app-pformgaugesymmetries}

Under the local transformations

\begin{equation}
  \label{eq:pformgaugetrans}
  \delta_{\Lambda}V
  =
  d\Lambda\,,
\end{equation}

\noindent
where $\Lambda$ is an arbitrary $p$-form, the action
Eq.~(\ref{eq:CStheoryaction}) is invariant up to a total derivative

\begin{equation}
  \label{eq:BVLambdadef}
  \delta_{\Lambda}S
  =
  \int d\left[\gamma \underbrace{G\wedge \cdots
      \wedge G}_\text{N-1 times}\wedge V\wedge d\Lambda \right]
  \equiv  -\int_{\mathcal{M}}d\mathbf{B}(V,\Lambda)\,.
\end{equation}

$\mathbf{B}(V,\Lambda)$ is defined up to a total derivative. Our choice
ensures that $\mathbf{B}(V,\Lambda)$ vanishes for \textit{Killing} (or
\textit{reducibility}) \textit{parameters} $\Lambda$ such that
$\delta_{\Lambda}V=0$ \cite{Barnich:2001jy,Barnich:2003xg}.

Particularizing Eq.~(\ref{eq:generalvariationCStheory}) for these
transformations we find an alternative expression for $\delta_{\Lambda}S$

\begin{equation}
  \delta_{\Lambda} S
  =
  \int \left\{ \mathbf{E}\wedge d\Lambda
    +d\mathbf{\Theta}(e,V,\delta_{\Lambda} V) \right\}
  =
  \int
    d\left[\mathbf{\Theta}(e,V,\delta_{\Lambda} V)
      +\mathbf{E}\wedge \Lambda \right]\,,
\end{equation}

\noindent
after use of the Noether identity

\begin{equation}
    \label{eq:pformNoetheridentity}
d\mathbf{E}=0\,.
\end{equation}

Comparing both expressions we arrive at

\begin{subequations}
  \begin{align}
    d\mathbf{J}[\Lambda]
    & = 0\,,
    \\
    & \nonumber \\
    \mathbf{J}[\Lambda]
    & \equiv
      \mathbf{\Theta}(e,V,\delta_{\Lambda} V)
    +\mathbf{E}\wedge \Lambda
      +\mathbf{B}(V,\Lambda)
      \nonumber \\
    & \nonumber \\
    & =
      d\mathbf{Q}[\Lambda]\,,
    \\
    \label{eq:QLambda}
    \mathbf{Q}[\Lambda]
    & =
     -\Lambda\wedge\left[\star G -\gamma (N+1) \underbrace{G\wedge \cdots
      \wedge G}_\text{N-1 times}\wedge V\right]\,. 
  \end{align}
\end{subequations}

The Noether charge $\mathbf{Q}[\Lambda]$ is not closed. By definition,

\begin{equation}
  d\mathbf{Q}[\Lambda]
  =
  \mathbf{J}[\Lambda]
  =
      \mathbf{\Theta}(V,\delta_{\Lambda} V)
    +\mathbf{E}\wedge \Lambda
      +\mathbf{B}(V,\Lambda)\,.
\end{equation}

The second term on the right-hand side vanishes on-shell, while the first and
the third vanish identically when $\Lambda$ is a Killing parameter that we
will call $\kappa$. $\kappa$ must be a closed $p$-form. It is easy to see
that, if $\kappa$ is exact, the integral of $\mathbf{Q}[\kappa]$ over closed
surfaces $\Sigma^{d-2}$ must vanish on-shell. Thus, only harmonic $\kappa$s
give non-trivial charges.
\par
For each independent harmonic $p$-form
$\mathfrak{h}^{(p)}_{i}$ we can define an independent charge which, up to
normalization, is given by the integral over a closed $(d-2)$-surface
$\Sigma^{d-2}$

\begin{equation}
  \label{eq:QLambdaharmonic}
  \mathcal{Q}_{i}
  =
\frac{1}{16\pi G_{N}^{(d)}}  \int_{\Sigma^{d-2}} \mathbf{Q}[\mathfrak{h}^{(p)}_{i}]
  =
-\frac{1}{16\pi G_{N}^{(d)}}  \int_{\Sigma^{d-2}}\mathfrak{h}^{(p)}_{i}
  \wedge\left[\star G -\gamma (N+1) \underbrace{G\wedge \cdots
      \wedge G}_{\text{N-1 times}}\wedge V\right]\,.
\end{equation}

Observe that the value of $\mathcal{Q}_{i}$ does not change when we add to the
harmonic form $\mathfrak{h}^{(p)}_{i}$ an exact $p$-form $de^{(p-1)}$ because,
by assumption, $\partial\Sigma^{d-2}=\emptyset$.

The on-shell closure of $\mathbf{Q}[\mathfrak{h}^{(p)}_{i}]$ guarantees that 
the $\mathcal{Q}_{i}$s satisfy Gauss laws. 

Except for $p=0$, the manifold has to be topologically non-trivial for these
charges to exist. If that is the case, Hodge's theorem tells us that we can
rewrite them in the form

\begin{equation}
  \mathcal{Q}_{i}
  =
-\frac{1}{16\pi G_{N}^{(d)}}  \int_{\mathcal{C}^{d-p-2}}
  \left[\star G -\gamma (N+1) \underbrace{G\wedge \cdots
      \wedge G}_{\text{N-1 times}}\wedge V\right]\,,
\end{equation}

\noindent
where where $\mathcal{C}^{d-p-2}$ is the $(d-p-2)$-cycle dual to the harmonic
$p$-form $\mathfrak{h}^{(p)}_{i}$ in $\Sigma^{d-2}$.

\subsubsection{Magnetic charges}
\label{sec-magneticchargesCS}

In the standard (``electric'') formulation of the theory in terms of the
$(p+1)$-form field $V$, one can define off-shell conserved $(d-2)$-form
(``magnetic'') charges which are not associated to any manifest local symmetry
in topologically non-trivial manifolds that admit harmonic $(d-p-4)$-forms.
It is possible to relate them to the gauge symmetry of the (``magnetic'')
$(d-p-3)$-form $\tilde{V}$ dual to $V$, but a democratic formulation of the
theory is needed to make this fact manifest,\footnote{See, for instance,
  Ref.~\cite{Fernandez-Melgarejo:2023kwk} and references therein.}  since in
general there is no way to rewrite the action Eq.~(\ref{eq:CStheoryaction})
in terms of the $(d-p-3)$-form $\tilde{V}$ only.

For each harmonic $(d-p-4)$-form $\mathfrak{h}^{(d-p-4)}_{i}$ we can construct
a magnetic $(d-2)$-form charge

\begin{equation}
  \label{eq:magneticcharged-2form}
  \mathbf{P}[\mathfrak{h}^{(d-p-4)}]
  \equiv
   \frac{1}{16\pi G_{N}^{(d)}} \mathfrak{h}^{(d-p-4)} \wedge G\,,
\end{equation}

\noindent
which is trivially closed off-shell. Notice that, if we can write $G=dV$
globally, this charge is automatically a total derivative and its integral over
closed $(d-2)$-dimensional surfaces will vanish.

One may construct more general closed charges, of ranks different from
$(d-2)$, ($p+q+2$, say) using harmonic $q$-forms in the obvious way:

\begin{equation}
  \label{eq:Rcharges}
  \mathbf{R}^{(p+q+2)}[\mathfrak{h}^{(q)}]
  \equiv
   \frac{1}{16\pi G_{N}^{(d)}} \mathfrak{h}^{(q)} \wedge G\,.
\end{equation}

A particularly interesting value of $q$ is the one for which $p+q+2=d-1$,
\textit{i.e.}~$q=d-p-3$. This is the rank of the potential dual to $V$:
$\tilde{p}=d-p-4$ is the dimension of the objects dual to the $p$-branes whose
worldvolumes $V$ couples to and the dual field coupling to them would be a
$\tilde{p}+1=(d-p-3)$-form potential.

Although $V$ cannot be completely replaced by $\tilde{V}$ in the type of theories
we are considering, it is interesting to study the field strength and gauge
symmetries of the latter.

The equation of motion of $V$, Eq.~(\ref{eq:EVCS}), can be written as a total
derivative

\begin{equation}
    \mathbf{E}
    =
      -d\left[\star G
      -(N+1)\gamma \underbrace{G\wedge \cdots \wedge G}_\text{N-1 times}\wedge
      V\right]\,,  
\end{equation}

\noindent
that can be locally solved by identifying the expression in brackets with
$d\tilde{V}$. Then, we can write

\begin{equation}
  \star G
  =
  d\tilde{V}
      +(N+1)\gamma \underbrace{G\wedge \cdots \wedge G}_\text{N-1 times}\wedge
      V
      \equiv \tilde{G}\,.
\end{equation}

$\tilde{G}$ is the $(\tilde{p}+2)$-form field strength of the
$(\tilde{p}+1)$-form $\tilde{V}$ and it is gauge invariant because $\star G$
is. The gauge transformations that leave it invariant are

\begin{subequations}
  \begin{align}
    \label{eq:deltaLambdatilde}
    \delta_{\tilde{\Lambda}}\tilde{V}
    & =
      d\tilde{\Lambda}\,,
    \\
    & \nonumber \\
    \label{eq:deltaLambdaontildeV}
    \delta_{\Lambda}\tilde{V}
    & =
      -(N+1)\gamma \underbrace{G\wedge \cdots \wedge G}_\text{N-1 times}\wedge
      \Lambda\,,
  \end{align}
\end{subequations}

\noindent
where $\tilde{\Lambda}$ is an arbitrary $\tilde{p}$-form.

The equation of motion now becomes the Bianchi identity of $\tilde{G}$

\begin{equation}
  \label{eq:tildeGBianchi}
  d\tilde{G}
  -(N+1)\gamma \underbrace{G\wedge \cdots \wedge G}_\text{N
    times}\,,
  =
  0\,,
\end{equation}

\noindent
and the Bianchi identity of $G$ becomes the equation of motion of $\tilde{V}$

\begin{equation}
  \label{eq:tildeVequationofmotion}
  d\star \tilde{G}
  =
  0\,.
\end{equation}

It is not difficult to see that the magnetic $(d-2)$-form charge
Eq.~(\ref{eq:magneticcharged-2form}) is nothing but the electric charge
associated to $\tilde{V}$. As for the $\mathbf{R}$ charges in
Eq.~(\ref{eq:Rcharges}), we will find an interpretation for it in
Section~\ref{sec-globalhigher1}.

\subsection{Local Lorentz transformations}
\label{sec-locallorentzCS}

The action Eq.~(\ref{eq:CStheoryaction}) is exactly invariant under local
Lorentz transformations which only act on the Vielbein

\begin{equation}
  \delta_{\sigma}e^{a}
  =
  \sigma^{a}{}_{b}e^{b}\,,
  \hspace{1cm}
  \sigma^{ab}
  =
  -\sigma^{ba}\,,
\end{equation}

\noindent
and on the fields derived from it

\begin{equation}
  \delta_{\sigma}\omega^{ab}
  =
  \mathcal{D}\sigma^{ab}\,,
  \hspace{1cm}
  \delta_{\sigma}R^{ab}
  =
  -2\sigma^{[a}{}_{c}R^{b]c}\,.
\end{equation}

Particularizing Eq.~(\ref{eq:generalvariationCStheory}) to this symmetry we
find the Noether identity

\begin{equation}
  \label{eq:LorentzNoetheridentity}
  \mathbf{E}^{[a}\wedge e^{b]}
  =
  0\,,
\end{equation}

\noindent
which asserts the symmetry of the energy-momentum tensor, and the closed
$(d-1)$-form current

\begin{equation}
  \mathbf{J}[\sigma]
  =
  -\star (e^{a}\wedge e^{b})\wedge \mathcal{D}\sigma_{ab}
  =
  d\mathbf{Q}_{L}[\sigma]\,,
\end{equation}

\noindent
with the Noether $(d-2)$-form charge

\begin{equation}
  \label{eq:LorentzNoetherchargeCS}
  \mathbf{Q}_{L}[\sigma]
  =
\frac{(-1)^{d-1}}{16\pi G_{N}^{(d)}} \star (e^{a}\wedge e^{b})\sigma_{ab}\,.
\end{equation}

As we have stressed elsewhere, this charge is closed for covariantly constant
Lorentz parameters $\mathcal{D}\sigma^{ab}=0$, which is the condition that
they must satisfy to leave invariant the spin connection. It is fortunate that
the invariance of the Vielbein is not required, as it is impossible to
fulfill this condition.

On the other hand, for particular, not necessarily covariantly-constant,
values of $\sigma^{ab}$ such as the \textit{Killing bivector}

\begin{equation}
  \label{eq:Killingbivectordef}
  P_{k}{}^{ab}
  \equiv
  \nabla^{a}k^{b}\,,
\end{equation}

\noindent
where $k$ is a Killing vector, the above charge plays an important role in
black-hole thermodynamics: $\mathbf{Q}_{L}[P_{k}]$ is the standard Komar
$(d-2)$-form \cite{Komar:1958wp} which is on-shell closed in pure gravity
owing to the identity

\begin{equation}
  \label{eq:integrabilityKVE}
  \mathcal{D}P_{k}{}^{ab}
  =
  -\imath_{k}R^{ab}\,.
\end{equation}

In the presence of matter, the Komar $(d-2)$-form is no longer
closed,\footnote{We will derive an on-shell closed \textit{generalized Komar
    $(d-2)$-form charge} later.} but its integral at spatial infinity still
gives (up to normalization) the value of the conserved gravitational charge
associated to $k$. Notice that in asymptotically-flat spacetimes the Killing
bivector is always covariantly constant there.

The integral of the standard Komar $(d-2)$-form over the horizon $\mathcal{H}$
gives the product of the Hawking temperature and the Bekenstein-Hawking
entropy. Notice that, if that section is the bifurcation surface
$\mathcal{BH}$ and $k$ is the Killing vector normal to the horizon (so that
$k\stackrel{\mathcal{BH}}{=}0$) the Killing bivector is always covariantly
constant there as well).

\subsection{GCTs}
\label{sec-gctsCS}

As discussed in Refs.~\cite{Elgood:2020svt,Elgood:2020nls},\footnote{See
  Ref.~\cite{Prabhu:2015vua} for a mathematically more rigorous formulation in
  terms of principal bundles which, unfortunately, does not include the gauge
  symmetry of $(p+1)$-form potentials with $p>0$.} when acting on fields with
gauge freedoms, GCTs induce gauge transformations. Thus in the infinitesimal
GCT generated by a vector field $\xi$, denoted by $\delta_{\xi}$, the Lie
derivative of the field with respect to $\xi$ must be supplemented by
\textit{induced} or \textit{compensating} gauge transformations (here
$\delta_{\sigma}$ and $\delta_{\Lambda}$) with gauge parameters
$\sigma_{\xi}{}^{a}{}_{b}$ and $\Lambda_{\xi}$ that depend on the fields as
well as on $\xi$.  Strictly speaking, these gauge parameters are only fully
determined when $\xi$ generates a symmetry of all the fields and
$\delta_{\xi}$ annihilates them all. In that case, we use the notation $\xi=k$
since, in particular, the vector field that generates the symmetry must be a
Killing vector of the metric, which does not have any gauge freedoms beyond
GCTs.\footnote{This is no longer true in Supergravity theories, where
  compensating supersymmetry transformations must also be included in
  $\delta_{\xi}$. \cite{Vandyck,Bandos:2023zbs,Bandos:2024pns,Bandos:2024rit}}

The parameters of the compensating local Lorentz transformations are given by 
\cite{kn:Lich,kn:Kos,kn:Kos2,Vandyck,Hurley:cf,Ortin:2002qb,Fatibene2011}

\begin{equation}
  \label{eq:sigmakdef}
  \sigma_{k}{}^{a}{}_{b}
  =
  \imath_{k}\omega^{a}{}_{b} -P_{k}{}^{a}{}_{b}\,,
\end{equation}

\noindent
where $P_{k}{}^{a}{}_{b}$ is the \textit{Lorentz momentum map}, defined by the
\textit{Lorentz momentum map equation}~(\ref{eq:integrabilityKVE}) which is
always\footnote{But we do not know in general if there are more solutions.}
satisfied by the so-called \textit{Killing bivector}\footnote{The Lorentz
  momentum map equation becomes the integrability condition for the Killing
  vector equation.} Eq.~(\ref{eq:Killingbivectordef}).

The parameters of the compensating $p$-form gauge transformation are given by

\begin{equation}
  \label{eq:Lambdak}
  \Lambda_{k}
  =
  \imath_{k}V -P_{k}\,,
\end{equation}

\noindent
where $P_{k}$ is the $p$-\textit{form} (``Maxwell'', for $p=0$)
\textit{momentum map} defined by the momentum map equation\footnote{In
  topologically non-trivial spacetimes this equation may include additional
  terms. In Appendix~\ref{app-generalized} we show that those terms are
  associated to generalized symmetric ansatzs of the kind used in boson star
  solutions \cite{Ballesteros:2024prz}. Thus, we will not consider them here.}

\begin{equation}
  \label{eq:pformmomentummap}
  \imath_{k}G+dP_{k} 
  =
  0\,.
\end{equation}

Then, the $\delta_{\xi}$ transformations take the general form 

\begin{equation}
  \label{eq:deltaxigeneral}
  \delta_{\xi}
  \equiv
  -\mathcal{L}_{\xi} +\delta_{\sigma_{\xi}} +\delta_{\Lambda_{\xi}}\,,
\end{equation}

\noindent
and their action on the fields of this theory 

\begin{subequations}
  \begin{align}
    \delta_{\xi}e^{a}
    & =
      -\left(\mathcal{D}\xi^{a}+P_{\xi}{}^{a}{}_{b}e^{b}\right)\,,
    \\
    & \nonumber \\
    \delta_{\xi}V
    & =
      -\left(\imath_{\xi}G+dP_{\xi}\right)\,,
  \end{align}
\end{subequations}

\noindent
is guaranteed to vanish when $\xi=k$, a parameter that generates a symmetry of
the field configuration under consideration.

Let us now derive the Noether charge associated to GCTs, also known as the
Noether--Wald charge. Under the transformations $\delta_{\xi}$, the action is
only invariant up to a total derivative with two terms: one due to the Lie
derivative of the fields and another one due to the induced $p$-form gauge
transformation:

\begin{equation}
  \delta_{\xi}S
  =
  -\int d\imath_{\xi}\mathbf{L}
  -\int d\mathbf{B}(e,V,\Lambda_{\xi})\,.
\end{equation}

On the other hand, particularizing Eq.~(\ref{eq:generalvariationCStheory}) to
the $\delta_{\xi}$ transformation, integrating by parts and using the Noether
identities Eqs.~(\ref{eq:pformNoetheridentity}) and
(\ref{eq:LorentzNoetheridentity}) plus the Noether identity

\begin{equation}
 \mathcal{D}\mathbf{E}_{a}
 -\mathbf{E}\wedge \imath_{a}G
 =
 0\,,
\end{equation}

\noindent
(which states that the covariant divergence of energy-momentum tensor vanishes
on-shell, a property which is usually (and wrongly) called ``conservation of
the energy momentum tensor'') we are left with the alternative expression for
$\delta_{\xi} S$

\begin{subequations}
  \begin{align}
  \delta_{\xi} S
   & =
  \int  d\mathbf{\Theta}'(e,V,\xi)\,.
    \\
    & \nonumber \\
    \label{eq:ThetaprimeCS}
    \mathbf{\Theta}'(e,V,\xi)
    & \equiv
      \mathbf{\Theta}(e,V,\delta_{\xi}e, \delta_{\xi} V)
    -\mathbf{E}_{a}\xi^{a}
     -\mathbf{E}\wedge P_{\xi}\,.
  \end{align}
\end{subequations}

Now, combining these two expressions for $\delta_{\xi} S$ we get

\begin{subequations}
  \begin{align}
    d\mathbf{J}[\xi]
    & = 0\,,
    \\
    & \nonumber \\
    \label{eq:JxiCS}
    \mathbf{J}[\xi]
    & \equiv
      \mathbf{\Theta}'(e, V,\xi)
      +\imath_{\xi}\mathbf{L}
      +\mathbf{B}(V,e,\Lambda_{\xi})
      \nonumber \\
    & \nonumber \\
    & =
      d\mathbf{Q}_{NW}[\xi]\,,
    \\
    & \nonumber \\
    \label{eq:NoetherWaldchargeCStheory}
    \mathbf{Q}_{NW}[\xi]
    & =
      -\star (e^{b}\wedge e^{c})P_{\xi\, bc}
      +P_{\xi}\wedge
      \left[\star G -\gamma (N+1) \underbrace{G\wedge \cdots
      \wedge G}_\text{N-1 times}\wedge V\right]
    \nonumber \\
    & \nonumber \\
    & \hspace{.5cm}
      +\gamma \underbrace{G\wedge \cdots
      \wedge G}_\text{N-1 times} \wedge V\wedge \imath_{\xi}V\,.
  \end{align}
\end{subequations}

Is the Noether--Wald charge  $\mathbf{Q}_{NW}[\xi]$ closed? According to its
definition, Eq.~(\ref{eq:JxiCS}),

\begin{equation}
  d\mathbf{Q}_{NW}[\xi]
=
    \mathbf{J}[\xi]
    \equiv
      \mathbf{\Theta}(e, V,\delta_{\xi} e,\delta_{\xi} V)
          -\mathbf{E}_{a}\xi^{a}
      -\mathbf{E}\wedge P_{\xi}
      +\imath_{\xi}\mathbf{L}
      +\mathbf{B}(V,\Lambda_{\xi})\,,  
\end{equation}

\noindent
which in general does not vanish. The first term in the right-hand side of the
above equation vanishes for $\xi=k$, a Killing parameter (in particular, a
Killing vector) that generates transformations that annihilate the fields. The
second and third terms vanish on-shell and we are left with

\begin{equation}
  \label{eq:dQk}
  d\mathbf{Q}_{NW}[k]
  \doteq 
  \imath_{k}\mathbf{L}
  +\mathbf{B}(V,\Lambda_{k})\,.  
\end{equation}

This equation, which expresses the failure of the Noether-Wald charge to be a
closed $(d-2)$-form, can be directly used to derive Smarr formulas
\cite{Smarr:1972kt}, following Ref.~\cite{Liberati:2015xcp}. However, as
discussed in Refs.~\cite{Ortin:2021ade,Cerdeira:2025elp},\footnote{See also
  Ref.~\cite{Golshani:2024fry}, based on the results of
  Ref.~\cite{Adami:2024gdx}.} the sum of these two terms must be a total
derivative simply because the left-hand side of the identity is. If we manage
to rewrite the right-hand side as a total derivative different from
$d\mathbf{Q}_{NW}[k]$, call it $d\boldsymbol{\omega}_{k}$,

\begin{equation}
  \label{eq:generalconstructiongeneralizedKomarcharge}
\mathbf{K}[k]
\equiv
\boldsymbol{\omega}_{k}-\mathbf{Q}_{NW}[k]\,,
\end{equation}

\noindent
will be an on-shell closed $(d-2)$-form charge that we can call the
\textit{generalized Komar charge}. The derivation of the Smarr formula from an
on-shell closed generalized Komar charge along the lines of
Refs.~\cite{Bardeen:1973gs,Carter:1973rla,Magnon:1985sc,Bazanski:1990qd,Kastor:2008xb,Kastor:2010gq}
is simpler, cleaner and physically more transparent. The procedure outlined
above can be used to derive more or less systematically generalized Komar
charges for many theories
(see, \textit{e.g.}~Refs.~\cite{Mitsios:2021zrn,Meessen:2022hcg,Ortin:2022uxa}).

In absence of the term $\mathbf{B}(V,e,\Lambda_{k})$, a prescription for
finding $\boldsymbol{\omega}_{k}\neq \mathbf{Q}_{NW}[k]$ has been given in
Ref.~\cite{Ortin:2024mmg} which is based on the observation that
Eq.~(\ref{eq:dQk}) has been obtained by computing $\imath_{k}\mathbf{L}$ first
and then using the equations of motion. In this order, these operations give
$d\mathbf{Q}_{NW}[k]$. In the reverse order, though, they give a formally different
expression, $d\boldsymbol{\omega}_{k}$, which, on-shell, should be
identical. By construction, the difference between $d\boldsymbol{\omega}_{k}$
and $d\mathbf{Q}_{NW}[k]$, which we have called $d\mathbf{K}[k]$ must vanish
on-shell, which proves that the generalized Komar charge defined as
$\mathbf{K}[k] = \boldsymbol{\omega}_{k}-\mathbf{Q}_{NW}[k]$ is closed on-shell.

The calculation of $\mathbf{L}$ on-shell and of $\boldsymbol{\omega}_{k}$ is
greatly simplified when there is a global transformation of the fields of the
theory that rescales $\mathbf{L}$ \cite{Cerdeira:2025elp}. The
Einstein--Hilbert term in Eq.~(\ref{eq:CStheoryaction}) scales with weight
$(d-2)$ under

\begin{equation}
  \label{eq:deltasea}
  \delta^{s}_{\lambda}e^{a}
  =
  \lambda e^{a}\,.
\end{equation}

The kinetic term of $V$ does not transform with weight $(d-2)$ under
$\delta^{s}_{\lambda}$ unless it also rescales $V$ with weight $p+1$, that is

\begin{equation}
  \label{eq:deltasV}
\delta^{s}_{\lambda} V = (p+1)\lambda V\,.
\end{equation}

Then, the CS term rescales with weight $(d-2)$ only for $N=2$ unless we also
rescale the dimensionful (for $N\neq 2$) coupling constant $\gamma$, which can
be done if we promote it to a scalar field and add a Lagrange-multiplier term
that enforces the constraint $d\gamma=0$. We will not pursue this possibility
here. As mentioned before, this means that the dimension $d$ has to be odd and $p$ even. 

Thus, for $N=2$, particularizing for these transformations
Eq.~(\ref{eq:generalvariationCStheory}), we find after setting $\lambda=1$

\begin{equation}
  \begin{aligned}
  (d-2)\mathbf{L}
  & =
  \mathbf{E}_{a}\wedge e^{a}
  +(p+1)\mathbf{E}\wedge V
    +d\mathbf{\Theta}(e,V,\delta^{s}_{\lambda} e,\delta^{s}_{\lambda} V)
    \\
    & \\
    & \doteq
      d\mathbf{\Theta}(e,V,\delta^{s}_{\lambda} e,\delta^{s}_{\lambda} V)
    \\
    & \\
    & =
      d\left[-(p+1) V\wedge \star G\right]\,,
  \end{aligned}
\end{equation}

\noindent
and\footnote{Notice that $N=2$ implies $p+1 =(d-2)/3$. Furthermore, since $p$
  must be even, $V\wedge V=0$.}

\begin{subequations}
  \begin{align}
    \mathbf{L}
    & \doteq
      d\mathbf{J}^{0}\,,
    \\
    & \nonumber \\
    \label{eq:J0CS}
    \mathbf{J}^{0}
    &  =
      -\tfrac{1}{3} V\wedge \star G\,.
  \end{align}
\end{subequations}

Since, by assumption, $\delta_{k}$, defined in Eq.~(\ref{eq:deltaxigeneral}),
annihilates all the field of the theory, it also annihilates
$\mathbf{J}^{0}$, which leads to


\begin{equation}
  \begin{aligned}
    \imath_{k}\mathbf{L}+\mathbf{B}(V,\Lambda_{k})
    & \doteq 
      -d\imath_{k}\mathbf{J}^{0}
      +\delta_{\Lambda_{k}}\mathbf{J}^{0}
      +\mathbf{B}(V,\Lambda_{k})
    \\
    & \\
    & \doteq
      d\boldsymbol{\omega}_{k}\,,
  \end{aligned}
\end{equation}

\noindent
where

\begin{equation}
  \label{eq:omegak}
  \boldsymbol{\omega}_{k}
  =
  \tfrac{1}{3} \tilde{P}_{k}\wedge G
  +\tfrac{1}{3}P_{k} \wedge\star G
  -3\gamma P_{k}\wedge  G\wedge V+\gamma G\wedge V\wedge \imath_{k}V\,,
\end{equation}

\noindent
after use of the magnetic momentum map equation
Eq.~(\ref{eq:magneticmomentummapequationCS}), which assumes the absence of
harmonic $(\tilde{p}+1)$-forms.

Then, the generalized Komar $(d-2)$-form charge is

\begin{equation}
  \label{eq:generalizedKomarCS}
  \mathbf{K}[k]
  =
  \star (e^{a}\wedge e^{b})P_{k\, ab}
  +\tfrac{1}{3} \tilde{P}_{k}\wedge G
  -\tfrac{2}{3}P_{k} \wedge\star G\,.  
\end{equation}

This expression is manifestly gauge invariant but not all the terms that occur
in it have the form of products of the integrands of conserved charges and
potentials that satisfy generalized zeroth laws in black-hole spacetimes, which
is the most convenient form for deriving Smarr formulas.  However, it can be rewritten in
a completely equivalent form that exhibits that structure that was first found
in Refs.~\cite{Elgood:2020mdx,Elgood:2020nls}:

\begin{equation}
  \label{eq:generalizedKomarCS2}
  \mathbf{K}[k]
  =
\frac{1}{16\pi G_{N}^{(d)}}\left\{  \star (e^{a}\wedge e^{b})P_{k\, ab}
  +\tfrac{1}{3} \left(\tilde{P}_{k}-6\gamma P_{k}\wedge V\right) \wedge G
  -\tfrac{2}{3}P_{k} \wedge\left(\star G -3\gamma G\wedge V\right)
  \right\}\,,  
\end{equation}

\noindent
or, using the definitions of the Lorentz, electric and magnetic charges
defined, respectively, in Eqs.~(\ref{eq:LorentzNoetherchargeCS}),
(\ref{eq:QLambdaharmonic}) and (\ref{eq:magneticcharged-2form}) and for
particular values of the Lorentz parameter and (not harmonic) $p$- and
$\tilde{p}$-forms 

\begin{equation}
      \label{eq:generalizedKomarCS3}
  \mathbf{K}[k]
  =
  \mathbf{Q}_{L}[P_{k}]
  +\tfrac{1}{3} \mathbf{P}[\tilde{P}_{k}-6\gamma P_{k}\wedge V]
  +\tfrac{2}{3}  \mathbf{Q}[P_{k}]\,.  
\end{equation}

Indeed, if $k$ is the Killing vector normal to the event horizon, in the
bifurcation surface $\mathcal{BH}$ in which $k=0$, using the momentum map
equations

\begin{subequations}
  \begin{align}
  \mathcal{D}P_{k}{}^{ab}
  & =
  -\imath_{k}R^{ab}
  \stackrel{\mathcal{BH}}{=}
    0\,,
    \\
    & \nonumber \\
  dP_{k}
  & =
  -\imath_{k}G
  \stackrel{\mathcal{BH}}{=}
  0\,,
    \\
    & \nonumber \\
  d\left(\tilde{P}_{k}-6\gamma P_{k}\wedge V  \right)
  & \stackrel{\mathcal{BH}}{=}  
  d\tilde{P}_{k}-6\gamma P_{k}\wedge G
  =
  -\imath_{k}\star G
  \stackrel{\mathcal{BH}}{=}  
  0\,.  
  \end{align}
\end{subequations}

Finally, let us consider the effect of the transformations
Eqs.~(\ref{eq:Ctransformations}), (\ref{eq:CtransformationsontildeP}) and
(\ref{eq:Dtransformations}), associated to the ambiguities of the solutions of
the momentum map equations, on the generalized Komar charge
Eq.~(\ref{eq:generalizedKomarCS}):

\begin{equation}
  \begin{aligned}
  \left(\delta_{C}+\delta_{D}\right)
    \mathbf{K}[k]
  & =
\frac{1}{16\pi G_{N}^{(d)}}\left\{\tfrac{1}{3} D \wedge G 
    -\tfrac{2}{3}C \wedge\left(\star G -3\gamma G\wedge V\right)\right\}
    \\
    & \\
    & =
\tfrac{1}{3}  \mathbf{P}[D]
    +\tfrac{2}{3} \mathbf{Q}[C]\,,
  \end{aligned}
\end{equation}

\noindent
where $ \mathbf{P}[D]$ and $ \mathbf{Q}[C]$ are the magnetic and electric
$(d-2)$-form charges defined in Eqs.~(\ref{eq:magneticcharged-2form}) and
Eq.~(\ref{eq:QLambda}), respectively, associated to the closed $(d-p-4)$- and
$p$-forms $D$ and $C$. Only the harmonic component of these closed forms will
give a non-vanishing contribution when integrated on a closed
$(d-2)$-surface. Judicious choices of $D$ and $C$ can be used to make
$ \mathbf{K}[k]$ give the ADM mass when integrated over the $(d-2)$-sphere at
spatial infinity \cite{Zatti:2024vqv} in asymptotically-flat spacetimes. In
asymptotically-KK spacetimes it will be necessary to combine $ \mathbf{K}[k]$
with another conserved charge \cite{Barbagallo:2025fkg}. We will discuss this
point in more detail in the context of the dimensional reduction of minimal
5-dimensional supergravity to 4 dimensions.  The addition of other closed
$(d-2)$-form charges to the generalized Komar charge will not modify the Smarr
formula, but the derivation will be far more transparent when the charge that
we integrate at infinity only gives the ADM mass.

What happens when the manifolds we consider admit harmonic
$(\tilde{p}+1)$-forms and the magnetic momentum map equation takes the form
Eq.~(\ref{eq:magneticmomentummapequationCS2})? It is not difficult to see that
$\boldsymbol{\omega}_{k}$ and, therefore, the generalized Komar charge
$\mathbf{K}[k]$ in Eq.~(\ref{eq:generalizedKomarCS}) gets an additional term
(ignoring again the overall normalization term $16\pi G_{N}^{(d)}$)

\begin{equation}
 +\tfrac{1}{3}V \wedge \mathfrak{h}^{(\tilde{p}+1)}\,.
\end{equation}

In this case is convenient to introduce the notation

\begin{equation}
  \label{eq:Kkh}
  \mathbf{K}_{\mathfrak{h}}[k]
  =
  \mathbf{K}[k] +\tfrac{1}{3}V \wedge \mathfrak{h}^{(\tilde{p}+1)}\,,
\end{equation}

\noindent
for the generalized Komar charge, where $\mathbf{K}[k]$, the charge defined in
Eq.~(\ref{eq:generalizedKomarCS}), is no longer on-shell closed. If $V$ is not
globally defined, $\mathbf{K}_{\mathfrak{h}}[k]$ will not be globally defined,
either. Furthermore, $\mathbf{K}_{\mathfrak{h}}[k]$ is only gauge-invariant up
to a total derivative. 

It is possible to work with $\mathbf{K}_{\mathfrak{h}}[k]$ if one takes into
account all these subtleties but there is a better, globally defined
expression that does not assume that, globally, $G=dV$, namely

\begin{equation}
  \label{eq:generalizedKomarcharged5h}
  d\mathbf{K}[k]
  +\tfrac{1}{3} G\wedge \mathfrak{h}^{(\tilde{p}+1)}
  =
  0\,,
\end{equation}

\noindent
where, again, $\mathbf{K}[k]$ is the charge given by the globally defined and
gauge-invariant expression Eq.~(\ref{eq:generalizedKomarCS}). This is the
equation obtained in Ref.~\cite{Gibbons:2013tqa} in the context of
5-dimensional supergravity, in which $\tilde{p}+1=p+2=2$ and, in some
supersymmetric solutions,

\begin{equation}
  \mathfrak{h}^{(2)}
  =
  3^{1/2}\alpha G\,,
\end{equation}

\noindent
for some constant $\alpha$.




In Section~\ref{sec-massandangularmomentum}, we are going to show that, in
minimal 5-dimensional supergravity with KK boundary conditions, for
configurations admitting a Killing vector $l$

\begin{equation}
  \mathfrak{h}^{(2)}
  =
  \tfrac{1}{\sqrt{3}}\mathbf{Q}_{l\, -}\,,
\end{equation}

\noindent
where $\mathbf{Q}_{l\, -}$ is the 4-dimensional on-shell-closed 2-form scalar
charge associated a particular isometry of the scalar manifold and to the
isometry of the solutions generated by $l$ and it is explicitly given in
Eq.~(\ref{eq:Ql-}).

Finally, notice that the second term in
Eq.~(\ref{eq:generalizedKomarcharged5h}) has the form of the topologically
closed $\mathbf{R}^{(d-1)}[\mathfrak{h}^{(\tilde{p}+1)}]$ charges defined in
Eq.~(\ref{eq:Rcharges}) and mentioned in
Section~\ref{sec-magneticchargesCS}. We will show that this charge is
associated to a higher-form symmetry in Section~\ref{sec-globalhigher1} (see
Eq.~(\ref{eq:dualhigherformcurrent}) and also the discussion in
Appendix~\ref{app:magneticmomentummap}).

\subsection{Global higher-form symmetries I}
\label{sec-globalhigher1} 

If the manifold admits harmonic $(p+1)$-forms $\mathfrak{h}^{(p+1)}_{i}$, the
theory Eq.~(\ref{eq:CStheoryaction}) is also invariant under global
transformations of the form

\begin{equation}
  \label{eq:globalhigherformp}
  \delta_{\eta}V
  =
  \eta^{i} \mathfrak{h}^{(p+1)}_{i}\,,
\end{equation}

\noindent
for infinitesimal, constant, parameters $\eta^{i}$, up to a total derivative

\begin{equation}
  \delta_{\eta}S
  =
\frac{1}{16\pi G_{N}^{(d)}}  \int d\left[\eta^{i}\gamma
    \underbrace{G\wedge \cdots \wedge G}_\text{N-1 times}\wedge
    V\wedge \mathfrak{h}^{(p+1)}_{i}
  \right]\,.
\end{equation}

Thus, the $(d-1)$-form Noether currents

\begin{equation}
  \mathbf{J}_{\eta_{i}}
  =
\frac{1}{16\pi G_{N}^{(d)}}\left\{
  \star G
    -\gamma (N+1) \underbrace{G\wedge \cdots
      \wedge G}_\text{N-1 times}\wedge V
    \right\}\wedge \mathfrak{h}^{(p+1)}_{i}\,,
\end{equation}

\noindent
are conserved (closed) on-shell.

This current is similar to the $(d-2)$-form charge
$\mathbf{Q}[\mathfrak{h}^{(p)}_{i}]$ defined in Eqs.~(\ref{eq:QLambda}) and
(\ref{eq:QLambdaharmonic}) with the harmonic the $p$-form
$\mathfrak{h}^{(p)}_{i}$ replaced by a harmonic $(p+1)$-form
$\mathfrak{h}^{(p+1)}_{i}$.

Even though we do not have an action for the dual $(\tilde{p}+1)$-form field
$\tilde{V}$ introduced in Section~\ref{sec-magneticchargesCS}, it is not
difficult to see that, if the manifolds on which we work have harmonic
$(\tilde{p}+1)$-forms $\mathfrak{h}^{(\tilde{p}+1)}_{i}$ the equations of motion of
that field Eq.~(\ref{eq:tildeVequationofmotion}) will be invariant under
further global higher-form symmetries

\begin{equation}
  \delta_{\tilde{\eta}} \tilde{V}
  =
  \tilde{\eta}^{i}\mathfrak{h}^{(\tilde{p}+1)}_{i}\,,
\end{equation}

\noindent
and it is not difficult to guess that the on-shell\footnote{On-shell from the
  $\tilde{V}$ point of view, off-shell from the $V$ point of view.} closed
$(d-1)$-form currents associated to them will be

\begin{equation}
  \label{eq:dualhigherformcurrent}
  \tilde{\mathbf{J}}_{\tilde{\eta}_{i}}
  =
  \frac{1}{16\pi G_{N}^{(d)}}\mathfrak{h}^{(\tilde{p}+1)}_{i}\wedge \star \tilde{G}
  =
  \frac{1}{16\pi G_{N}^{(d)}}\mathfrak{h}^{(\tilde{p}+1)}_{i}\wedge G
  =
  \mathbf{R}^{(d-1)}[\mathfrak{h}^{(\tilde{p}+1)}]\,.
\end{equation}

These are, precisely, the $(d-1)$-form $\mathbf{R}$-charges
Eq.~(\ref{eq:Rcharges}) mentioned in Section~\ref{sec-magneticchargesCS}.

\subsection{Global higher-form symmetries II}
\label{sec-globalhigher2}

We have seen in Section~\ref{sec-gctsCS} that in the  $N=2$ case there is a
global transformation $\delta^{s}_{\lambda}$ defined in Eqs.~(\ref{eq:deltasea})
and (\ref{eq:deltasV}) that rescales the action with weight $(d-2)$. With
a second, independent, transformation that rescales the action we could
define a symmetry of the action.

Let us first focus on the EH term. It is a well-established fact that, with
standard boundary conditions and for $d>2$, it does not have any global
symmetries \cite{Torre:1993jm,Anderson:1994eg}.  However, when the topology of
the manifolds we are considering admits a harmonic 1-form $\mathfrak{h}^{(1)}$
and the metrics under consideration admit a Killing vector $k$ (KK boundary
conditions), there is a second global transformation of the Vielbein

\begin{equation}
  \label{eq:deltahea}
   \delta^{h}_{\epsilon}e^{a}
     =
     -\epsilon  \mathfrak{h}^{(1)} \imath_{k} e^{a}\,,
\end{equation}

\noindent
that also rescales the EH term

\begin{equation}
  \delta^{h}_{\epsilon}S_{\rm EH}
  =
  -\epsilon S_{\rm EH}\,,
\end{equation}

\noindent
and which will allow us to define a symmetry of the EH term.

It is not difficult to derive this result using\footnote{The transformation of
  the curvature tensor follows from the Palatini identity
  $\delta R^{ab} = \mathcal{D}\delta\omega^{ab}$ and the transformation of the
  connection
  \begin{equation}
   \delta^{h}_{\epsilon}\omega^{ab}
     =
      -\epsilon P_{k}{}^{ab}\mathfrak{h}^{(1)}\,,
    \end{equation}
    where $P_{k}{}^{ab}$ has been defined in Eq.~(\ref{eq:Killingbivectordef})
    and satisfies Eq.~(\ref{eq:integrabilityKVE}).
  }

\begin{equation}
        \delta^{h}_{\epsilon}R^{ab}
     =
  -\epsilon \mathfrak{h}^{(1)} \wedge \imath_{k}R^{ab}\,.
\end{equation}

\noindent
Ignoring the normalization factor $(16\pi G_{N}^{(d)})^{-1}$

\begin{equation}
  \begin{aligned}
    \delta^{h}_{\epsilon}S_{\rm EH}
    & =
      -\epsilon \int_{\mathcal{M}} 
  \frac{(-1)^{d-1}}{(d-2)!}\varepsilon_{a_{1}\cdots a_{d-2}bc}
      \left\{
      (d-2) \mathfrak{h}^{(1)} \imath_{k} e^{a_{1}}
      \wedge \cdots \wedge e^{a_{d-2}}
      \wedge R^{bc}
      \right.
    \\
    & \\
    & \hspace{.5cm}
      \left.
      +e^{a_{1}}\wedge \cdots \wedge e^{a_{d-2}}
      \wedge  \mathfrak{h}^{(1)} \wedge \imath_{k}R^{bc} 
      \right\}
    \\
    & \\
    & =
      -\epsilon \int_{\mathcal{M}} 
  \frac{(-1)^{d-1}}{(d-2)!}\varepsilon_{a_{1}\cdots a_{d-2}bc}
      \left\{
       \mathfrak{h}^{(1)}\wedge \imath_{k} \left(e^{a_{1}}
      \wedge \cdots \wedge e^{a_{d-2}}\right)
      \wedge R^{bc}
      \right.
    \\
    & \\
    & \hspace{.5cm}
      \left.
      +(-1)^{d}\mathfrak{h}^{(1)}\wedge e^{a_{1}}\wedge \cdots \wedge e^{a_{d-2}}
      \wedge \imath_{k}R^{bc} 
      \right\}
    \\
    & \\
    & =
      -\epsilon \int_{\mathcal{M}}  \mathfrak{h}^{(1)}\wedge \imath_{k}\mathbf{L}
    \\
    & \\
    & =
      -\epsilon S_{\rm EH}\,,
  \end{aligned}
\end{equation}

\noindent
assuming the normalization\footnote{If $\imath_{k}\mathfrak{h}^{(1)}=0$ we get
  a symmetry. With KK boundary conditions in a single direction, though,
  generically we will have $\imath_{k}\mathfrak{h}^{(1)}\neq 0$.}

\begin{equation}
  \label{eq:kh1normalization}
  \imath_{k}\mathfrak{h}^{(1)}
  =
  1\,.
\end{equation}

Thus, the EH action is exactly invariant under a combination of these two
transformations with $\lambda=\epsilon/(d-2)$ \cite{Gomez-Fayren:2024cpl}

\begin{equation}
  \label{eq:deltaepsilonea}
  \delta_{\epsilon}
  \equiv
  \delta^{h}_{\epsilon}
  +\frac{1}{d-2}\delta^{s}_{\epsilon}\,,
  \hspace{1cm}
  \delta_{\epsilon}S_{\rm EH}
  =
  0\,.
\end{equation}

In order to extend the symmetry Eq.~(\ref{eq:deltaepsilonea}) to the
action Eq.~(\ref{eq:CStheoryaction}) we must find a generalization of the
transformation $\delta^{h}_{\epsilon}$ acting on $(p+1)$-forms $V$.  A natural
proposal is

\begin{equation}
  \label{eq:deltahV}
  \delta^{h}_{\epsilon} V
  =
  -\epsilon \mathfrak{h}^{(1)}\wedge \imath_{k}V \,.
\end{equation}

On the field strength, using KK boundary
conditions\footnote{\label{foot:KKassumption} In
  particular, we assume that
  \begin{equation}
    \label{eq:LkVassumption}
  \mathcal{L}_{k}V
  =
  0\,,
  \,\,\,\,\,\,\,
  \Rightarrow
  \,\,\,\,\,\,\,
    \delta_{k}V
=
          0\,,
          \,\,\,\,
          \text{and}
          \,\,\,\,
      P_{k}
      = 
      \imath_{k}V\,.
    \end{equation}
    See Section~\ref{sec-KKreduction} and the discussion around
    Eq.~(\ref{eq:LkV=0}).
  }

\begin{subequations}
  \begin{align}
  \delta^{h}_{\epsilon} G
& =
 -\epsilon \mathfrak{h}^{(1)}\wedge \imath_{k}G\,,
    \\
    & \nonumber \\
  \delta^{h}_{\epsilon} \star G
& =
 -\epsilon \mathfrak{h}^{(1)}\wedge \imath_{k} \star G\,,
  \end{align}
\end{subequations}

\noindent
and it is not difficult to arrive to the result that, under the
transformations Eq.~(\ref{eq:deltahea}) and (\ref{eq:deltahV}) the whole
action scales homogeneously

\begin{equation}
  \delta^{h}_{\epsilon} S
   =
    -\epsilon\int \mathfrak{h}^{(1)}\wedge \imath_{k}\mathbf{L}
 =
  -\epsilon S\,,
\end{equation}

\noindent
and the whole action Eq.~(\ref{eq:CStheoryaction}) with $N=2$ is invariant
under the global transformations

\begin{subequations}
  \label{eq:deltaepsilonp+1forms}
  \begin{align}
  \delta_{\epsilon}e^{a}
    & \equiv
      -\epsilon \left(\mathfrak{h}^{(1)}\imath_{k}e^{a}
      -\tfrac{1}{d-2}e^{a}\right)\,,
    \\
    & \nonumber \\
    \delta_{\epsilon}V
    & \equiv
      -\epsilon \left(\mathfrak{h}^{(1)}\wedge \imath_{k}V
      -\tfrac{1}{3}V\right)\,.
  \end{align}
\end{subequations}

The on-shell closed Noether $(d-1)$-form current associated to this symmetry
is

\begin{equation}
  \begin{aligned}
    \mathbf{J}_{\epsilon}
    & =
      \star (e^{a}\wedge e^{b}) P_{k\, ab} \wedge\mathfrak{h}^{(1)}
      -\left(\mathfrak{h}^{(1)}\wedge \imath_{k}V
      -\tfrac{1}{3}V\right)\wedge\left[ -\star G +2\gamma G \wedge V\right]
    \\
    & \\
    & =
      \left[ \star (e^{a}\wedge e^{b}) P_{k\, ab}
      -P_{k}\wedge\left(\star G -2\gamma G\wedge V\right)\right]
      \wedge\mathfrak{h}^{(1)}
      -\tfrac{1}{3}V\wedge\star G
    \\
    & \\
    & =
      -\mathbf{Q}_{NW}[k] \wedge\mathfrak{h}^{(1)}
      +\mathbf{J}^{0}\,,
  \end{aligned}
\end{equation}

\noindent
where $\mathbf{Q}_{NW}[k]$ is the Noether-Wald charge associated to the
Killing vector $k$ and $\mathbf{J}^{0}$ is the $(d-1)$-form in
Eq.~(\ref{eq:J0CS}) whose exterior derivative gives the on-shell Lagrangian.

\section{Minimal 5-dimensional Supergravity}
\label{sec-thetheory}

As we have mentioned in the preceding section, the bosonic sector of minimal
5-dimensional supergravity is a particular case of the theory
Eq.~(\ref{eq:CStheoryaction}) with

\begin{equation}
  d=5\,,
  \hspace{.5cm}
  p=0\,,
  \hspace{.5cm}
  N=2\,,
  \hspace{.5cm}
 \gamma=\tfrac{1}{3^{3/2}}\,. 
\end{equation}

Thus, the action is

\begin{equation}
\label{eq:minimalN1d5action}
\begin{aligned}
  S[e^{a},V]
  & =
  \frac{1}{16\pi G_{N}^{(5)}} \int 
  \left[\star (e^{a}\wedge e^{b}) \wedge R_{ab}
    -\tfrac{1}{2}G\wedge \star G
        +\tfrac{1}{3^{3/2}} G\wedge G \wedge V
      \right]
        \\
  & \\
  & \equiv
  \int \mathbf{L}\,,
\end{aligned}
\end{equation}

\noindent
and the equations of motion and presymplectic potential are 

\begin{subequations}
  \begin{align}
    \label{eq:Ea}
  \mathbf{E}_{a}
  & =
     \imath_{a}\star (e^{c}\wedge e^{d})\wedge R_{cd}
      +\tfrac{1}{2}\left(\imath_{a}G\wedge \star G
      -G\wedge \imath_{a}\star G
      \right)\,,
    \\
    & \nonumber \\
    \label{eq:E}
    \mathbf{E}
    & =
      -d\star G+\tfrac{1}{3^{1/2}}G\wedge G\,,
    \\
    & \nonumber \\
   \label{eq:Theta}
    \mathbf{\Theta}(e,V,\delta e,\delta V)
    & =
    -\star (e^{a}\wedge e^{b})\wedge \delta \omega_{ab}
    +\left(\star G
    -\tfrac{2}{3^{3/2}}G\wedge V\right)\wedge \delta V\,,
 \end{align}
 \end{subequations}

\noindent
where we have omitted the overall normalization factor of
$(16\pi G_{N}^{(5)})^{-1}$ in the right-hand side of all these expressions.

We can use the general results obtained in the previous section to write the
definition of the electric, magnetic and Lorentz 3-form charges associated to
functions $f$, 1-forms $\omega^{(1)}$ and Lorentz parameters
$\sigma^{ab}$:

\begin{subequations}
  \label{eq:5formgaugecharges}
  \begin{align}
    \label{eq:electriccharge3form}
        \mathbf{Q}[f]
    & =
      -\frac{1}{16\pi G_{N}^{(5)}}
      f\left(\star G -\tfrac{1}{3^{1/2}}G\wedge V\right)\,,
    \\
    & \nonumber \\
    \label{eq:magneticcharge3form}
    \mathbf{P}[\omega^{(1)}]
    & =
    \frac{1}{16\pi G_{N}^{(5)}} \omega^{(1)}\wedge G\,,
    \\
    & \nonumber \\
        \label{eq:lorentzharge3form}
    \mathbf{Q}_{L}[\sigma]
    & =
    \frac{1}{16\pi G_{N}^{(5)}} \star (e^{a}\wedge e^{b})\sigma_{ab}\,.
  \end{align}
\end{subequations}

These three charges are closed when the parameters are (covariantly) closed
and the first (but not the second nor the third) also needs the equations of
motion to be satisfied. The charges are non-trivial (not exact) when the
parameters are closed, but not exact. In this case, this is only relevant for
the magnetic charge, which will only be non-trivial for $\omega^{(1)}$
harmonic. The values of the charges in a given compact spatial region are
given by the integrals over the closed 3-dimensional boundary $\Sigma^{3}$

With KK boundary conditions we always have a harmonic 1-form at our disposal,
$\mathfrak{h}^{(1)}$ and if the integration domain includes the U$(1)$ fiber
of length $2\pi \ell$

\begin{equation}
  \label{eq:magneticcharge5d}
  P[\mathfrak{h}^{(1)}]
  =
  \frac{1}{16\pi G_{N}^{(5)}}\int_{\Sigma^{3}}  \mathfrak{h}^{(1)}\wedge G
  =
  \frac{\ell }{8 G_{N}^{(5)}} \int_{\Sigma^{2}} G\,,
\end{equation}

\noindent
where $\Sigma^{2}$ is the 2-cycle dual to $\mathfrak{h}^{(1)}$ in $\Sigma^{3}$
and belongs to the 4-dimensional spacetime. Thus, this magnetic charge must be
the standard 4-dimensional magnetic charge of the 4-dimensional gauge vector
coming from $V$, which (we must stress it) is not the KK vector field, that
originates in the 5-dimensional metric.

The generalized Komar 3-form charge associated to a Killing vector $k$ has the
general form Eq.~(\ref{eq:generalizedKomarCS3})

\begin{equation}
      \label{eq:generalizedKomar5d}
  \mathbf{K}[k]
  =
  \mathbf{Q}_{L}[P_{k}]
  +\tfrac{2}{3}  \mathbf{Q}[P_{k}]
    +\tfrac{1}{3} \mathbf{P}[\tilde{P}_{k}-\tfrac{2}{3^{1/2}} P_{k}\wedge V]\,,
\end{equation}

\noindent
where the charges involved are now given in Eqs.~(\ref{eq:5formgaugecharges}).
This charge is on-shell closed in absence of harmonic 2-forms in the magnetic
momentum map Eq.~(\ref{eq:magneticmomentummapequationCS2}). As we have already
mentioned, with KK boundary conditions the magnetic momentum map equation of
minimal 5-dimensional supergravity does include an on-shell closed but not
exact 2-form. Then, we can either say that the above generalized Komar charge
is not closed on-shell but satisfies instead
Eq.~(\ref{eq:generalizedKomarcharged5h}) or we can say that the on-shell
closed generalized Komar charge is

\begin{equation}
  \label{eq:Kkhd5}
  \mathbf{K}_{\mathfrak{h}}[k]
  =
  \mathbf{K}[k] +\tfrac{1}{3}V \wedge \mathfrak{h}^{(2)}\,,
\end{equation}

\noindent
taking into account the issues mentioned in Section~\ref{sec-gctsCS}.

\subsection{Global symmetries}
\label{sec-globalhigherformsymmetriesd5}

Minimal 5-dimensional supergravity with KK boundary conditions has the two
(higher-form) global symmetries\footnote{Since there is a single harmonic
  1-form, there is only one symmetry of each kind.} discussed in
Section~\ref{sec-globalhigher1} and \ref{sec-globalhigher2} whose
transformations we have labeled $\delta_{\eta}$

\begin{equation}
  \label{eq:deltaeta5d}
  \delta_{\eta}V
  =
  \eta \mathfrak{h}^{(1)}\,,
\end{equation}

\noindent
and $\delta_{\epsilon}$

\begin{subequations}
  \label{eq:deltaepsilon5d}
  \begin{align}
  \delta_{\epsilon} e^{a}
  & =
  -\epsilon\left(\mathfrak{h}^{(1)} \imath_{k} e^{a}
    -\tfrac{1}{3}e^{a}\right)\,,
    \\
    & \nonumber \\
    \delta_{\epsilon} V
    & =
      -\epsilon \left(\mathfrak{h}^{(1)}\wedge \imath_{k}V
      -\tfrac{1}{3}V\right)\,.
\end{align}
\end{subequations}

The on-shell closed 4-form Noether currents associated to these global
symmetries take the form 

\begin{subequations}
  \begin{align}
  \label{eq:Jeta}
      \mathbf{J}_{\eta}
  & =
\frac{1}{16\pi G_{N}^{(5)}}\left\{
    \star G  -\tfrac{1}{3^{1/2}}G \wedge V \right\}\wedge \mathfrak{h}^{(1)}
    \nonumber \\
    & \nonumber \\
    & = -\mathbf{Q}[1]\wedge \mathfrak{h}^{(1)}\,,
    \\
    & \nonumber \\
  \label{eq:Jepsilon}
    \mathbf{J}_{\epsilon}
    & =
      \left[ \star (e^{a}\wedge e^{b}) P_{k\, ab}
      -P_{k}\wedge\left(\star G -\tfrac{2}{3^{3/2}} G\wedge V\right)\right]
      \wedge\mathfrak{h}^{(1)}
      -\tfrac{1}{3}V\wedge\star G
      \nonumber 
    \\
    & \nonumber \\
    & =
      -\mathbf{Q}_{NW}[k] \wedge\mathfrak{h}^{(1)}
      +\mathbf{J}^{0}\,,
  \end{align}
\end{subequations}

\noindent
where $\mathbf{Q}[1]$ is the electric charge 3-form for $f=1$,
$\mathbf{Q}_{NW}[k]$ is the Noether-Wald 3-form for the Killing vector $k$ and
$\mathbf{J}^{0}$ is the 4-form in Eq.~(\ref{eq:J0CS}) whose exterior
derivative gives the on-shell Lagrangian.

\subsubsection{Associated scalar charges}

As shown in Ref.~\cite{Ballesteros:2023iqb}, for field configurations
invariant under the GCT generated by a vector field $l$ (which, in particular,
will be a Killing vector), in general it is possible to construct
on-shell-closed $(d-2)$-form charges starting with on-shell closed
$(d-1)$-form currents. Let us see how this comes about in the current, KK,
context.

In the KK context, the 5-dimensional Killing vector $l$ will be of the form
\cite{Gibbons:1985ac,Gomez-Fayren:2023wxk,Barbagallo:2025fkg}

\begin{equation}
  \label{eq:5dKillingvectorKK}
l = \underline{l} -\chi_{\underline{l}}k\,,  
\end{equation}

\noindent
where $\underline{l}$ is assumed to be 4-dimensional Killing vector, $k$ is
the Killing vector that generates shifts along the compact dimension and the
function $\chi_{l}$ can be interpreted as the parameter of a compensating
gauge transformation of the KK vector field $A$ which satisfies\footnote{This
  equation follows from the 5-dimensional Killing vector equation
  $\delta_{l}g_{\mu\nu}=0$ satisfied by $l$.}

\begin{equation}
  \delta_{\underline{l}}A
  =
  -\mathcal{L}_{\underline{l}}A +\delta_{\chi_{\underline{l}}}A 
  =
  -\mathcal{L}_{\underline{l}}A +d\chi_{\underline{l}}
 =
  0\,.
\end{equation}

This condition and the condition that $l$ is null over the horizon imply
that $ \chi_{\underline{l}}$ is given by

\begin{equation}
  \label{eq:chil1}
  \chi_{\underline{l}}
  =
  \imath_{\underline{l}}A -P_{\underline{l}}\,,
\end{equation}

\noindent
where $P_{\underline{l}}$ satisfies

\begin{equation}
  \label{eq:chil2}
  \imath_{\underline{l}}F(A) +dP_{\underline{l}}=0\,,
  \hspace{1cm}
  P_{\underline{l}}\stackrel{\mathcal{H}}{=} 0\,.  
\end{equation}

Since our convention is to choose the additive constants in momentum maps such
that they vanish or do not contribute at infinity \cite{Zatti:2024vqv}, we are
going to use in Eq.~(\ref{eq:chil1}) $\bar{P}_{\underline{l}}$

\begin{equation}
  \label{eq:chil3}
  \chi_{\underline{l}}
  =
  \imath_{\underline{l}}A -\bar{P}_{\underline{l}}\,,
  \hspace{1cm}
  \bar{P}_{\underline{l}}
  \equiv
  P_{\underline{l}}-P_{\underline{l}}(\mathcal{H})\,,
\end{equation}

\noindent
with $P_{\underline{l}}$ vanishing at infinity and
$P_{\underline{l}}(\mathcal{H})$ the constant value of $P_{\underline{l}}$
over the horizon.

It can be argued that, under those gauge transformations

\begin{equation}
  \delta_{\chi_{\underline{l}}} \mathfrak{h}^{(1)}
  =
  -d\chi_{\underline{l}}\,,
\end{equation}

\noindent
and, therefore,

\begin{equation}
  \delta_{l}\mathfrak{h}^{(1)}
  =
  -\mathcal{L}_{l} \mathfrak{h}^{(1)}
  +\delta_{\chi_{\underline{l}}} \mathfrak{h}^{(1)}
  =
  0\,.
\end{equation}

Since, by hypothesis, the GCT generated by $l$ leaves invariant all the
fields, and, as we have just seen, it also leaves invariant
$\mathfrak{h}^{(1)}$, it will leave invariant $\mathbf{J}_{\eta}$ and we can
write\footnote{$\delta_{l}\mathbf{Q}[1]$ also vanishes, but we have to keep it
  and rewrite it in order to get a non-trivial result.}

\begin{equation}
  \begin{aligned}
    0
    & =
      \delta_{l}\mathbf{J}_{\eta}
    \\
    & \\
  & =
      -\delta_{l}\mathbf{Q}[1]\wedge \mathfrak{h}^{(1)}
    \\
    & \\
    & \doteq
      d\left(\imath_{l}\mathbf{Q}[1]\wedge \mathfrak{h}^{(1)} \right)
      -\delta_{\chi_{\underline{l}}}\mathbf{Q}[1]\wedge \mathfrak{h}^{(1)}\,,
  \end{aligned}
\end{equation}

\noindent
where we have used $d\mathbf{Q}[1] \doteq 0$ and $d\mathfrak{h}^{(1)}=0$.
Operating and using the electric momentum map Eq.~(\ref{eq:pformmomentummap})
we find the on-shell closed 3-form charge

\begin{equation}
  \mathbf{Q}_{\eta\, l}
  =
\left(\imath_{l}\mathbf{Q}[1] +\frac{1}{16\pi G_{N}^{(5)}} \chi_{l}G \right)\wedge  \mathfrak{h}^{(1)}\,.
\end{equation}

Operating and using the electric and magnetic momentum map
Eqs.~(\ref{eq:pformmomentummap}) and (\ref{eq:magneticmomentummapequationCS2})
we arrive to 

\begin{equation}
  \label{eq:Qetal} 
   \mathbf{Q}_{\eta\, l}
  =
 \mathfrak{h}^{(2)}\wedge  \mathfrak{h}^{(1)}\,.
\end{equation}

As mentioned in the paragraph below
Eq.~(\ref{eq:magneticmomentummapequationCS2}) the harmonic 2-form
$\mathfrak{h}^{(2)}$ may be associated to different on-shell closed 2-forms of
the theory. A detailed calculation leads to
Eq.~(\ref{eq:dimensionalreductionmomentummap}) and, in particular to
Eq.~(\ref{eq:h2versusQl-}) which we reproduce here for the sake of convenience

\begin{equation}
  \mathfrak{h}^{(2)}
  =
  \tfrac{e^{-2\phi_{\infty}}(16\pi G_{N}^{(4)})}{\sqrt{3}}\mathbf{Q}_{l\, -}\,,
\end{equation}

\noindent
where $\mathbf{Q}_{l\, -}$ is the 4-dimensional on-shell-closed 2-form scalar
charge associated a particular isometry of the scalar manifold and to the
isometry of the solutions generated by $l$ and it is explicitly given in
Eq.~(\ref{eq:Ql-}). An equivalent result is obtained in the dimensional
reduction of $\mathbf{Q}_{\eta\, l}$ Eq.~(\ref{eq:Qepsilonreduction}).

We can do the same starting with $\mathbf{J}_{\epsilon}$, but it is convenient
to massage the expression Eq.~(\ref{eq:Jepsilon}) a bit first. Using
the general expression Eq.~(\ref{eq:generalconstructiongeneralizedKomarcharge})

\begin{equation}
      \mathbf{J}_{\epsilon}
      =
      \mathbf{K}[k] \wedge\mathfrak{h}^{(1)}
      -\boldsymbol{\omega}_{k}\wedge\mathfrak{h}^{(1)}
      +\mathbf{J}^{0}\,,
\end{equation}

\noindent
where $\boldsymbol{\omega}_{k}$, given in Eq.~(\ref{eq:omegak}), takes the
form 

\begin{equation}
  \boldsymbol{\omega}_{k}
  =
  \tfrac{1}{3} \tilde{P}_{k}\wedge G
  +\tfrac{1}{3}P_{k} \star G
  -\tfrac{2}{3^{3/2}} P_{k} G\wedge V\,,
\end{equation}

\noindent
once we take into account Eq.~(\ref{eq:LkVassumption}). Using
Eq.~(\ref{eq:kh1normalization}), we can write

\begin{equation}
  \label{eq:otraexpresionparaJepsilon}
  \mathbf{J}_{\epsilon}
  =
  \mathbf{K}[k] \wedge\mathfrak{h}^{(1)}
  +d\left(\tfrac{1}{3}\tilde{P}_{k}\wedge V\wedge \mathfrak{h}^{(1)}  \right)
  +\imath_{k}\left( \mathbf{J}^{0}\wedge \mathfrak{h}^{(1)}\right) \,.
\end{equation}

The second term in the right-hand side of this expression is an exact 4-form
while the third is closed under the assumptions we have made:\footnote{Notice
  that $\mathbf{J}^{0}\wedge \mathfrak{h}^{(1)}$ is a volume form.}

\begin{equation}
  d\imath_{k}\left( \mathbf{J}^{0}\wedge \mathfrak{h}^{(1)}\right)
  =
  -\imath_{k}d\left( \mathbf{J}^{0}\wedge \mathfrak{h}^{(1)}\right) 
  =
  0\,.
\end{equation}

Since conserved (closed) 4-form currents are defined up to closed 4-forms, we
can redefine $\mathbf{J}_{\epsilon}$ removing the second and third terms in
Eq.~(\ref{eq:otraexpresionparaJepsilon})

\begin{equation}
  \label{eq:Jepsilon2}
  \mathbf{J}_{\epsilon}
  \equiv
  \mathbf{K}[k] \wedge\mathfrak{h}^{(1)}\,,
\end{equation}

\noindent
which is, qualitatively, the same result obtained in 
Ref.~\cite{Barbagallo:2025fkg}. Then, following the same steps as in this
reference, we find the on-shell closed 3-form charge

\begin{equation}
  \label{eq:Qepsilonl}
  \mathbf{Q}_{\epsilon\, l}
  =
  \imath_{l}\mathbf{J}_{\epsilon} -\chi_{\underline{l}}\mathbf{K}[k]
  =
  \imath_{l}\mathbf{K}[k]\wedge \mathfrak{h}^{(1)}\,,
\end{equation}

\noindent
that we can combine with other on-shell conserved charges such as the
generalized Komar 3-form charge, for instance, to obtain new on-shell
conserved charges with the properties we want.

Comparing this expression with Eq.~(\ref{eq:Qepsilonreduction}) we also find
the relation

\begin{equation}
  \imath_{l}\mathbf{K}[k]
  =
   \tfrac{2}{3}\mathbf{Q}_{l\, 1}\,,
\end{equation}

\noindent
where $\mathbf{Q}_{l\, 1}$ is the 4-dimensional on-shell closed 2-form scalar
charge in Eq.~(\ref{eq:Ql1}).

Notice that, if we allow for the possibility of harmonic 2-forms in the
magnetic momentum map equation (\ref{eq:magneticmomentummapequationCS2}), to
include their effect we just need to replace in the above expression the
generalized Komar charge $\mathbf{K}[k]$ by the locally-defined
$\mathbf{K}_{\mathfrak{h}}[k]$ defined in Eq.~(\ref{eq:Kkh}).

Finally, it is interesting to compute the algebra of these global
symmetries. The commutator $[\delta_{\eta},\delta_{\epsilon}]$ only has a
non-trivial action on the gauge field and it gives

\begin{equation}
  [\delta_{\eta},\delta_{\epsilon}]
  =
  -\tfrac{2}{3}\delta_{\eta'}\,,
  \hspace{1cm}
  \eta'
  =
  \eta\epsilon\,,
\end{equation}

\noindent
\textit{i.e.}~they generate a closed subalgebra.

It is trivial to see that these higher-form symmetries commute with GCTs.

\subsection{Black-object thermodynamics: asymptotically-flat case}

Although the thermodynamics of asymptotically-flat 5-dimensional black holes
and black rings has been extensively studied in, for instance,
Refs.~\cite{Gauntlett:1998fz,Emparan:2004wy,Copsey:2005se,Astefanesei:2005ad,Rogatko:2005aj,Rogatko:2006xh,Compere:2007vx,Gibbons:2013tqa,Kunduri:2018qqt}
and we are ultimately interested in the asymptotically-KK case, it is
convenient to rederive the generalized zeroth laws, Smarr formulas and first
law for the asymptotically-flat case using the same techniques we are going to
use in the asymptotically-KK one not just as a warm-up exercise but as a way
of making manifest the effects of the boundary conditions on the calculations
and results.

\subsubsection{Generalized zeroth laws}

The zeroth law of black-hole thermodynamics states that the surface gravity
$\kappa$ and, hence, the Hawking temperature $T=\kappa/(2\pi)$ are constant
over the event horizon \cite{Bardeen:1973gs}. This is a purely kinematical
(geometrical) result that can be proven using the properties of Killing
horizons without the use of any particular field equations for the
gravitational and matter fields \cite{Racz:1995nh}. Thus, it is valid in the
theory we are considering for black holes or rings, with asymptotically-flat
or asymptotically-KK boundary conditions.

In the 4-dimensional Einstein-Maxwell theory, the fact that the corotating
electrostatic potential is also constant over the horizon is commonly referred
to as the \textit{generalized zeroth law} of black-hole thermodynamics. A
similar statement can be made for the corotating magnetostatic potential and,
in other dimensions and for $p$-form field of different ranks one can
enunciate \textit{generalized restricted zeroth laws} stating that the
potentials dual to the conserved charges associated to those fields are
constant over the bifurcation surface, whose existence we are going to assume
\cite{Elgood:2020svt,Elgood:2020mdx,Elgood:2020nls}.  While it would be
desirable to lift the restriction of these statements so that they are
unrestrictedly valid over the whole event horizon, it will be enough for
our purposes that they are satisfied just over the bifurcations surface.  We
are going to consider the general case here.

Let us start with the corotating electrostatic potential. If $l$ is the
Killing vector that defines the event horizon of a stationary black object
(hole, ring, brane...) as a Killing horizon, it follows from the Maxwell
momentum map equation~(\ref{eq:pformmomentummap}) that the corotating
electrostatic $p$-form potential coincides with the momentum map associated to
$l$, $P_{l}$. Since, by definition of the bifurcation surface $\mathcal{BH}$,
$l \stackrel{\mathcal{BH}}{=} 0$ it follows that

\begin{equation}
  dP_{l}
    = -\imath_{l}G
      \stackrel{\mathcal{BH}}{=} 0\,,  
\end{equation}

\noindent
Notice that this result would not hold true in presence of a harmonic
$(p+1)$-form modifying the momentum map equation, which we have explicitly
excluded in the case of minimal 5-dimensional supergravity.

Following Refs.~\cite{Copsey:2005se,Compere:2007vx}, the above result allows
us to write the $p$-form $P_{l}$ on $\mathcal{BH}$ as a linear combination of
the harmonic $p$-forms $\mathfrak{h}^{(p)}_{m}$ up to an exact $p$-form $de$:

\begin{equation}
  \label{eq:p-formHodgedecomposition}
  P_{l}
  \stackrel{\mathcal{BH}}{=}
  \Phi^{m}_{\mathcal{H}}\mathfrak{h}^{(p)}_{m} +de\,.
\end{equation}

The exact $p$-form will play no role whatsoever since it does not give rise to
non-trivial electric charges charges. The coefficients
$\Phi^{m}_{\mathcal{H}}$, are constant by definition over $\mathcal{BH}$ and
we can say that they satisfy generalized zeroth laws restricted to
$\mathcal{BH}$. They will be identified with the thermodynamical
(``chemical'') potentials dual to the electric charges $Q_{m}$ obtained by
integrating $\mathbf{Q}[\mathfrak{h}^{(p)}_{m}]$ over $\mathcal{BH}$. In
stationary black-hole or -ring spacetimes all the sections of the horizon have
the same topology and we can define potentials $\Phi^{m}$ in each of them
which are constant on each of them, but it remains to be proven that all of
them take the same value $\Phi^{m}_{\mathcal{H}}$ so they are constant over
the whole event horizon and satisfy unrestricted generalized zeroth laws. The
restricted ones will be enough for our purposes, though.

The $p=0$ case is special because $P_{l}$ is a function and there is always a
harmonic function on any surface $\mathcal{BH}$: $1$. The single coefficient
$\Phi_{\mathcal{H}}$ is the value of $P_{l}$ over $\mathcal{BH}$, which we
will normalize by requiring $P_{l}=0$ at infinity or, at least that the term
in which $P_{l}$ occurs gives no contribution at infinity. Furthermore, in the
$p=0$ case since, by definition $\imath_{l}dP_{l}=0$, one can conclude that

\begin{equation}
  \label{eq:0-formHodgedecomposition}
P_{l} \stackrel{\mathcal{H}}{=}  \Phi_{\mathcal{H}}\,,
\end{equation}

\noindent
which is the generalized zeroth law for this potential.

For $p\neq 0$ it is worth stressing that

\begin{enumerate}
\item A non-spherical and topologically non-trivial horizon such as that of a
  black ring or the horizon of a black hole with KK boundary conditions is
  necessary to obtain non-vanishing charges $Q_{m}$  defined on it.
\item Such a horizon may not be smoothly deformable into spatial infinity
  (S$^{d-2}_{\infty}$ in the asymptotically-flat case), which means that the
  Gauss theorem may not apply to those charges.
\item Actually, if the topologies of the spatial sections of the horizon are
  different from those of spatial infinity it may be impossible to define the
  same charges at infinity. In the minimal supergravity case, the usual
  understanding is that these are not conserved, but no-conserved, dipole-like
  charges \cite{Emparan:2004wy}, even though they occur in the first law of
  thermodynamics and in the Smarr formula.
\end{enumerate}

Next, let us consider the $\tilde{p}$-form
$\tilde{P}_{l}-6\gamma P_{l}\wedge V$ that occurs in the magnetic term of the
generalized Komar $(d-2)$-form charge
Eq.~(\ref{eq:generalizedKomarCS2}). Using the previous result, it is not
difficult to see that it is on-shell closed over the bifurcation surface as
well. Indeed, using the magnetic momentum map
equation~(\ref{eq:magneticmomentummapequationCS}) and the Maxwell momentum map
equation~(\ref{eq:pformmomentummap})

\begin{equation}
    d\left(\tilde{P}_{l}-6\gamma P_{l}\wedge V \right)
    \doteq
      -\imath_{l}\star G 
      +6\gamma \imath_{l}G \wedge V
      \stackrel{\mathcal{BH}}{=} 0\,.  
\end{equation}

Notice that this result would not hold true in presence of a harmonic 2-form
in the magnetic momentum map equation and, therefore we are going to exclude
its presence in our analysis of asymptotically-flat black-object
thermodynamics.

As in the electric case, the above result allows us to write

\begin{equation}
  \label{eq:1-formHodgedecomposition}
\tilde{P}_{l}-6\gamma P_{l}\wedge V
  \stackrel{\mathcal{BH}}{=}
  \tilde{\Phi}^{i}_{\mathcal{H}}\mathfrak{h}^{(\tilde{p})}_{i}+de\,, 
\end{equation}

\noindent
where the $\mathfrak{h}^{(\tilde{p})}_{i}$ are the harmonic $\tilde{p}$-forms
of $\mathcal{BH}$ and the $\tilde{\Phi}^{i}_{\mathcal{H}}$ are constant
coefficients that will be interpreted as the thermodynamical potentials dual
to magnetic (or dipole) charges $P_{i}$ obtained by integrating the
topologically closed $(d-2)$-forms
$\mathbf{P}[\mathfrak{h}^{(\tilde{p})}_{i}]$. The same remarks we made for the
electric charges and their potentials can be repeated here verbatim: in
minimal 5-dimensional supergravity $\tilde{p}=1$ and, in black rings, these
magnetic (``dipole'' \cite{Emparan:2004wy}) charges will be the only ones
associated to the non-trivial topology of the horizon.

\subsubsection{Smarr formulas}

The form Eq.~(\ref{eq:generalizedKomar5d}) of the Komar charge makes it easy
to use these generalized, restricted, zeroth laws when integrated over
$\mathcal{BH}$. Thus, it is very convenient to derive from it Smarr formulas
\cite{Smarr:1972kt} for the stationary black holes or black rings of the
theory following the idea of Refs.~\cite{Bardeen:1973gs,Carter:1973rla}
(implemented in more general theories in
Refs.~\cite{Magnon:1985sc,Bazanski:1990qd,Kastor:2008xb,Kastor:2010gq,Liberati:2015xcp,Ortin:2021ade})\footnote{Other
  methods have been used in the literature to derive Smarr formulas such as
  the one explained in the excellent lecture notes
  Ref.~\cite{Townsend:1997ku}, applied in the current context in
  Ref.~\cite{Gibbons:2013tqa} or the method used in Ref.~\cite{Heusler:1997am}
  based on the reduction of the action to that of a $\sigma$-model. See also
  Ref.~\cite{Chrusciel:2012jk}.}.

Let $l$ be the Killing vector that defines the event horizon as a Killing
horizon in a stationary, asymptotically-flat black-hole or black-ring
spacetime. Such spacetimes have an asymptotically timelike and two axial
Killing vectors which, in adapted coordinates, take the form
$\partial_{t},\partial_{\phi_{1}},\partial_{\phi_{2}}$ and

\begin{equation}
  \label{eq:5dKillingvectorAF}
  l
  =
  \partial_{t}-\Omega^{1}\partial_{\phi_{1}}-\Omega^{2}\partial_{\phi_{2}} \,,
\end{equation}

\noindent
where $\Omega^{1}$ and $\Omega^{2}$ are the angular velocities of the horizon
around the two independent axes of rotation \cite{Gauntlett:1998fz}.

Let $\mathbf{K}[l]$ be the on-shell closed generalized Komar 3-form charge
Eq.~(\ref{eq:generalizedKomar5d}) evaluated on that Killing vector. If we
integrate both sides of the equation $d\mathbf{K}[l]\doteq 0$ over a
hypersurface $\Sigma^{4}$ with boundaries at spatial infinity and at the
bifurcation surface

\begin{equation}
  \label{eq:favoritehypersurfaces}
  \partial \Sigma^{4}
  =
  S^{3}_{\infty}  \cup \mathcal{BH}\,,
\end{equation}

\noindent
and apply Stokes' theorem, we find the identity

\begin{equation}
  \label{eq:identityafterstokesd5}
  \int_{S^{3}_{\infty}} \mathbf{K}[l]
  \doteq
  \int_{\mathcal{BH}} \mathbf{K}[l]\,.
\end{equation}

Let us evaluate the integral at spatial infinity first. It is the sum of three
terms, the first of which is the standard Komar integral and gives

\begin{equation}
  \frac{1}{16\pi G_{N}^{(5)}}
  \int_{S^{3}_{\infty}}\star (e^{a}\wedge e^{b})P_{l\, ab}
  =
 \tfrac{2}{3}M -\Omega^{1} J_{1} -\Omega^{2} J_{2}\,,
\end{equation}

\noindent
where $J_{1}$ and $J_{2}$ are the angular momenta along the two independent
rotation axes.

In simple situations (electrically charged, static black holes) $\imath_{l}G$
vanishes at infinity and the electric momentum map $P_{l}$ becomes a constant
that we can always set to zero, since $P_{l}$ is defined up to a constant. In
more complex situations it has been shown that the integral of
$P_{l}\mathbf{Q}$ can be made to vanish by an adequate choice of integration
constant \cite{Zatti:2024vqv} and, therefore, the second integral at infinity
will vanish as well.

Since there are no harmonic 1-forms on a 3-sphere, according to
Eq.~(\ref{eq:1-formHodgedecomposition}) and because of the Bianchi identity
$dG=0$, the third term is the integral of an exact 3-form and vanishes
identically.

As it is well known, due to the property

\begin{equation}
  \label{eq:Pabproperty}
P_{l}{}^{ab} \stackrel{\mathcal{BH}}{=} \kappa n^{ab}\,,  
\end{equation}

\noindent
where $\kappa$ is the surface gravity of the Killing horizon of $l$ and
$n^{ab}$ is the binormal to it, normalized $n^{ab}n_{ab}=-2$, the integral of
the first term of $\mathbf{K}[l]$ over $\mathcal{BH}$ gives

\begin{equation}
  \frac{1}{16\pi G_{N}^{(5)}}
  \int_{\mathcal{BH}}\star (e^{a}\wedge e^{b})P_{l\, ab}
  =
  \frac{\kappa \mathcal{A}_{\mathcal{H}}}{16\pi G_{N}^{(5)}}
  =
  TS \,,
\end{equation}

\noindent
where $\kappa$ is the surface gravity, $\mathcal{A}_{\mathcal{H}}$ is the area
(actually, volume) of the spatial sections of the horizon, $T=\kappa/(2\pi)$
is the Hawking temperature and $S=\mathcal{A}_{\mathcal{H}}/(4G_{N}^{(5)})$
the Bekenstein-Hawking entropy.

The generalized zeroth law says that, over the horizon, the Maxwell momentum
map $P_{l}$, which we can identify with the corotating electrostatic
potential $\Phi$, is the constant $\Phi_{\mathcal{H}}$. Then, the second term
is $\tfrac{2}{3}\Phi_{\mathcal{H}}$ times the integral of the electric 3-form
charge $\mathbf{Q}[1]$. The Gauss law ensures that we get the same value
integrating over $\mathcal{BH}$ or over $S^{3}_{\infty}$, namely $Q$, that is

\begin{equation}
  \int_{\mathcal{BH}} \tfrac{2}{3} \mathbf{Q}[P_{l}]
  = \tfrac{2}{3} \Phi_{\mathcal{H}} \int_{\mathcal{BH}} \mathbf{Q}[1]
  = \tfrac{2}{3} \Phi_{\mathcal{H}} \int_{S^{3}_{\infty}} \mathbf{Q}[1]
  = \tfrac{2}{3} \Phi_{\mathcal{H}} Q\,.
\end{equation}

According to the generalized, restricted, zeroth law, the 1-form
$\tilde{P}_{l}-\tfrac{2}{3^{1/2}} P_{l}V$ admits the decomposition
Eq.~(\ref{eq:1-formHodgedecomposition}) and, following our previous
discussion, the integral decomposes into

\begin{equation}
  \int_{\mathcal{BH}} \tfrac{1}{3}
  \mathbf{P}[\tilde{P}_{l}-\tfrac{2}{3^{1/2}}P_{l}V]
=
\tfrac{1}{3} \tilde{\Phi}^{i}_{\mathcal{H}}
\int_{\mathcal{BH}}\mathbf{P}[\mathfrak{h}^{(1)}_{i}]
=
\tfrac{1}{3} \tilde{\Phi}^{i}_{\mathcal{H}}P_{i}\,.
\end{equation}

As we have discussed before, only for horizons with non-trivial topology (not
black holes) is this term different from zero. Black rings have horizons with
topology $S^{1}\times S^{2}$, admitting a single harmonic 1-form associated to
the $S^{1}$ that gives rise to a magnetic or dipole charge on the horizon that
cannot be seen (not even defined) at infinity. The charge will vanish if
$G=dV$ globally.

The Smarr formula for all these objects can be written in the unified form

\begin{equation}
  \label{eq:5dSmarrforumula}
  M
  =
  \tfrac{3}{2}\left(ST
  +\Omega^{1} J_{1} +\Omega^{2} J_{2}\right)
  +\Phi_{\mathcal{H}} Q +\tfrac{1}{2}\tilde{\Phi}_{\mathcal{H}}P\,. 
\end{equation}

\subsubsection{First law}
\label{sec-firstlawminimal5dsugra}

The first law of black-hole (-ring) mechanics can be obtained by integrating
the exterior derivative of the on-shell-closed $3$-form $\mathbf{W}[l]$ over
the same spacelike hypersurface we used in the derivation of the Smarr formula
\cite{Wald:1993nt,Iyer:1994ys}. Here we are going to review the construction
of $\mathbf{W}[l]$ in detail.

Denoting by $\varphi$ all the fields of the theory, the symplectic potential
$4$-form is defined as \cite{Lee:1990nz,DePaoli:2018erh}

\begin{equation}
  \label{eq:symplecticform}
  \omega(\varphi,\delta_{1}\varphi,\delta_{2}\varphi)
  \equiv
  \delta_{1}\mathbf{\Theta}(\varphi,\delta_{2}\varphi)
  -\delta_{2}\mathbf{\Theta}(\varphi,\delta_{1}\varphi)
  -\mathbf{\Theta}(\varphi,[\delta_{1},\delta_{2}]\varphi)\,,
\end{equation}

\noindent
where $\delta_{1,2}$ are arbitrary variations. We choose $\delta_{1}=\delta$
(arbitrary, infinitesimal variations of the fields) and
$\delta_{2}=\delta_{\xi}$ (infinitesimal GCTs of the fields which, as we have
discussed, include induced gauge transformations). In most of the literature,
the last term has been ignored because, if all the fields $\varphi$ are world
tensors, $\delta_{\xi}= -\mathcal{L}_{\xi}$ and

\begin{equation}
[\delta,\mathcal{L}_{\xi}]=0\,.
\end{equation}

\noindent
When the fields have some gauge freedom,
$\delta_{\xi}= -\mathcal{L}_{\xi}+\delta_{\Lambda_{\xi}}$ and

\begin{equation}
  \label{eq:commutatorofdeltas}
  [\delta,\delta_{\xi}]
  =
  [\delta,\delta_{\Lambda_{\xi}}]
  =
  \delta_{\delta\Lambda_{\xi}}\,,
\end{equation}

\noindent
\textit{i.e.}~a gauge transformation with parameter $\delta\Lambda_{\xi}$. In
general there are several gauge symmetries and we have to consider all of them.

Thus, our starting point is 

\begin{equation}
  \label{eq:symplecticpotentialdef}
  \omega(\varphi,\delta\varphi,\delta_{\xi}\varphi)
  =
  \delta \mathbf{\Theta}(\varphi,\delta_{\xi}\varphi)
  -\delta_{\xi}\mathbf{\Theta}(\varphi,\delta\varphi)
  -\mathbf{\Theta}(\varphi,\delta_{\delta\Lambda_{\xi}}\varphi)\,.
\end{equation}

On-shell we can replace $\mathbf{\Theta}$ by $\mathbf{\Theta}'$, which
contains the terms proportional to the equations of motion which are left in
the variation of the action under GCTs after use of the Noether identities. In
the case of the higher-rank-form theories with Chern--Simons terms we have
been considering, $\mathbf{\Theta}'$ is given in Eqs.~(\ref{eq:ThetaCS}) and
(\ref{eq:ThetaprimeCS}).  We can consider simultaneously the $d=11,p=2$ and
$d=5,p=0$ cases (both have $N=2$ and differ in the values of $\gamma$).

Using the definition of the Noether--Wald  $(d-2)$-form
charge Eq.~(\ref{eq:JxiCS}), we find

\begin{equation}
  \begin{aligned}
  \omega(\varphi,\delta\varphi,\delta_{\xi}\varphi)
  & \doteq
    \delta\left[d\mathbf{Q}_{NW}[\xi]-\imath_{\xi}\mathbf{L}
    -\mathbf{B}(V,e,\Lambda_{\xi})\right]
    \\
    & \\
    & \hspace{.5cm}
    -\left(-\mathcal{L}_{\xi}+\delta_{\Lambda_{\xi}}\right)
    \mathbf{\Theta}(\varphi,\delta\varphi)
  -\mathbf{\Theta}(\varphi,\delta_{\delta\Lambda_{\xi}}\varphi)
  \end{aligned}
\end{equation}

\noindent
where $\mathbf{B}(V,e,\Lambda)$ is defined in Eq.~(\ref{eq:BVLambdadef}). By
definition of the Noether--Wald charge and Lie derivative and using
$[\delta,d]=0$ and $\delta \xi =0$

\begin{equation}
  \label{eq:symplecticpotentialCS2}
  \begin{aligned}
  \omega(\varphi,\delta\varphi,\delta_{\xi}\varphi)
  & \doteq
    d\delta\mathbf{Q}_{NW}[\xi]-\imath_{\xi}\delta \mathbf{L}
    -\delta \mathbf{B}(V,e,\Lambda_{\xi})
    \\
    & \\
    & \hspace{.5cm}
    +d\imath_{\xi}\mathbf{\Theta}(\varphi,\delta\varphi)
      +\imath_{\xi}d\mathbf{\Theta}(\varphi,\delta\varphi)
      -\delta_{\Lambda_{\xi}} \mathbf{\Theta}(\varphi,\delta\varphi)
      -\mathbf{\Theta}(\varphi,\delta_{\delta\Lambda_{\xi}}\varphi)
    \\
    & \\
  & \doteq
    d\left[\delta\mathbf{Q}_{NW}[\xi]
    +\imath_{\xi}\mathbf{\Theta}(\varphi,\delta\varphi)\right]
    \\
    & \\
    & \hspace{.5cm}
    -\delta \mathbf{B}(V,e,\Lambda_{\xi})
      -\delta_{\Lambda_{\xi}} \mathbf{\Theta}(\varphi,\delta\varphi)
      -\mathbf{\Theta}(\varphi,\delta_{\delta\Lambda_{\xi}}\varphi)\,.
  \end{aligned}
\end{equation}

Using Eq.~(\ref{eq:commutatorofdeltas}), the last three terms of this
expression combine into

\begin{equation}
  \begin{aligned}
    \label{eq:esacombinacion}
& 2\star (e^{c}\wedge e^{b})\wedge \sigma_{\xi}^{a}{}_{c} \delta \omega_{ab}
      +\star ( e^{a}\wedge e^{b})\wedge \delta \mathcal{D}\sigma_{\xi\, ab}
    \\
    & \\
    & \hspace{.5cm}
      +d\delta \Lambda_{\xi}\wedge\left[ \star G -3\gamma G\wedge V \right]
          +d\left[\gamma \delta V\wedge V\wedge d\Lambda_{\xi}\right]\,.
  \end{aligned}
\end{equation}

Then, since

\begin{equation}
  \delta \mathcal{D}\sigma_{\xi\, ab}
  =
  \mathcal{D} \delta \sigma_{\xi\, ab} -2 \delta \omega_{[a|}{}^{c}\sigma_{\xi\, c|b]}\,,
\end{equation}

\noindent
the first term in Eq.~(\ref{eq:esacombinacion}) cancels and, integrating by
parts and using the equation of motion of $V$, we obtain a total derivative
that we can substitute in Eq.~(\ref{eq:symplecticpotentialCS2})

\begin{equation}
    \label{eq:Wxidef}
  \begin{aligned}
  \omega(\varphi,\delta\varphi,\delta_{\xi}\varphi)
  & \doteq
    d\left\{\delta\mathbf{Q}_{NW}[\xi]
    +\imath_{\xi}\mathbf{\Theta}(\varphi,\delta\varphi)
    - \star ( e^{a}\wedge e^{b})\wedge \delta \sigma_{\xi\, ab}
    \right.
    \\
    & \\
    & \hspace{.5cm}
      \left.
    +\delta \Lambda_{\xi}\wedge\left[ \star G -3\gamma G\wedge V \right]
      +\gamma \delta V\wedge V\wedge d\Lambda_{\xi}\right\}
    \\
    & \\
    & \equiv d\mathbf{W}[\xi]\,.
  \end{aligned}
\end{equation}

The symplectic potential vanishes if $\xi=l$, where $\delta_{l}\varphi=0$ for
all the fields of the theory. Our definitions of $\delta_{\xi}$ taking into
account the gauge freedoms of the fields ensure that $\delta_{l}\varphi=0$ is
a well-defined, gauge-independent statement. Then, the 2-form $\mathbf{W}[l]$
is closed on-shell

\begin{equation}
    d\mathbf{W}[l]
    \doteq
      0\,.  
\end{equation}

Our next task is to find an explicit expression for $\mathbf{W}[l]$ using
Eqs.~(\ref{eq:NoetherWaldchargeCStheory}), (\ref{eq:ThetaCS}),
(\ref{eq:sigmakdef}) and (\ref{eq:Lambdak}). We find that  $\mathbf{W}[l]$ can be
written in the form 

\begin{equation}
  \begin{aligned}
        \mathbf{W}[l]
    & =
      -P_{l\, bc} \delta\star (e^{b}\wedge e^{c})
      -\imath_{l}\star (e^{a}\wedge e^{b})\wedge \delta \omega_{ab}
          \\
    & \\
    & \hspace{.5cm}
      +P_{l} \wedge \delta  \left(\star G -3\gamma G\wedge V\right)
      -\left(\tilde{P}_{l} -6\gamma P_{l}\wedge V\right)\wedge \delta G
      -3\gamma \delta V\wedge V\wedge \imath_{l}G
          \\
    & \\
    & \hspace{.5cm}
      +d\left\{\gamma \delta V \wedge V\wedge \imath_{l}V
      +\delta V\wedge
      \left(\tilde{P}_{l}-6\gamma  P_{l} \wedge V\right)\right\}\,.
  \end{aligned}
\end{equation}

This result (which actually comes multiplied by the standard normalization
factor $(16\pi G_{N}^{(d)})^{-1}$) is valid for all the theories under
consideration (in particular minimal 5-dimensional and 11-dimensional
supergravity) independently of the boundary conditions. We can ignore the
total derivative in the third line. Also, we can rewrite the three terms
second line in the alternative simple form

\begin{equation}
      P_{l} \wedge \delta  \star G 
      -\tilde{P}_{l}\wedge \delta G\,,  
\end{equation}

\noindent
up to another total derivative that we will also ignore. This form is more
convenient to deal with the integration at infinity and it is manifestly
invariant under gauge transformations $\delta_{\Lambda}$

Let us now apply this result to the case of minimal 5-dimensional supergravity
for asymptotically-flat black holes and rings. Integrating $d\mathbf{W}[l]=0$
over the same spacelike hypersurface we used to derive the Smarr formula and
applying Stokes' theorem, we find an identity similar to that in
Eq.~(\ref{eq:identityafterstokesd5})

\begin{equation}
  \int_{S^{3}_{\infty}} \mathbf{W}[l]
  \doteq
  \int_{\mathcal{BH}} \mathbf{W}[l]\,.  
\end{equation}

Let us first consider the integral at spatial infinity.  In this case $P_{l}$
is a function and, just as we did in the derivation of the Smarr formula, we
can choose the integration constant in the solution of the momentum map
equation so that $P_{l}$ or the integral of the product $P_{l}\delta \star G$
vanishes on the 3-sphere at spatial infinity \cite{Zatti:2024vqv}. Since the
magnetic momentum map $\tilde{P}_{l}$ is defined up to a closed 1-form, we
should be able to make the contribution of the term
$\tilde{P}_{l}\wedge \delta G$ vanish at infinity as well. Thus, 

\begin{equation}
  \begin{aligned}
  \int_{S^{3}_{\infty}} \mathbf{W}[l]
  & \doteq
  \frac{1}{16\pi G_{N}^{(5)}}
  \int_{S^{3}_{\infty}} \left\{  
      -P_{l\, bc} \delta\star (e^{b}\wedge e^{c})
      -\imath_{l}\star (e^{a}\wedge e^{b})\wedge \delta \omega_{ab}
    \right\}
    \\
    & \\
    & =
    \delta M -\Omega^{1}\delta J_{1}-\Omega^{2}\delta J_{2}\,.
  \end{aligned}
\end{equation}

Over the bifurcation surface, taking into account
$l \stackrel{\mathcal{BH}}{=} 0$ and the generalized, restricted, zeroth laws
satisfied by the function $P_{l}$ and by the 1-form
$\tilde{P}_{l} -\tfrac{2}{3^{1/2}} P_{l}V$ and the Hodge decompositions
Eq.~(\ref{eq:0-formHodgedecomposition}) and
(\ref{eq:1-formHodgedecomposition}) (for black rings only) plus the general
property Eq.~(\ref{eq:Pabproperty}) and the zeroth law satisfied by $\kappa$
we have

\begin{equation}
  \begin{aligned}
  \int_{\mathcal{BH}} \mathbf{W}[l]
  & =
  -\frac{\kappa}{16\pi G_{N}^{(5)}}
  \delta \int_{\mathcal{BH}} n_{bc} \star (e^{b}\wedge e^{c})
  +\Phi_{\mathcal{H}}  \delta  \int_{\mathcal{BH}}\mathbf{Q}[1]
 +\tilde{\Phi}_{\mathcal{H}} \delta \int_{\mathcal{BH}}
    \mathbf{P}[\mathfrak{h}^{(1)}]
    \\
    & \\
    & =
      \frac{\kappa \delta \mathcal{A}}{4 G_{N}^{(5)}}
  +\Phi_{\mathcal{H}}  \delta Q +\tilde{\Phi}_{\mathcal{H}} \delta P\,,
  \end{aligned}
\end{equation}

\noindent
which leads to the first law

\begin{equation}
  \delta M
  =
  T\delta S +\Omega^{1}\delta J_{1}+\Omega^{2}\delta J_{2}
  +\Phi_{\mathcal{H}}  \delta Q +\tilde{\Phi}_{\mathcal{H}} \delta P\,,
\end{equation}

\noindent
where the last term is only present for black rings.

These results change in a non-trivial way with KK boundary conditions. We will
study the KK case in Section~\ref{sec-5vs4dthermodynamics} because it
naturally leads to the thermodynamics of 4-dimensional black holes.

\section{The T$^{3}$ model of $ \mathcal{N}=2,d=4$ Supergravity }
\label{sec-thet3model}

The so-called ``T$^{3}$ model of $\mathcal{N}=2,d=4$ Supergravity'' describes
the supergravity multiplet interacting with a single vector supermultiplet in
a particular way to be described below that distinguishes it from other models
with the same field content and duality group, such as the axidilaton
model. The bosonic sector of this model consists of the Vierbein $e^{a}$, the
graviphoton $A^{0}$, which is the gauge field of the supergravity multiplet,
the gauge field of the vector supermultiplet, $A^{1}$ and a complex scalar
conventionally denoted by $t$. Since $t$ parametrizes a
SL$(2,\mathbb{R})/$SO$(2)$ coset space and the duality group
SL$(2,\mathbb{R})$ acts on $t$ through fractional-linear transformations, and
because of the couplings of $t$ to the gauge fields, it is customary to write

\begin{equation}
t=\chi+ie^{-2\phi}\,,   
\end{equation}

\noindent
where $\chi$ is the axion field and $\phi$ the dilaton field.

The action of this model is a particular case of the generic 4-dimensional
ungauged supergravity model recently considered in this context in
Ref.~\cite{Ballesteros:2023iqb}

\begin{equation}
\label{eq:genericd4action}
\begin{aligned}
  S[e,\phi,A]
  & =
\frac{1}{16\pi G_{N}^{(4)}}\int 
  \left[ -\star (e^{a}\wedge e^{b}) \wedge R_{ab}
    +\tfrac{1}{2}g_{xy}d\phi^{x}\wedge \star d\phi^{y}
  \right.
  \\
  & \\
  & \hspace{.5cm}
  \left.
    -\tfrac{1}{2}I_{\Lambda\Sigma}F^{\Lambda}\wedge \star F^{\Sigma}
    -\tfrac{1}{2}R_{\Lambda\Sigma}F^{\Lambda}\wedge F^{\Sigma}
\right]
\\
& \\
& \equiv
\int \mathbf{L}\,,
\end{aligned}
\end{equation}

\noindent
where $F^{\Lambda}=dA^{\Lambda}$ are Abelian gauge field strengths (2-forms)
and where the  scalar dependent
matrices  $R_{\Lambda\Sigma}$ and $I_{\Lambda\Sigma}$ are, respectively, the
real and imaginary parts of the \textit{period matrix}

\begin{equation}
  \mathcal{N}_{\Lambda\Sigma}
  \equiv
  R_{\Lambda\Sigma}+iI_{\Lambda\Sigma}\,.
\end{equation}

For this particular model, $\Lambda=0,1$, and

\begin{subequations}
  \begin{align}
  g_{xy}d\phi^{x}\wedge \star d\phi^{y}
  & =
    3\frac{dt\wedge \star dt^{*}}{(\Im\mathrm{m}\, t)^{2}}
    =
    12 d\phi\wedge \star d\phi
    +3e^{4\phi}d\chi\wedge \star d\chi \,,
    \\
    & \nonumber \\
    \left( \mathcal{N}_{\Lambda\Sigma} \right)
    & =
      \tfrac{1}{2}\left(
      \begin{array}{lr}
         t^{*\, 2}(t^{*}+3t)  & 3t^{*}(t^{*}+t) \\
                                          & \\
        3t^{*}(t^{*}+t) & 3(t+3t^{*})\\
      \end{array}
    \right)\,,
    \\
    & \nonumber \\
        \left( R_{\Lambda\Sigma} \right)
    & =
      \left(
      \begin{array}{lr}
         2\chi^{3} & 3\chi^{2} \\
                       & \\
      3\chi^{2}  & 6\chi\\
      \end{array}
    \right)\,,
    \\
    & \nonumber \\
    \left( I_{\Lambda\Sigma} \right)
    & =    
      -e^{-2\phi}\left(
      \begin{array}{lr}
        e^{-4\phi}+3\chi^{2} & 3\chi \\
                             & \\
        3\chi  & 3 \\
      \end{array}
    \right)\,.
  \end{align}
\end{subequations}

\noindent
so the action takes the particular form

\begin{equation}
\label{eq:t3action}
\begin{aligned}
  S_{T^{3}}[e,\phi,A]
  & =
\frac{1}{16\pi G_{N}^{(4)}}\int 
  \left[ -\star (e^{a}\wedge e^{b}) \wedge R_{ab}
    +\tfrac{3}{2}e^{4\phi}d\chi\wedge \star d\chi
    +6d\phi\wedge \star d\phi
  \right.
  \\
  & \\
  & \hspace{.5cm}
  \left.
    +\tfrac{1}{2}e^{-2\phi}\left(e^{-4\phi}+3\chi^{2}\right) F^{0}\wedge \star F^{0}
    +3e^{-2\phi} \chi F^{0}\wedge \star F^{1}
    +\tfrac{3}{2}e^{-2\phi} F^{1}\wedge \star F^{1}
  \right.
  \\
  & \\
  & \hspace{.5cm}
  \left.
    -\chi^{3}F^{0}\wedge F^{0}
    -3\chi^{2}F^{0}\wedge F^{1}
    -3\chi F^{1}\wedge F^{1}
    \right]
  \\
  & \\
  & \equiv
    \int \mathbf{L}_{T{3}}\,.
\end{aligned}
\end{equation}

It is, however, quite convenient to work with the generic form of the action
Eq.~(\ref{eq:genericd4action}), particularizing the fields and interactions to
the T$^{3}$ model case only when strictly needed. Thus, we define the
equations of motion and the presymplectic potential through the arbitrary
infinitesimal variations of the action

\begin{equation}
  \delta S_{T^{3}}
  =
  \int\left\{
    \mathbf{E}_{a}\wedge \delta e^{a} + \mathbf{E}_{x}\delta\phi^{x}
    +\mathbf{E}_{\Sigma}\wedge \delta A^{\Sigma}  
    +d\mathbf{\Theta}_{T^{3}}(\varphi,\delta\varphi)
    \right\}\,.
\end{equation}

\noindent
They are given by

\begin{subequations}
  \begin{align}
    \label{eq:Ead4}
    \mathbf{E}_{a}
    & =
      \imath_{a}\star(e^{b}\wedge e^{c})\wedge R_{bc}
      +\tfrac{1}{2}g_{xy}\left(\imath_{a}d\phi^{x} \star d\phi^{y}
      +d\phi^{x}\wedge \imath_{a}\star d\phi^{y}\right)
      \nonumber \\
    & \nonumber \\
    & \hspace{.5cm}
      -\tfrac{1}{2}\left(\imath_{a}F^{\Lambda}\wedge F_{\Lambda}
      -F^{\Lambda}\wedge \imath_{a} F_{\Lambda}\right)\,,
    \\
    & \nonumber \\
    \mathbf{E}_{x}
    & =
      -g_{xy}\left\{d\star d\phi^{y}
      +\Gamma_{zw}{}^{y}d\phi^{z}\wedge\star d\phi^{w} \right\}
      \nonumber \\
    & \nonumber \\
    & \hspace{.5cm}
      -\tfrac{1}{2}\partial_{x}I_{\Lambda\Sigma} F^{\Lambda}\wedge\star
      F^{\Sigma}
      -\tfrac{1}{2}\partial_{x}R_{\Lambda\Sigma} F^{\Lambda}\wedge F^{\Sigma}\,,
    \\
    & \nonumber \\
    \label{eq:ELambda}
    \mathbf{E}_{\Lambda}
    & =
      d F_{\Lambda}\,,
    \\
    & \nonumber \\
      \label{eq:Thetad4general}
    \mathbf{\Theta}_{T^{3}}(\varphi,\delta\varphi)
    & =
      -\star (e^{a}\wedge e^{b})\wedge \delta \omega_{ab}
      +g_{xy}\star d\phi^{x}\delta\phi^{y}
      -F_{\Lambda}\wedge \delta A^{\Lambda}\,.
  \end{align}
\end{subequations}

\noindent
where we have defined the dual 2-form field strength

\begin{equation}
  \label{eq:dualfieldstrengthsdef}
  F_{\Lambda}
  \equiv
  I_{\Lambda\Sigma}\star F^{\Sigma}+R_{\Lambda\Sigma}F^{\Sigma}\,.
\end{equation}

For the T$^{3}$ model, the scalar equations of motion and the presymplectic
potential take the explicit form

\begin{subequations}
  \begin{align}
    \mathbf{E}_{\chi}
    & =
      -3d\left(e^{4\phi}\star d\chi\right)
      +3e^{-2\phi}\chi F^{0}\wedge \star F^{0}
      +3e^{-2\phi} F^{0}\wedge \star F^{1}
      \nonumber \\
    & \nonumber \\
    & \hspace{.5cm}
      -3\chi^{2}F^{0}\wedge F^{0}
      -6\chi F^{0}\wedge F^{1}
      -3F^{1}\wedge F^{1}\,,
    \\
    & \nonumber \\
    \mathbf{E}_{\phi}
    & =
      -12d\star d\phi +6e^{4\phi}d\chi\wedge \star d\chi
          -3e^{-2\phi}\left(e^{-4\phi}+\chi^{2}\right) F^{0}\wedge \star F^{0}
      \nonumber \\
    & \nonumber \\
    & \hspace{.5cm}
    -6e^{-2\phi} \chi F^{0}\wedge \star F^{1}
    -3e^{-2\phi} F^{1}\wedge \star F^{1}\,,
    \\
    & \nonumber \\
      \label{eq:Thetad5explicit}
    \mathbf{\Theta}(\varphi,\delta\varphi)
    & =
      -\star (e^{a}\wedge e^{b})\wedge \delta \omega_{ab}
      +12\star d\phi \delta \phi +3 e^{4\phi}\star d\chi\delta \chi
      -F_{\Lambda}\wedge \delta A^{\Lambda}\,,    
  \end{align}
\end{subequations}

\noindent
while the dual gauge field strengths are given by

\begin{subequations}
  \begin{align}
    F_{0}
    & =
      -e^{-2\phi}\left(e^{-4\phi}+3\chi^{2}\right) \star F^{0}
    -3e^{-2\phi} \chi \star F^{1}
    +2\chi^{3} F^{0}
      +3\chi^{2}F^{1}\,,
    \\
    & \nonumber \\
    F_{1}
    & =
    -3e^{-2\phi} \chi \star F^{0}
    -3e^{-2\phi} \star F^{1}
    +3\chi^{2}F^{0}
    +6\chi  F^{1}\,.      
  \end{align}
\end{subequations}

It is convenient to collect the gauge field strengths and their duals
and the (left-hand side of the) Maxwell equations (\ref{eq:ELambda}) and the
(left-hand side of the) Bianchi identities

\begin{equation}
  \mathbf{E}^{\Lambda}
  \equiv
  dF^{\Lambda}\,,
\end{equation}

\noindent
in two symplectic vectors of 2- and 3-forms

\begin{equation}
  \left(F^{M}\right)
  \equiv
  \left(
    \begin{array}{c}
      F^{\Lambda} \\ F_{\Lambda} \\
    \end{array}
  \right)\,,
  \hspace{1cm}
  \left(\mathbf{E}^{M}\right)
  \equiv
  \left(
    \begin{array}{c}
      \mathbf{E}^{\Lambda} \\ \mathbf{E}_{\Lambda} \\
    \end{array}
    \right)\,,
\end{equation}

\noindent
so we can write the manifestly symplectic-covariant expressions

\begin{subequations}
  \begin{align}
    \label{eq:Ead4inv}
    \mathbf{E}_{a}
    & =
      \imath_{a}\star(e^{b}\wedge e^{c})\wedge R_{bc}
      +\tfrac{1}{2}g_{xy}\left(\imath_{a}d\phi^{x} \star d\phi^{y}
      +d\phi^{x}\wedge \imath_{a}\star d\phi^{y}\right)
      \nonumber \\
    & \nonumber \\
    & \hspace{.5cm}
      +\tfrac{1}{2}\Omega_{MN}F^{M}\wedge \imath_{a}F^{N}
    \\
    & \nonumber \\
    \mathbf{E}^{M}
    & =
    dF^{M}\,,
  \end{align}
\end{subequations}

\noindent
where

\begin{equation}
  (\Omega_{MN})
  =
  \left(
    \begin{array}{rr}
      0 & \mathbb{1}_{2\times 2} \\
      -\mathbb{1}_{2\times 2} & 0 \\
    \end{array}
    \right)\,.
\end{equation}

\noindent
is the symplectic metric.

\subsection{Symmetries and conserved charges}

In this section we are going to study the local and global symmetries and
associated currents charges of this theory. We shall be very brief since they
are particular instances of those of the generic theory
Eq.~(\ref{eq:genericd4action}), studied in
Ref.~\cite{Ballesteros:2023iqb}\footnote{See also
  Ref.~\cite{Mitsios:2021zrn}.} to which we refer the reader to further
details.

\subsubsection{U$(1)^{2}$ gauge symmetry}

The action Eq.~(\ref{eq:genericd4action}) is exactly invariant under U$(1)^{2}$
gauge transformations

\begin{equation}
  \label{eq:U12guagetrans}
  \delta_{\Lambda}A^{\Sigma}
  =
  d\Lambda^{\Sigma}\,.
\end{equation}

The Noether 2-form charge associated to this invariance is

\begin{equation}
  \label{eq:QLambdad4}
  \mathbf{Q}[\Lambda]
  =
  -\frac{1}{16\pi G_{N}^{(4)}} \Lambda^{\Sigma}F_{\Sigma}\,, 
\end{equation}

\noindent
and, for the two independent Killing parameters\footnote{The signs are purely
  conventional.}  $\Lambda^{\Sigma}= \delta^{\Sigma}{}_{0}$ and
$\Lambda^{\Sigma}= \delta^{\Sigma}{}_{1}$ we can define two electric charges

\begin{equation}
  \label{eq:electricchargesd4def}
  \mathbf{Q}_{\Lambda}
  =
  -\frac{1}{16\pi G_{N}^{(4)}} F_{\Lambda}\,,
  \hspace{1cm}
  q_{\Lambda}
  =
  \int_{\Sigma^{2}}\mathbf{Q}_{\Lambda}\,.
\end{equation}

\noindent
satisfying Gauss laws on-shell.

\subsubsection{Magnetic charges}
\label{sec-magneticchargesd4}

The charges

\begin{equation}
  \label{eq:magneticchargesd4def}
  \mathbf{P}^{\Lambda}
  =
  -\frac{1}{16\pi G_{N}^{(4)}} F^{\Lambda}\,,
  \hspace{1cm}
  p^{\Lambda}
  =
  \int_{\Sigma^{2}}\mathbf{P}^{\Lambda}\,,
\end{equation}

\noindent
satisfy a Gauss law by virtue of the Bianchi identities. Observe that, for the
$p^{\Lambda}$ to be non-zero, $F^{\Lambda}$ cannot be globally exact on
$\Sigma^{2}$.

Electric $q_{\Lambda}$ and magnetic $p^{\Lambda}$ charges are naturally
combined into a symplectic vector of charges

\begin{equation}
  \label{eq:electricandmagneticchargesd4def}
  (\mathbf{Q}^{M})
  \equiv
  \left(
    \begin{array}{c}
      \mathbf{P}^{\Lambda} \\ \mathbf{Q}_{\Lambda} \\
    \end{array}
  \right)\,,
  \hspace{1cm}
  \mathcal{Q}^{M}
  =
  \int_{\Sigma^{2}}\mathbf{Q}^{M}\,,
  \hspace{1cm}
  (\mathcal{Q}^{M})
  \equiv
  \left(
    \begin{array}{c}
      p^{\Lambda} \\ q_{\Lambda} \\
    \end{array}
  \right)\,.
\end{equation}

\subsubsection{Local Lorentz transformations}

The theory is exactly invariant under the 4-dimensional version of the
transformations discussed in Section~\ref{sec-locallorentzCS} and the Noether
2-form charge associated to this symmetry has the form

\begin{equation}
  \mathbf{Q}[\sigma]
  =
-\frac{1}{16\pi G_{N}^{(4)}} \star (e^{a}\wedge e^{b})\sigma_{ab}\,.
\end{equation}

The same observations made for the $d$-dimensional one apply to this charge,
adapted to the 4-dimensional context.

\subsubsection{General coordinate transformations (GCTs)}

The only (small) differences with the $d$-dimensional case is the definition
of the parameters of the 2 induced gauge transformations, which now are
defined as

\begin{equation}
  \label{eq:LambdaLambdak}
  \Lambda^{\Lambda}_{k}
  =
  \imath_{k}A^{\Lambda}-P_{k}{}^{\Lambda}\,,  
\end{equation}

\noindent
and the definition of the symplectic vector of momentum maps $P_{k}{}^{M}$
containing the electric  $P_{k}{}^{\Lambda}$ and magnetic
 $P_{k\, \Lambda}$ momentum maps

\begin{equation}
  \label{eq:symplecticmomentummaps}
  \imath_{k}F^{M}+dP_{k}{}^{M}
    \doteq 
    0\,,
\end{equation}

\noindent
which follows from invariance, Bianchi identities and equations of motion
$\mathbf{E}^{M}=0$.

The electric and magnetic momentum maps are defined up to an additive constant
and, therefore, the expressions involving physical quantities should be
invariant under

\begin{equation}
  \label{eq:deltaCPkM}
  \delta_{C}P_{k}{}^{M}
  =
  C^{M}\,,
\end{equation}

\noindent
for arbitrary constants $C^{M}$.

A calculation very similar to the $d$-dimensional one (actually, simpler,
because the action is exactly gauge invariant), leads to the Noether-Wald
charge \cite{Ballesteros:2023iqb}

\begin{equation}
  \label{eq:NoetherWaldcharged4}
  \mathbf{Q}_{NW}[\xi]
  =
  \star (e^{a}\wedge e^{b})P_{\xi\, ab}+P_{\xi}{}^{\Lambda}F_{\Lambda}\,.
\end{equation}

Observe that, according to Ref.~\cite{Ortin:2024mmg}, since the action
Eq.~(\ref{eq:genericd4action}) is exactly invariant, we can obtain
$\mathbf{Q}_{NW}[k]$ on-shell computing first $\imath_{k}\mathbf{L}$
and then using
the equations of motion

\begin{equation}
  \mathcal{O}_{s}\imath_{k}\mathbf{L}
  =
  d\mathcal{O}_{s}\mathbf{Q}_{NW}[k]\,.
\end{equation}

The Komar charge can be computed by using first the
equations of motion on $\mathbf{L}$, taking the interior product with $k$
afterwards. The simplest way to do this is to compute the trace of the
Einstein equations (for instance, in the form in the form
Eq.~(\ref{eq:Ead4inv})). We find


\begin{equation}
  \mathbf{L}
  =
  -\tfrac{1}{2} e^{a}\wedge \mathbf{E}_{a}  -\tfrac{1}{2}F^{\Lambda}\wedge F_{\Lambda}\,,  
\end{equation}

\noindent
and, therefore,

\begin{equation}
  \imath_{k}\mathcal{O}_{s}\mathbf{L}
  =
  d\tfrac{1}{2}\left( P_{k}{}^{\Lambda}F_{\Lambda}
  +P_{k\, \Lambda} F^{\Lambda} \right)\,.  
\end{equation}

The (exterior derivative of the) Komar charge is given by

\begin{equation}
  [\imath_{k},\mathcal{O}_{s}]\mathbf{L}
  =
  d\mathbf{K}[k]\,,
\end{equation}

\noindent
and the on-shell closed generalized Komar charge is given by the manifestly
symplectic-invariant expression \cite{Mitsios:2021zrn}

\begin{equation}
  \label{eq:4dGKC}
  \mathbf{K}[k]
  =
  -\frac{1}{16\pi G_{N}^{(4)}}\star (e^{a}\wedge e^{b})P_{k\, ab}
  -\frac{1}{32\pi G_{N}^{(4)}}\Omega_{MN}P_{k}{}^{M}F^{N}\,.  
\end{equation}

The generalized Komar 2-form is manifestly gauge invariant. Under the
transformations Eq.~(\ref{eq:deltaCPkM}) of the momentum maps

\begin{equation}
\delta_{C}  \mathbf{K}[k]
=
-\frac{1}{32\pi G_{N}^{(4)}}\Omega_{MN}C^{M}F^{N}\,,
\end{equation}

\noindent
which is non-vanishing but closed on-shell. We get unambiguous relations
between physical quantities when we integrate $\mathbf{K}[k]$ over boundaries
of 3-dimensional hypersurfaces but, anyway, the ambiguity in each of the
integrals is removed by the boundary conditions imposed on the momentum maps
at infinity.

\subsubsection{Global symmetries and scalar charges}
\label{sec-globalsymmetriesd4}

As we have mentioned before, the complex scalar $t$ parametrizes a
SL$(2,\mathbb{R})/$SO$(2)$ coset space and SL$(2,\mathbb{R})$ acts on it
through fractional-linear transformations that leave the scalar kinetic term
in the action invariant. Infinitesimally, these transformations take the
form

\begin{equation}
  \delta_{\alpha}t
  =
  \alpha_{1}K_{1} +\alpha_{+}K_{+}+\alpha_{-}K_{-}\,,
\end{equation}

\noindent
where $\alpha_{1},\alpha_{+},\alpha_{-}$ are infinitesimal global parameters
and where

\begin{equation}
  K_{1} = t\,,
  \hspace{1cm}
  K_{+} = t^{2}\,,
  \hspace{1cm}
  K_{-} = 1\,,
\end{equation}

\noindent
are the single, complex component of the holomorphic Killing
vectors.\footnote{Defining
  \begin{equation}
    K_{2} \equiv \tfrac{1}{2}(K_{-}-K_{+})\,,
    \hspace{1cm}
    K_{3} \equiv \tfrac{1}{2}(K_{-}+K_{+})\,,
  \end{equation}
  it is not difficult to see that they satisfy the
  $\mathfrak{so}(2,1)\sim \mathfrak{sl}(2)$ Lie algebra
  \begin{equation}
    [K_{m},K_{n}]
    =
    +\varepsilon_{mnq}\eta^{qp}K_{p}\,,
    \hspace{1cm}
    \varepsilon_{123}=+1\,,
    \hspace{1cm}
    (\eta^{qp})= \mathrm{diag}(++-)\,.
  \end{equation}
 }

 These transformations act non-trivially on the scalar matrices
 $I_{\Lambda\Sigma}$ and $R_{\Lambda\Sigma}$ and their transformations are
 compensated in the equations of motion extended by the Bianchi identities by
 symplectic transformations of the gauge fields\footnote{Defining
  \begin{equation}
    T_{2} \equiv \tfrac{1}{2}(T_{-}-T_{+})\,,
    \hspace{1cm}
    T_{3} \equiv \tfrac{1}{2}(T_{-}+T_{+})\,,
  \end{equation}
  it is not difficult to see that these matrices satisfy the
  $\mathfrak{so}(2,1)\sim \mathfrak{sl}(2)$ Lie algebra
  \begin{equation}
    [T_{m},T_{n}]
    =
    -\varepsilon_{mnq}\eta^{qp}T_{p}\,.
  \end{equation}
}

\begin{equation}
  \delta_{\alpha}A^{M}
  =
  \left(
    \alpha_{1}T_{1}{}^{M}{}_{N}+\alpha_{+}T_{+}{}^{M}{}_{N} +\alpha_{-}T_{-}{}^{M}{}_{N}
  \right)A^{N}\,,
\end{equation}

\noindent
with

\begin{subequations}
  \begin{align}
    \left(T_{1}{}^{M}{}_{N}\right)
    & =
      \tfrac{1}{2}\left(
      \begin{array}{cccc}
        -3   &    &   &   \\
             & -1 &   &   \\
             &    & 3 &   \\
             &    &   & 1 \\
      \end{array}
    \right)\,,
    \\
    & \nonumber \\
    \left(T_{+}{}^{M}{}_{N}\right)
    & =
      \left(
      \begin{array}{cccc}
           & 3   &     &       \\
           &     &     & -\frac{2}{3}  \\
        0  &     &     &       \\
           &     & -3  &       \\
      \end{array}
    \right)\,,
    \\
    & \nonumber \\
    \left(T_{-}{}^{M}{}_{N}\right)
    & =
      \left(
      \begin{array}{cccc}
          &    & 0 &   \\
      -1  &    &   &   \\
          &    &   & 1 \\
          & 6  &   &   \\
      \end{array}
    \right)\,.
  \end{align}
\end{subequations}

Some of these transformations ($\delta_{+}$) do not leave invariant the action
because they involve electric-magnetic duality transformations of the gauge
fields. In these conditions, the most efficient way to find the closed
Noether-Gaillard-Zumino (NGZ) 3-form currents \cite{Gaillard:1981rj} is to
contract the scalar equations of motion with the real Killing vectors
\cite{Bandos:2016smv}

\begin{equation}
  K_{\alpha}{}^{x}\mathbf{E}_{x}
  \doteq
  d\mathbf{J}_{\alpha}\,.
\end{equation}

The real Killing vectors of the SL$(2,\mathbb{R})/$SO$(2)$ coset space are

\begin{subequations}
  \label{eq:SL2KVFs}
  \begin{align}
    K_{1}
    & =
      \chi\partial_{\chi}-\tfrac{1}{2}\partial_{\phi}\,,
    \\
    & \nonumber \\
    K_{+}
    & =
      (\chi^{2}-e^{-4\phi})\partial_{\chi} -\chi\partial_{\phi}\,,
    \\
    & \nonumber \\
    K_{-}
    & =
      \partial_{\chi}\,,
  \end{align}
\end{subequations}

\noindent
and we find

\begin{subequations}
  \begin{align}
    K_{1}{}^{x}\mathbf{E}_{x}
    & =
      -3d\left(e^{4\phi}\chi\star d\chi -2\star d\phi\right)
      -\tfrac{3}{2} F^{0}\wedge F_{0} -\tfrac{1}{2}F^{1}\wedge F_{1}\,,
    \\
    & \nonumber \\
    K_{+}{}^{x}\mathbf{E}_{x}
      & =
        d\left[12\chi\star d\phi
        -3\left(\chi^{2}e^{4\phi}-1\right)\star d\chi\right]
      +3F^{1}\wedge F_{0}
      -\tfrac{1}{3}F_{1}\wedge F_{1}
        \\
    & \nonumber \\
    K_{-}{}^{x}\mathbf{E}_{x}
    & =
      d\left(-3e^{4\phi}\star d\chi\right)
      -F^{0}\wedge F_{1}
      -3F^{1}\wedge F^{1}\,,
  \end{align}
\end{subequations}

\noindent
that we can rewrite as

\begin{subequations}
  \begin{align}
    K_{1}{}^{x}\mathbf{E}_{x}
    & =
      d\mathbf{J}_{1}
      -\tfrac{3}{2} A^{0}\wedge \mathbf{E}_{0} -\tfrac{1}{2}A^{1}\wedge \mathbf{E}_{1}\,,
    \\
    & \nonumber \\
    \label{eq:K+xEx}
    K_{+}{}^{x}\mathbf{E}_{x}
    & \doteq
      d\mathbf{J}_{+}
      +3A^{1}\wedge \mathbf{E}_{0} -\tfrac{1}{3}A_{1}\wedge \mathbf{E}_{1}\,, 
    \\
    & \nonumber \\
    K_{-}{}^{x}\mathbf{E}_{x}
    & =
      d\mathbf{J}_{-}
      -A^{0}\wedge \mathbf{E}_{1}-3 A^{1}\wedge \mathbf{E}^{1}\,,
  \end{align}
\end{subequations}

\noindent
with

\begin{subequations}
  \begin{align}
    \label{eq:J1}
    \mathbf{J}_{1}
    & =
      \frac{1}{16\pi G_{N}^{(4)}}\left\{
      -3\left(e^{4\phi}\chi\star d\chi -2\star d\phi\right) 
      -\tfrac{3}{2} A^{0}\wedge F_{0} -\tfrac{1}{2}A^{1}\wedge F_{1}
      \right\}\,,
    \\
    & \nonumber \\
    \mathbf{J}_{+}
    & =
      \frac{1}{16\pi G_{N}^{(4)}}\left\{
12\chi\star d\phi -3\left(\chi^{2}e^{4\phi}-1\right)\star d\chi
      +3A^{1}\wedge F_{0} -\tfrac{1}{3}A_{1}\wedge F_{1}\right\}\,,
    \\
    & \nonumber \\
    \label{eq:J-}
\mathbf{J}_{-}
    & =
      \frac{1}{16\pi G_{N}^{(4)}}\left\{
      -3e^{4\phi}\star d\chi -A^{0}\wedge F_{1}-3 A^{1}\wedge F^{1}\right\}\,.
  \end{align}
\end{subequations}

Observe that we have used the local solution $F_{1}=dA_{1}$ of the Maxwell
equation $\mathbf{E}_{1}=0$ in order to find $\mathbf{J}_{+}$. Thus,
Eq.~(\ref{eq:K+xEx}) only holds strictly on-shell. This is related to the fact
that the symmetry generated by $K_{+},T_{+}$ does not leave invariant the
action. It only leaves invariant the equations of motion because it involves
electric-magnetic duality transformations, as we are going to see. The two
generators of symmetries of the action $K_{1},T_{1}$ and $K_{-},T_{-}$
generate a subalgebra

\begin{equation}
  \label{eq:subalgebrad4}
  [K_{1},K_{-}] =-K_{-}\,,
  \hspace{1cm}
  [T_{1},T_{-}] = T_{-}\,,
\end{equation}

\noindent
as is required for consistency.

Following Ref.~\cite{Ballesteros:2023iqb} (see also
Ref.~\cite{Pacilio:2018gom}) we can define 2-form charges using the on-shell
conservation of the 3-form currents plus the assumption of invariance of all
the fields of the theory under the GCTs generated by the Killing vector $l$

\begin{equation}
  d\mathbf{J}
  \doteq
  0\,,
  \hspace{1cm}
  \delta_{l}\mathbf{J}
  =
  \left( -\mathcal{L}_{l}+\delta_{\Lambda_{l}}\right)\mathbf{J}
  =
  0\,.
\end{equation}

Combining these two conditions and taking into account the value of the
induced gauge transformation parameter $\Lambda^{\Lambda}_{l}$ in
Eq.~(\ref{eq:LambdaLambdak}) we find

\begin{subequations}
  \begin{align}
    d\left(-\imath_{l}\mathbf{J}_{1}\right)
    -\tfrac{3}{2} \delta_{\Lambda_{l}}A^{0}\wedge F_{0}
    -\tfrac{1}{2}\delta_{\Lambda_{l}}A^{1}\wedge F_{1}
    & \doteq
      0\,,
          \\
    & \nonumber \\
    d\left(-\imath_{l}\mathbf{J}_{+}\right)
    +3\delta_{\Lambda_{l}}A^{1}F_{0} -\tfrac{1}{3}\delta_{\Lambda_{l}}A_{1}F_{1}
    & \doteq
      0\,,
    \\
    & \nonumber \\
    d\left(-\imath_{l}\mathbf{J}_{-}\right)
    -\delta_{\Lambda_{l}}A^{0}\wedge F_{1}-3 \delta_{\Lambda_{l}}A^{1}\wedge
    F^{1}
    & \doteq
      0\,.
  \end{align}
\end{subequations}

Operating and using the definitions of the electric and magnetic momentum maps
and restoring the normalization factor, we find

\begin{equation}
  d\mathbf{Q}_{l\, A}
  \doteq
  0\,,
\end{equation}

\noindent
for

\begin{subequations}
  \label{eq:scalarcharges}
  \begin{align}
    \label{eq:Ql1}
    \mathbf{Q}_{l\, 1}
    & =
  \frac{1}{16\pi G_{N}^{(4)}}
  \left\{
      3\left(e^{4\phi}\chi\imath_{l}\star d\chi -2\imath_{l}\star d\phi\right) 
      \right.
      \nonumber \\
    & \nonumber \\
    & \hspace{.5cm}
      \left.
      +\tfrac{3}{2} \left(P_{l}{}^{0} F_{0} +P_{l\, 0} F^{0}\right)
    +\tfrac{1}{2}\left(P_{l}{}^{1} F_{1} +P_{l\,1}F^{1}\right)\right\}\,,
    \\
    & \nonumber \\
    \mathbf{Q}_{l\, +}
    & =
  \frac{1}{16\pi G_{N}^{(4)}}
  \left\{
    -12\chi \imath_{l}\star d\phi +3(\chi^{2}e^{4\phi}-1)\imath_{l}\star d\chi
    -3P_{l}{}^{1}F_{0} -3P_{l\, 0}F^{1} +\tfrac{2}{3}P_{l\, 1}F_{1}
  \right\}\,,
    \\
    & \nonumber \\
    \label{eq:Ql-}
    \mathbf{Q}_{l\, -}
    & =
  \frac{1}{16\pi G_{N}^{(4)}}
  \left\{
      3e^{4\phi}\imath_{l}\star d\chi
    +\left(P_{l}{}^{0} F_{1}+P_{l\, 1} F^{0}\right)
    +6P_{l}{}^{1}F^{1}\right\}\,,
  \end{align}
\end{subequations}

\noindent
in full agreement with the general result of Ref.~\cite{Ballesteros:2023iqb}

\begin{equation}
  \label{eq:scalarchargesgeneral}
  \mathbf{Q}_{l\, A}
  =
  \frac{1}{16\pi G_{N}^{(4)}}
  \left\{
    K_{A\, x}\imath_{l}\star d\phi^{x}
    +T_{A\, MN}P_{l}{}^{M}F^{N}
  \right\}\,.
\end{equation}

\subsection{Black-hole thermodynamics}

\subsubsection{Generalized zeroth laws}

If $l$ is the Killing vector normal to the bifurcate horizon of a stationary
black hole, $l\stackrel{\mathcal{BH}}{=}0$ and, from the definition of the
momentum maps $P_{l}{}^{M}$ Eq.~(\ref{eq:symplecticmomentummaps}) it follows
that they can be identified with the (corotating) electric and magnetic
potentials $\Phi^{M}$ and that 

\begin{equation}
  d \Phi^{M}
  \stackrel{\mathcal{BH}}{=}
  0\,,
\end{equation}

\noindent
(restricted, generalized zeroth laws). Furthermore, because, from the momentum
map equation, $\imath_{l}dP_{l}{}^{M}=0$, it follows that

\noindent
(generalized zeroth laws).

\begin{equation}
  d\Phi^{M}
  \stackrel{\mathcal{H}}{=}
  0\,.
\end{equation}

\noindent
(generalized zeroth laws).

\subsubsection{Smarr formula for asymptotically-flat black holes}
\label{sec-Smarrd4}

The event horizon of asymptotically-flat, stationary black holes is the
Killing horizon of a Killing vector that, in adapted coordinate, can be
written as

\begin{equation}
  \label{eq:4dgenerator}
  l
  =
  \partial_{t}-\Omega\partial_{\varphi}\,,
\end{equation}

\noindent
where $\varphi$ is the angular coordinate around the axis of rotation and
$\Omega$ is the angular velocity of the horizon. Assuming that the horizon is
bifurcate ($l\stackrel{\mathcal{BH}}{=}0$), we can choose a spacelike
hypersurface $\Sigma^{3}$ with boundary
$\partial\Sigma^{3}= \mathcal{BH}\cup S^{2}_{\infty}$ and, integrating
$d\mathbf{K}[l]\doteq 0$, where $\mathbf{K}[l]$ is the 4-dimensional
generalized Komar charge in Eq.~(\ref{eq:4dGKC}), over $\Sigma^{3}$ and
applying Stokes' theorem we arrive to

\begin{equation}
  \int_{\mathcal{BH}}\mathbf{K}[l]
  \doteq
  \int_{S^{2}_{\infty}}\mathbf{K}[l]\,.  
\end{equation}

The electric and magnetic momentum maps are defined up to an additive constant
which gives the only contribution of the electric and magnetic terms 
to the integral over $S^{2}_{\infty}$ \cite{Zatti:2024vqv}

\begin{equation}
     -\tfrac{1}{32\pi G_{N}^{(4)}}P_{l}{}^{M}
     \int_{S^{2}_{\infty}}F_{M}
     =
     \tfrac{1}{2}\Phi_{\infty}^{M}\int_{S^{2}_{\infty}}\mathbf{Q}_{M}
     =
     \tfrac{1}{2}\Phi_{\infty}^{M}\mathcal{Q}_{N}
     =
     0\,,
\end{equation}

\noindent
where we have used the definitions of electric and magnetic charges
Eqs.~(\ref{eq:electricchargesd4def}), (\ref{eq:magneticchargesd4def}) and
(\ref{eq:electricandmagneticchargesd4def}) and our choice
$\Phi_{\infty}^{M}=0$.

The remaining term is the standard Komar integral, 

\begin{equation}
  -\frac{1}{16\pi G_{N}^{(4)}}  \int_{S^{2}_{\infty}}\star (e^{a}\wedge e^{b})P_{l\, ab}
  =
  \tfrac{1}{2}M -\Omega J \,,
\end{equation}

\noindent
where $J$ is the angular momentum associated to $\varphi$, and

\begin{equation}
  \int_{S^{2}_{\infty}}\mathbf{K}[l]
  =
  \tfrac{1}{2}M -\Omega J\,.
\end{equation}

Let us now consider the integral over the bifurcation sphere.  Using the
generalized zeroth laws and the Gauss laws satisfied by the electric and
magnetic charges, the electric and magnetic terms give

\begin{equation}
  -\tfrac{1}{32\pi G_{N}^{(4)}}\Phi_{\mathcal{BH}}^{M}
  \int_{\mathcal{BH}}F_{M}
  =
  \tfrac{1}{2}
  \Phi_{\mathcal{H}}^{M}\mathcal{Q}_{M}\,.
\end{equation}

The standard Komar term gives

\begin{equation}
      -\frac{1}{16\pi G_{N}^{(4)}}  \int_{\mathcal{BH}}\star
      (e^{a}\wedge e^{b})P_{l\, ab}
      =
      \frac{\kappa \mathcal{A}_{\mathcal{H}}}{8\pi G_{N}^{(4)}}
      = TS\,,
\end{equation}

\noindent
where $\kappa$ is the surface gravity of the horizon, $A$ is its area,
$T=\frac{\kappa}{2\pi}$ is the Hawking temperature and
$S=\frac{\mathcal{A}_{\mathcal{H}}}{4G_{N}^{(4)}}$ is its Bekenstein-Hawking
entropy.

Thus, we obtain the symplectic-invariant Smarr formula

\begin{equation}
  \label{eq:4dSmarrformula}
  M= 2TS+2\Omega J+\Phi_{\mathcal{H}}^{M}\mathcal{Q}_{M}\,.  
\end{equation}

\subsubsection{No-hair theorem}
\label{sec-4dnohairtheorem}

Since the 2-forms
Eqs.~(\ref{eq:scalarcharges}),(\ref{eq:scalarchargesgeneral}) are closed
on-shell, the scalar charges satisfy Gauss laws. In a stationary,
asymptotically-flat black hole with bifurcate horizon, integrating over the
same hypersurface $\Sigma^{3}$ we used to derive the Smarr formula in
Section~\ref{sec-Smarrd4}, we find 

\begin{equation}
  \int_{\mathcal{BH}} \mathbf{Q}_{l\, A}
  \doteq
  \int_{S^{2}_{\infty}}\mathbf{Q}_{l\, A}\,.  
\end{equation}

The integral on the right-hand side gives the scalar charges
$\mathcal{Q}_{A}$, by definition. The integral in the left-hand side gives

\begin{equation}
  \int_{\mathcal{BH}} \mathbf{Q}_{l\, A}
  =
  -T_{A\, MN}\Phi_{\mathcal{H}}^{M}\mathcal{Q}^{N}\,,
\end{equation}

\noindent
which implies that, in presence of a bifurcate horizon, the scalar charges are
constrained to depend on the rest of the charges according to the formula
Ref.~\cite{Ballesteros:2023iqb}

\begin{equation}
  \mathcal{Q}_{A}
  =
  -T_{A\, MN} \Phi_{\mathcal{H}}^{M}\mathcal{Q}^{N}\,,
\end{equation}

\noindent
which means that these black holes only admit secondary scalar
hair.\footnote{This expression takes into account that we have ser
  $\Phi_{\infty}^{M}=0$.}

For the scalar charges of the T$^{3}$ model Eqs.~(\ref{eq:scalarcharges}), we
find the explicit relations

\begin{subequations}
  \label{eq:scalarchargesintegrated}
  \begin{align}
    \label{eq:scalarchargeQ1}
    \mathcal{Q}_{1}
    & =
      -\tfrac{3}{2} \Phi_{\mathcal{H}}^{0} q_{0}
      -\tfrac{3}{2}\Phi_{\mathcal{H}\, 0} p^{0}
      -\tfrac{1}{2}\Phi_{\mathcal{H}}^{1} q_{1}
      -\tfrac{1}{2}\Phi_{\mathcal{H}\,1}p^{1}\,,
    \\
    & \nonumber \\
    \mathcal{Q}_{+}
    & =
      3\Phi_{\mathcal{H}}^{1}q_{0}
      +3\Phi_{\mathcal{H}\, 0}p^{1}
      -\tfrac{2}{3}\Phi_{\mathcal{H}\,1}q_{1}\,,
    \\
    & \nonumber \\
    \label{eq:scalarchargeQ-}
    \mathcal{Q}_{-}
    & =
      -\Phi_{\mathcal{H}}^{0} q_{1}
      -\Phi_{\mathcal{H}\, 1} p^{0}
    -6\Phi_{\mathcal{H}}^{1}p^{1}\,.
  \end{align}
\end{subequations}

The conventional definition of the dilaton and axion scalar charges
$\Sigma^{\phi}$ and $\Sigma^{\chi}$ based on their asymptotic behavior would
be

\begin{subequations}
  \label{eq:conventionaldefinitionscalarcharges}
  \begin{align}
    \phi
    & \sim
      \phi_{\infty}+\frac{G_{N}^{(4)}\Sigma^{\phi}}{r}
      +\mathcal{O}\left(\frac{1}{r^{2}}\right)\,,
    \\
    & \nonumber \\
    \chi
    & \sim \chi_{\infty} +\frac{G_{N}^{(4)}\Sigma^{\chi} }{r}
      +\mathcal{O}\left(\frac{1}{r^{2}}\right)\,.
  \end{align}
\end{subequations}

\noindent
These charges can be obtained from the integrals at infinity

\begin{subequations}
  \label{eq:covariantdefinitionscalarcharges}
  \begin{align}
    \Sigma^{\phi}
    & =
      \frac{1}{4\pi G_{N}^{(4)}}\int_{S^{2}_{\infty}}\imath_{l}\star d\phi
      =
      \tfrac{1}{6}\int_{S^{2}_{\infty}}\left(\chi \mathbf{Q}_{l\, -}-\mathbf{Q}_{l\, 1}\right)\,,
    \\
    & \nonumber \\
    \Sigma^{\chi}
    & =
      \frac{1}{4\pi G_{N}^{(4)}}\int_{S^{2}_{\infty}}\imath_{l}\star d\chi
      =
      \frac{e^{-4\phi_{\infty}}}{3}\int_{S^{2}_{\infty}}\mathbf{Q}_{l\, -}\,,
  \end{align}
\end{subequations}

\noindent
and are related to the charges $\mathcal{Q}_{A}$ by

\begin{equation}
  \label{eq:SigmaxversusQA}
  \Sigma^{x}
  =
  4\mathcal{Q}_{A} g^{AB}K_{B}{}^{x}(\phi_{\infty})\,.  
\end{equation}

Using the values (\ref{eq:scalarchargesintegrated}) of the charges
$\mathcal{Q}_{A}$, of the inverse Killing metric $g^{AB}$ in
Eq.~(\ref{eq:gABmetric}) and of the Killing vectors $K_{A}{}^{x}$
Eqs.~(\ref{eq:SL2KVFs}) we find that the scalar charges $\Sigma^{\phi}$ and
$\Sigma^{\chi}$ are given by

\begin{subequations}
  \label{eq:valuesscalarcharges}
  \begin{align}
    \Sigma^{\phi}
    & =
      \Phi_{\mathcal{H}}^{0}\left(q_{0} -\tfrac{2}{3}\chi_{\infty}q_{1}\right)
      +\Phi_{\mathcal{H}\, 0}p^{0}
     +\Phi_{\mathcal{H}}^{1}\left(\tfrac{1}{3}q_{1} -4\chi_{\infty}p^{1}\right)   
     +\Phi_{\mathcal{H}\, 1}\left(\tfrac{1}{3}p^{1}
      -\tfrac{2}{3}\chi_{\infty}p^{0}\right)\,,
    \\
    & \nonumber \\
    \Sigma^{\chi}
    & =
      \Phi_{\mathcal{H}}^{0}\left[-2\chi_{\infty}q_{0}
      +\tfrac{2}{3}\left(\chi_{\infty}^{2}-e^{-4\phi_{\infty}}\right)q_{1}\right]
      +\Phi_{\mathcal{H}\, 0}\left(-2\chi_{\infty}p^{0}-2p^{1}\right)
      \nonumber \\
    & \nonumber \\
    & \hspace{.5cm}
      +\Phi_{\mathcal{H}}^{1}\left[-\tfrac{2}{3}\chi_{\infty}q_{1}
      +4\left(\chi_{\infty}^{2}-e^{-4\phi_{\infty}}\right)p^{1}
      -2q_{0}\right]   
      \nonumber \\
    & \nonumber \\
    & \hspace{.5cm}
      +\Phi_{\mathcal{H}\, 1}\left[-\tfrac{2}{3}\chi_{\infty}p^{1}
      +\tfrac{2}{3}\left(\chi_{\infty}^{2}-e^{-4\phi_{\infty}}\right)p^{0}
      +\tfrac{4}{9}q_{1}\right]\,.
  \end{align}
\end{subequations}

\subsubsection{First law}

In this section we will outline the detailed derivation of the first law
carried out in Ref.~\cite{Ballesteros:2023iqb} making some small corrections.
Since we are going to follow the same procedure we employed in the
5-dimensional case (Section~\ref{sec-firstlawminimal5dsugra}), we shall be
brief.

We start by defining the symplectic 3-form
Eq.~(\ref{eq:symplecticpotentialdef}) for $\delta_{1}=\delta$ and
$\delta_{2}=\delta_{\xi}$ defined as as indicated there and replacing
$\mathbf{\Theta}_{T^{3}}(\varphi,\delta_{\xi}\varphi)$ in
Eq.~(\ref{eq:Thetad4general}) by

\begin{equation}
  \mathbf{\Theta}_{T^{3}}'(\varphi,\delta_{\xi}\varphi)
  \equiv
  \mathbf{\Theta}_{T^{3}}(\varphi,\delta_{\xi}\varphi)
  +\mathbf{E}_{a}\xi^{a}+\mathbf{E}_{\Sigma}P_{\xi}{}^{\Sigma}\,,
\end{equation}

\noindent
because we are going to derive on-shell identities.
$ \mathbf{\Theta}_{T^{3}}'(\varphi,\delta_{\xi}\varphi)$ is then replaced
by\footnote{The Lagrangian is invariant up to a total derivative under GCTs,
  but exactly under gauge transformations.}

\begin{equation}
d\mathbf{Q}_{NW}[\xi]-\imath_{\xi}\mathbf{L}\,,  
\end{equation}

\noindent
and we end up with the analog of Eq.~(\ref{eq:symplecticpotentialCS2})

\begin{equation}
  \label{eq:symplecticpotentialT3-2}
  \omega(\varphi,\delta\varphi,\delta_{\xi}\varphi)
   \doteq
    d\left[\delta\mathbf{Q}_{NW}[\xi]
    +\imath_{\xi}\mathbf{\Theta}(\varphi,\delta\varphi)\right]
      -\delta_{\Lambda_{\xi}} \mathbf{\Theta}(\varphi,\delta\varphi)
      -\mathbf{\Theta}(\varphi,\delta_{\delta\Lambda_{\xi}}\varphi)\,.
\end{equation}

The last two terms combine into a total derivative and

\begin{equation}
  \label{eq:symplecticpotentialT3-3}
  \begin{aligned}
  \omega(\varphi,\delta\varphi,\delta_{\xi}\varphi)
   & \doteq
    d\left[\delta\mathbf{Q}_{NW}[\xi]
      +\imath_{\xi}\mathbf{\Theta}_{T^{3}}(\varphi,\delta\varphi)
      +\star (e^{a}\wedge e^{b})\delta\sigma_{\xi\, ab}
     +F_{\Sigma}\wedge \delta \Lambda_{\xi}^{\Sigma}\right]
    \\
    & \\
    & \equiv
      d\mathbf{W}[\xi]\,.
  \end{aligned}
\end{equation}

When $\xi=l$ generates a symmetry of the configurations under consideration
(so that, in particular, it is a Killing vector) 

\begin{equation}
  d\mathbf{W}[l]
  \doteq
  0\,.
\end{equation}

Now we just have to compute the explicit form of $\mathbf{W}[l]$ using the
explicit form of the Noether--Wald 2-form Eq.~(\ref{eq:NoetherWaldcharged4})
and the explicit form of the presymplectic potential 3-form
Eq.~(\ref{eq:Thetad4general})

\begin{equation}
  \begin{aligned}
    \mathbf{W}[l]
    & \doteq
      \delta\star (e^{a}\wedge e^{b})P_{l\, ab}
      -\imath_{l}\star (e^{a}\wedge e^{b})\wedge \delta \omega_{ab}
      +P_{l}{}^{M}\delta F_{M}
      +g_{xy}\imath_{l} \star d\phi^{x}\delta\phi^{y}\,,
  \end{aligned}
\end{equation}

\noindent
up to a total derivative.

Following Ref.~\cite{Ballesteros:2023iqb}, when the scalar metric
$g_{xy}(\phi)$ is the left-invariant metric of a Riemannian symmetric space,
we can use the identity\footnote{The metric $g_{AB}$ in the Lie algebra and
  its inverse are
  \begin{equation}
    \label{eq:gABmetric}
  \left(g_{AB}\right)
  =
  3\left(
    \begin{array}{ccc}
     1   &    &    \\
         &    & -2 \\
         & -2 &    \\
    \end{array}
  \right)\,,
  \hspace{1cm}
    \left(g^{AB}\right)
  =
\tfrac{1}{3}\left(
    \begin{array}{ccc}
     1   &               &               \\
         &               & -\tfrac{1}{2} \\
         & -\tfrac{1}{2} &               \\
    \end{array}
  \right)\,.
\end{equation}
} $g_{xy}=g^{AB}K_{A\, x} K_{B\, y}$ in the last term,
which can be rewritten in the form 

\begin{equation}
  g_{xy}\imath_{l}\star d\phi^{x}\delta\phi^{y}
  =
  g^{AB} \imath_{l}\star \hat{K}_{A} K_{B\, y}\delta\phi^{y}
  =
  \left(\mathbf{Q}_{A}[l]-T_{A\, MN}P_{k}{}^{M}F^{N}\right)\delta^{A}\,.
\end{equation}

\noindent
where $\mathbf{Q}_{A}[l]$ is the scalar charge associated to the Killing
vector $l$ and the target-space Killing vector $K_{A}$,
$\hat{K}_{A}= K_{A}{}^{z}g_{zw}\partial_{\mu}\phi^{w}dx^{\mu}$ is the pullback
of the target-space 1-form dual to $K_{A}$ and where we have defined
variations of the scalars in the direction of the Killing vectors

\begin{equation}
  \label{eq:deltaAdef}
\delta^{A} \equiv g^{AB}K_{B\, y}\delta\phi^{y}\,.  
\end{equation}

The final result for $\mathbf{W}[l]$  is

\begin{equation}
  \begin{aligned}
  \mathbf{W}[l]
  & \equiv
  \delta\star (e^{a}\wedge e^{b})P_{l\, ab}
  -\imath_{l}\star (e^{a}\wedge e^{b})\wedge \delta \omega_{ab}
    +P_{l}{}^{M}\delta F_{M}
    \\
    & \\
    & \hspace{.5cm}
      +\left(\mathbf{Q}_{A}[l]
      -T_{A\, MN}P_{k}{}^{M}F^{N}\right)\delta^{A}\,,
  \end{aligned}
\end{equation}

\noindent
in full agreement with Ref.~\cite{Ballesteros:2023iqb}.  Thus, it leads to the
same first law, first proposed in Ref~\cite{Gibbons:1996af}, which we just
quote from that reference, setting all $\Phi^{M}_{\infty}=0$:

\begin{equation}
    \label{eq:firstlaw2}
  \delta M
=
\frac{\kappa \delta A_{\mathcal{H}}}{8\pi G_{N}^{(4)}}
+\Omega\delta J   +\Phi^{M}_{\mathcal{H}}\delta Q_{M}
  -\mathcal{Q}_{A}\delta^{A}_{\infty}\,.
\end{equation}

The last term, involving the scalar charges can be expressed in terms of the
electric and magnetic charges and their thermodynamical potentials using the
no-hair theorem reviewed in Section~\ref{sec-4dnohairtheorem}

\begin{equation}
  \mathcal{Q}_{A}
  =
  -T_{A\, MN}\Phi^{M}_{\mathcal{H}}Q^{N}\,, 
\end{equation}

\noindent
or using the scalar charges defined through the asymptotic expansions

\begin{equation}
  \phi^{x}
  \sim
  \phi^{x}_{\infty} +\frac{G_{N}^{(4)}\Sigma^{x}}{r}\,,
\end{equation}

\noindent
which are related to the former by Eq.~(\ref{eq:SigmaxversusQA}).  In terms of
these

\begin{equation}
  \mathcal{Q}_{A}\delta^{A}_{\infty}
  =
  \tfrac{1}{4}\Sigma^{x}g_{xy}(\phi_{\infty})\delta \phi^{x}_{\infty}\,,
\end{equation}

\noindent
and, for the particular case of the T$^{3}$ model

\begin{equation}
  \mathcal{Q}_{A}\delta^{A}_{\infty}
  =
    \tfrac{3}{4}\Sigma^{\chi}e^{4\phi_{\infty}}\delta \chi_{\infty}
    +3\Sigma^{\phi}\delta \phi_{\infty}\,.
\end{equation}

\section{Relating the 5- and 4-dimensional theories}
\label{sec-relation5d4d}

Minimal 5-dimensional Supergravity and the T$^{3}$ model are related by
dimensional reduction. In the next section we are going to review this
reduction in full detail so that we can use in the next two sections to relate
the conserved charges and the thermodynamics of 5- and 4-dimensional black
objects of these theories.

\subsection{Kaluza-Klein dimensional reduction of minimal 5-dimensional supergravity }
\label{sec-KKreduction}

The dimensional reduction of minimal 5-dimensional Supergravity is an
extension of the dimensional reduction of pure 5-dimensional Einstein gravity
revisited in Ref.~\cite{Gomez-Fayren:2023wxk}. We will use the conventions and
results of that reference here, reviewing them briefly under the light of
Ref.~\cite{Gomez-Fayren:2024cpl} and extending them afterwards.\footnote{The
  dimensional reduction of the bosonic sector of minimal 5-dimensional
  supergravity over a circle can also be found in Ref.~\cite{Ortin:2015hya}.}

As usual, we will denote all 5-dimensional quantities and objects
(fields, indices...) with hats, to distinguish them from the 4-dimensional
(unhatted) ones.

We assume that the metric admits an isometry generated by a spacelike Killing
vector\footnote{The underlined index $\underline{z}$ is a ``world'' index
  (associated to the coordinate basis). The not-underlined index $z$ which we
  will use later is a tangent-space Lorentz, index.}

\begin{equation}
  \hat{k}
  =
  \hat{k}^{\hat{\mu}}\partial_{\hat{\mu}}
  =
  \partial_{\underline{z}}\,,
\end{equation}

\noindent
and the coordinates we are using $(x^{\hat{\mu}})=(x^{\mu},x^{4}\equiv z)$ are
coordinates adapted to the isometry. Furthermore, we assume that the orbits of
$\hat{k}$ are periodic, which means that 5\textsuperscript{th} coordinate is
periodic

\begin{equation}
z\sim z+2\pi \ell\,,  
\end{equation}

\noindent
where $\ell$ is some length scale. The components of the 5-dimensional metric

\begin{equation}
  ds_{(5)}^{2}
  =
  \hat{g}_{\hat{\mu}\hat{\nu}}dx^{\hat{\mu}}x^{\hat{\nu}}\,.
\end{equation}

\noindent
are related to those of the 4-dimensional metric $g_{\mu\nu}$, KK gauge field
$A_{\mu}$ and KK scalar $k$ by the well-known relations

\begin{subequations}
  \begin{align}
 g_{\mu\nu}  
  & =
  \hat{g}_{\mu\nu}
  -\hat{g}_{\mu\underline{z}}\hat{g}_{\nu\underline{z}}/\hat{g}_{\underline{z}\underline{z}}\,,
    \\
    & \nonumber \\
    A_{\mu}
    & =
      \hat{g}_{\mu\underline{z}}/\hat{g}_{\underline{z}\underline{z}}\,,
    \\
    & \nonumber \\
    k^{2}
  & = 
   -\hat{g}_{\underline{z}\underline{z}}\,,
  \end{align}
\end{subequations}

\noindent
whose inverse is

\begin{equation}
  ds_{(5)}^{2}
   =
   ds_{(4)}^{2} -k^{2}\left(dz+A\right)^{2}\,,
   \hspace{1cm}
   ds_{(4)}^{2}
  =
  g_{\mu\nu}dx^{\mu}dx^{\nu}\,.
\end{equation}

The dimensional reduction of the 5-dimensional gauge field $\hat{V}$ is
based on the assumption that the GCT generated by Killing vector $\hat{k}$
leaves it invariant, that is

\begin{equation}
  \delta_{\hat{k}}\hat{V}
  =
  -\mathcal{L}_{\hat{k}}\hat{V} +\delta_{\hat{\Lambda}_{\hat{k}}}\hat{V}
  =
  0\,,
\end{equation}

\noindent
where the parameter of the induced gauge transformations is defined in
Eq.~(\ref{eq:Lambdak}). However, it is always performed in a gauge in which
none of its components depends on $z$, \textit{i.e.},

\begin{equation}
  \label{eq:partialzV=0}
  \partial_{\underline{z}}\hat{V}_{\hat{\mu}}
  =
  0\,.
\end{equation}

Since we are working in adapted coordinates, this implies that

\begin{equation}
  \label{eq:LkV=0}
  \mathcal{L}_{\hat{k}}\hat{V}
  =
  0\,,  
\end{equation}

\noindent
which implies 

\begin{equation}
  \label{eq:PkinKKtheory2}
  \hat{P}_{\hat{k}}
  =
  \imath_{\hat{k}}\hat{V}\,.  
\end{equation}

The 5-dimensional gauge field $\hat{V}$ can be decomposed into a 4-dimensional
gauge field $V=V_{\mu}dx^{\mu}$ with field strength $G=dV$ and a 4-dimensional
scalar $l$ as follows:

\begin{subequations}
 \begin{align}
   \hat{V}_{\underline{z}}
   & =  l\,,
   \\
& \nonumber  \\
   \hat{V}_{\mu}
   & =
     V_{\mu}+lA_{\mu}\,,
 \end{align}
\end{subequations}

\noindent
and the inverse relations are

\begin{subequations}
  \begin{align}
    l
    & =
      \hat{V}_{\underline{z}}\,,
    \\
& \nonumber \\
    V_{\mu}
    & =
      \hat{V}_{\mu} -\hat{V}_{\underline{z}}
\hat{g}_{\mu\underline{z}}/\hat{g}_{\underline{z}\underline{z}}\,.
  \end{align}
\end{subequations}

Notice that, due to our gauge choices and definitions, the scalar $l$ is
identical to the 5-dimensional Maxwell momentum map associated to $\hat{k}$

\begin{equation}
\label{eq:PkinKKtheory3}
  l
  =
\hat{P}_{\hat{k}}\,.  
\end{equation}

As we have done in the previous sections, it is convenient to use
differential-form language. The decomposition of the 5-dimensional fields
reads in this language

\begin{subequations}
  \label{eq:formreductionformulae}
  \begin{align}
    \hat{e}^{a}
    & =
      e^{a}\,,
    \\
    & \nonumber \\
    \hat{e}^{z}
    & =
      k(dz+A)\,,
    \\
    & \nonumber \\
    \hat{\star} (\hat{e}^{\hat{a}}\wedge \hat{e}^{\hat{b}})\wedge
    \hat{R}_{\hat{a}\hat{b}}
    & =
      dz\wedge\left\{  -k\star (e^{a}\wedge e^{b})\wedge R_{ab}
      +\tfrac{1}{2}k^{3}F\wedge \star F
      +d\left[2\star dk\right]\right\}\,,
    \\
    & \nonumber \\
    \hat{V}
    & =
      V+l(dz+A)\,,
    \\
    & \nonumber \\
    \hat{G}
    & =
      G+lF +dl\wedge (dz+A)\,,
    \\
    & \nonumber \\
    \hat{\star}\hat{G}
    & =
      -k\star (G+lF)\wedge (dz+A) +k^{-1}\star dl\,.      
  \end{align}
\end{subequations}

Using these relations, the action of minimal 5-dimensional supergravity
Eq.~(\ref{eq:minimalN1d5action}) takes the form

\begin{equation}
  \begin{aligned}
    S[\hat{e},\hat{V}]
    & =
  \frac{1}{16\pi G_{N}^{(5)}} \int 
        dz\wedge \left\{
        -k\star (e^{a}\wedge e^{b})\wedge R_{ab}
        +\tfrac{1}{2}k^{3}F\wedge \star F
        \right.
    \\
      & \\
      & \hspace{.5cm}
        \left.
        +\tfrac{1}{2}k^{-1}dl\wedge \star dl
        +\tfrac{1}{2}k(G+lF)\wedge \star (G+lF)
        \right.
    \\
      & \\
      & \hspace{.5cm}
        \left.
        +\tfrac{1}{3^{3/2}}\left[l(G+lF)\wedge (G+lF)
        -2dl\wedge (G+lF)\wedge V \right]
        \right.
    \\
      & \\
      & \hspace{.5cm}
        \left.
        +d\left[2\star dk\right]
        \right\}\,.
  \end{aligned}
\end{equation}

In order to rewrite this action in a more standard fashion, we integrate by
parts the second term in the third line

\begin{equation}
  \begin{aligned}
    -2dl \wedge (G+lF)\wedge V 
    & =
d\left[-2l  G \wedge V -l^{2} F\wedge V \right]   
+2l  G \wedge G +l^{2} F\wedge G\,,
  \end{aligned}
\end{equation}

\noindent
which leaves us with

\begin{equation}
  \begin{aligned}
    S[\hat{e},\hat{V}]
    & =
  \frac{1}{16\pi G_{N}^{(5)}} \int 
        kdz\wedge \left\{
        -\star (e^{a}\wedge e^{b})\wedge R_{ab}
        +\tfrac{1}{2}k^{-2}dl\wedge \star dl
        \right.
    \\
      & \\
      & \hspace{.5cm}
        \left.
        +\tfrac{1}{2}(k^{2}+l^{2})F\wedge \star F
        +lG\wedge \star F
        +\tfrac{1}{2}G\wedge \star G
        \right.
    \\
      & \\
      & \hspace{.5cm}
        \left.
        +\tfrac{1}{3^{3/2}}k^{-1}l^{3}F\wedge F
        +\tfrac{1}{3^{1/2}}k^{-1}l^{2}F\wedge G
        +\tfrac{1}{3^{1/2}}k^{-1}l  G \wedge G
        \right.
    \\
      & \\
      & \hspace{.5cm}
        \left.
        +k^{-1}d\left[2\star dk
        -\tfrac{2}{3^{3/2}}l  G \wedge V -\tfrac{1}{3^{3/2}}l^{2} F\wedge V
\right]
        \right\}\,.
  \end{aligned}
\end{equation}

The next step consists in rescaling the metric to the Einstein frame. If
$k_{\infty}$ is the asymptotic value of the KK scalar at infinity, related to
the physical radius of the compact dimension at infinity $R_{z}$ and to the
length scale $\ell$ by

\begin{equation}
  \label{eq:kinftyRzl}
k_{\infty} = R_{z}/\ell\,,  
\end{equation}

\noindent
the 4-dimensional Einstein-frame fields are defined as

\begin{subequations}
  \begin{align}
    g_{\mu\nu}
    & =
      \left(k/k_{\infty}\right)^{-1}g_{E\, \mu\nu}\,,
    \\
    & \nonumber \\
    e^{a}{}_{\mu}
    & =
      \left(k/k_{\infty}\right)^{-1/2}e_{E}{}^{a}{}_{\mu}\,,
    \\
    & \nonumber \\
    A_{\mu}
    & =
      k_{\infty}^{1/2}A_{E\, \mu}\,,
    \\
    & \nonumber \\
    V_{\mu}
    & =
      k_{\infty}^{1/2}V_{E\, \mu}\,,
  \end{align}
\end{subequations}

\noindent
and, in terms of them, and integrating over $z$, the action takes the form

\begin{equation}
  \begin{aligned}
    S[\hat{e},\hat{V}]
    & =
      \frac{2\pi \ell k_{\infty}}{16\pi G_{N}^{(5)}} \int 
      \left\{
      -\star_{E}(e_{E}{}^{a}\wedge e_{E}{}^{b})
      \wedge R_{E\, ab}
      \right.
    \\
    & \\
    & \hspace{.5cm}
      \left.
      +\tfrac{3}{2}d\log{k}\wedge \star_{E} d\log{k}
      +\tfrac{1}{2}k^{-2}dl\wedge \star_{E} dl
      \right.
    \\
    & \\
    & \hspace{.5cm}
      \left.
      +\tfrac{1}{2}k(k^{2}+l^{2})F_{E}\wedge \star_{E} F_{E}
      +klG_{E}\wedge \star_{E} F_{E}
      +\tfrac{1}{2}kG_{E}\wedge \star_{E} G_{E}
      \right.
    \\
    & \\
    & \hspace{.5cm}
      \left.
      +\tfrac{1}{3^{3/2}}l^{3}F_{E}\wedge F_{E}
      +\tfrac{1}{3^{1/2}}l^{2}F_{E}\wedge G_{E}
      +\tfrac{1}{3^{1/2}}l  G_{E} \wedge G_{E}
      \right.
    \\
    & \\
    & \hspace{.5cm}
      \left.
      +d\left[-\star_{E}d\log{k}
      -\tfrac{2}{3^{3/2}}l  G_{E} \wedge V_{E} -\tfrac{1}{3^{3/2}}l^{2} F_{E}\wedge V_{E}
      \right]
      \right\}\,.
  \end{aligned}
\end{equation}

Identifying (see Eq.~(\ref{eq:kinftyRzl}))

\begin{subequations}
  \begin{align}
    G_{N}^{(4)}
    & =
      (2\pi R_{z})^{-1}G_{N}^{(5)}\,,
    \\
    & \nonumber \\
    l & =
        -\sqrt{3}\chi\,,
    \\
    & \nonumber \\
    k & =
        e^{-2\phi}\,,
    \\
    & \nonumber \\
    A_{E}
    & =
      A^{0}\,,
    \\
    & \nonumber \\
    V_{E}
    & =
      -\sqrt{3}A^{1}\,.
  \end{align}
\end{subequations}

\noindent
this action is, up to a total derivative, the action of the T$^{3}$ model
Eq.~(\ref{eq:t3action}) described in Section~\ref{sec-thet3model}

\begin{equation}
  S[\hat{e},\hat{V}]
  =
  S_{T^{3}}[e,\phi,A] +\int d\left[-\star_{E}d\log{k}
      +2\chi F^{1}\wedge A^{1} +\chi^{2} F^{0}\wedge A^{1}
      \right]\,. 
\end{equation}

Taking into account all the intermediate steps one can find the direct
relation between the fields of minimal 5-dimensional supergravity and the
fields of the T$^{3}$ model (see Appendix~\ref{app-4-5fields}).

\subsection{Kaluza-Klein dimensional reduction of the global and local symmetries }
\label{sec-KKreductionsymmetries}

The definitions of the of the 4-dimensional fields defined in the previous
section are justified by the behavior of the under the 5-dimensional GCTs
that respect the $z$-independence of the metric. These are generated by
$z$-independent 5-dimensional vector fields $\hat{\xi}^{\hat{\mu}}(x)$ and
their action on the 4-dimensional fields can be interpreted as that of a
4-dimensional GCT generated by the 4-dimensional vector

\begin{equation}
\xi^{\mu}(x)\equiv \hat{\xi}^{\mu}(x)\,,
\end{equation}

\noindent
plus standard gauge transformations generated by the gauge parameter

\begin{equation}
  \Lambda^{0}(x) \equiv -k_{\infty}^{-1/2}\hat{\xi}^{\underline{z}}(x)\,,
\end{equation}

\noindent
acting on the  $A^{0}$ only as

\begin{equation}
\delta_{\chi}A^{0} = d\Lambda^{0}\,.  
\end{equation}

As stressed in Ref.~\cite{Gomez-Fayren:2024cpl}, $z\partial_{\underline{z}}$,
which also preserves the $z$-independence of the 5-dimensional metric, is not
a well-defined 5-dimensional vector field because it is not single-valued. The
global higher-form transformations studied in
Section~\ref{sec-globalhigherformsymmetriesd5}, turn out to have the same effect
on the 4-dimensional fields, though.

In the notation used in this section, the transformations in
Eqs.~(\ref{eq:deltaepsilonp+1forms}) read

\begin{subequations}
  \begin{align}
  \label{eq:highersymmetriesEHghats}
  \delta_{\epsilon}\hat{g}_{\hat{\mu}\hat{\nu}}
  & =
  -2 \epsilon \mathfrak{h}^{(1)}_{(\hat{\mu}}\hat{k}_{\hat{\nu})}
  + \tfrac{2}{3}\epsilon \hat{g}_{\hat{\mu}\hat{\nu}}\,,
    \\
    & \nonumber \\
      \label{eq:highersymmetriesEHAhats}
    \delta_{\epsilon}\hat{V}_{\hat{\mu}}
  & =
    -\epsilon \hat{k}^{\hat{\nu}}\hat{V}_{\hat{\nu}} \mathfrak{h}^{(1)}_{\hat{\mu}}
    +\tfrac{1}{3}\epsilon \hat{V}_{\hat{\mu}}\,,
  \end{align}
\end{subequations}

\noindent
and rescale the 4-dimensional fields with different weights

\begin{subequations}
  \begin{align}
    \delta_{\epsilon}A^{0}
    & =
      \epsilon A^{0}\,,
    \\
    & \nonumber \\
    \delta_{\epsilon}A^{1}
    & =
      \tfrac{1}{3}  \epsilon A^{1}\,,
    \\
    & \nonumber \\
    \delta_{\epsilon}\phi
    & =
      \tfrac{1}{3}\epsilon \,.
    \\
    & \nonumber \\
    \delta_{\epsilon}\chi
    & =
      -\tfrac{2}{3}\epsilon \chi\,.
        \end{align}
\end{subequations}

\noindent
leaving the Einstein-frame metric invariant.

This transformation is the transformation generated by the matrix $T_{1}$ and
the Killing vector $K_{1}$ in Section~\ref{sec-globalsymmetriesd4}

\begin{equation}
  \delta_{\alpha_{1}}A^{M}
  =
  \alpha_{1}T_{1}{}^{M}{}_{N}A^{N}\,,
  \hspace{1cm}
    \delta_{\alpha_{1}}\phi^{x}
  =
  \alpha_{1}K_{1}{}^{x}\,,
\end{equation}

\noindent
if the parameters are related by

\begin{equation}
  \alpha_{1}
  =
  -\tfrac{2}{3}\epsilon\,.
\end{equation}

In the theory at hand, the  higher-form symmetry
Eq.~(\ref{eq:globalhigherformp}) has the form

\begin{equation}
  \label{eq:deltaetatransd5hat}
  \delta_{\eta}\hat{V}
  =
  \eta \mathfrak{h}^{(1)}\,,
\end{equation}

\noindent
and acts on the 4-dimensional fields as

\begin{subequations}
  \begin{align}
    \delta_{\eta} \chi
    & =
      -\tfrac{1}{\sqrt{3}}\eta\,,
    \\
    & \nonumber \\
    \delta_{\eta}A^{1}
    & =
      \tfrac{1}{\sqrt{3}}\eta A^{0}\,,
  \end{align}
\end{subequations}

\noindent
leaving the rest of the fields invariant.

This transformation is the transformation generated by the matrix $T_{-}$ and
the Killing vector $K_{-}$ in Section~\ref{sec-globalsymmetriesd4}

\begin{equation}
  \delta_{\alpha_{-}}A^{M}
  =
  \alpha_{-}T_{-}{}^{M}{}_{N}A^{N}\,,
  \hspace{1cm}
    \delta_{\alpha_{-}}\phi^{x}
  =
  \alpha_{-}K_{-}{}^{x}\,,
\end{equation}

\noindent
if the parameters are related by

\begin{equation}
  \alpha_{-}
  =
  -\tfrac{1}{\sqrt{3}}\eta\,.
\end{equation}

As we might have expected, the higher-dimensional origin of the transformation
generated by the target-space Killing vector $K_{+}$ and the matrix $T_{+}$,
which, involving electric-magnetic duality, is only a symmetry of the
equations of motion and non-local, is unknown. Notice, though, that the scalar
charge associated to the transformation generated by $K_{+},T_{+}$,
$\mathcal{Q}_{+}$, is a linear combination of $\mathcal{Q}_{1}$ and
$\mathcal{Q}_{-}$ of $\Sigma^{\chi}$ and $\Sigma^{\phi}$ and that we do not
need to deal with it.

To end this section, let us consider the gauge transformations of $\hat{V}$

\begin{equation}
  \delta_{\hat{\Lambda}}\hat{V}
  =
  d\hat{\Lambda}\,.
\end{equation}

In the KK case, the dependence of $\hat{\Lambda}$ in $z$ can at most be
linear, with constant coefficients, if the condition
Eq.~(\ref{eq:partialzV=0}) is to be preserved, but, in that case, the
transformation is, actually, the global higher-form symmetry $\delta_{\eta}$
in Eq.~(\ref{eq:deltaetatransd5hat}). Thus, here we will just consider
transformations generated by $z$-independent gauge parameters
$\hat{\Lambda}(x)$ which only act on the 4-dimensional field $A^{1}$ as a
standard U$(1)$ gauge transformation Eq.~(\ref{eq:U12guagetrans}) and the
gauge parameter is

\begin{equation}
  \label{eq:Lambda1versusLambdahat}
  \Lambda^{1}
  =
  -\tfrac{1}{\sqrt{3}}k_{\infty}^{-1/2}\hat{\Lambda}\,.
\end{equation}

The proportionality factor is relevant to relate the electric charges, as we
are going to see.

\subsection{Kaluza-Klein dimensional reduction of the currents and charges}
\label{sec-KKreductioncharges}

Having identified the higher-dimensional origin of all the local and global
symmetries of the T$^{3}$ model (with the exception of the symmetry generated
by $K_{+},T_{+}$), we want to find the relation between the associated
conserved currents and charges in the 4- and 5-dimensional theories.

The reduction of 5-dimensional closed $r$-forms over a circle has an
interesting property: it gives two 4-dimensional closed forms of ranks $r$
and $r-1$

\begin{equation}
  \begin{aligned}
  \hat{\omega}^{(r)}
  & =
    \omega^{(r-1)}\wedge \mathfrak{h}^{(1)}+ \omega^{(r)}\,,
    \\
    & \\
  d\hat{\omega}^{(r)} =0\,,\,\,\,\,
  & \Rightarrow\,\,\,\,\, 
    d\omega^{(r-1)}=0\,\,\,\,\, \text{and}\,\,\,\,\, d\omega^{(r)}=0\,.
  \end{aligned}
\end{equation}

Thus, 5-dimensional conserved (closed) 4-form currents  give 4-dimensional
(uninteresting) conserved 4-forms and 3-form currents. Since $(d-1)$-form
currents are associated to global symmetries, we expect these 5- and
4-dimensional currents to be associated to the same symmetries in their 5- and
4-dimensional forms.

More interestingly, 5-dimensional conserved (closed) 3-form charges give
4-dimensional conserved 2-form charges and 3-form currents. Since conserved
$(d-2)$-form charges are associated to local symmetries, we expect these
charges to correspond to the 5- and 4-dimensional versions of the same local
symmetry. The relation between the 5-dimensional 3-form charges and the
4-dimensional 3-form currents is less clear and, as we are going to see, in
one case we are going to find a 4-dimensional 3-form current which must be
associated to an unknown or, at least, unaccounted for, global symmetry of the
4-dimensional theory.

The order in which we are going to study the dimensional reduction of currents
of charges, is meant to reduce the number of calculations. Thus, we start with
the reduction of the electric charge and of the Komar charge associated to the
KK Killing vector because we are going to use the results obtained there to
reduce the global currents and charges and the results of the latter will be
used in subsequent calculations.

\subsubsection{Local symmetries:  U$(1)$ gauge transformations, electric charge}

The 5-dimensional 3-form charge associated to the gauge symmetry
$\delta_{\hat{\Lambda}}$ is given by Eq.~(\ref{eq:electriccharge3form}) with
$f=1$. Then, using the decomposition Eqs.~(\ref{eq:formreductionformulae}) we
find

\begin{equation}
    \label{eq:Q1versusQhat}
  \begin{aligned}
    \hat{\mathbf{Q}}[1]
       & =
         -\frac{k_{\infty}^{-1/2}}{\sqrt{3}}\left[\mathbf{Q}_{1}
         +\frac{3}{16\pi G_{N}^{(4)}}d\left(\chi A^{1}\right)\right]\wedge
         \frac{\mathfrak{h}^{(1)}}{2\pi \ell}
    \\
       & \\
    & \hspace{.5cm}
         -\frac{1}{2\pi \ell \sqrt{3}}\left[\mathbf{J}_{-}
         +\frac{3}{16\pi G_{N}^{(4)}} d\left(\chi A^{1}\wedge A^{0}\right)\right]\,,
  \end{aligned}
\end{equation}

\noindent
where $\mathbf{Q}_{1}$ is defined in Eq.~(\ref{eq:electricchargesd4def}) and
$\mathbf{J}_{-}$ is defined in Eq.~(\ref{eq:J-}). Obviously, the coefficient
of $\mathbf{Q}_{1}$ follows from the relation
Eq.~(\ref{eq:Lambda1versusLambdahat}).

As we have anticipated, the decomposition of this on-shell closed
5-dimensional 3-form charge contains an on-shell closed 4-dimensional 2-form
charge and an on-shell closed 4-dimensional 3-form current. Both the charge
and the current include terms which are written as total derivatives of
products of 4-dimensional gauge fields $A^{0}$ and $A^{1}$ and the scalar
$\chi$ and which may not be globally defined (for instance, when $A^{0}$ and
$A^{1}$ describe magnetic monopoles). Thus, in general, these terms must be
considered as local expressions of closed but not exact forms and cannot be
ignored.

On the other hand, the presence of the scalar $\chi$ poses an important
problem. Let us consider, for example, a spherically-symmetric 4-dimensional
solution, so $\chi=\chi(r)$, with a gauge field $A^{1}$ describing a magnetic
monopole, so that, locally, $A^{1}\sim \cos{\theta}d\varphi$. From the
4-dimensional point of view, $F^{1}$ is a closed 2-form whose normalized
integral over 2-spheres always gives the same value for the magnetic charge,
independently of the radius of the sphere. If $\Sigma^{3}$ is the hypersurface
contained between two concentric spheres of radii $r_{+}$ and $r_{-}$, we have

\begin{equation}
  0
  =
  \int_{\Sigma^{3}}dF^{1}
  =
  \int_{S^{2}_{r_{+}}}F^{1}
  -\int_{S^{2}_{r_{-}}}F^{1}\,.
\end{equation}

These integrals can be calculated using the Wu-Yang \cite{Wu:1975es} or the
original Dirac \cite{Dirac:1931kp} description of the monopole obtaining the
same result. Let us see how this comes about in the Dirac description in which
$A^{1}\sim \cos{\theta}d\varphi$ globally, which implies the presence of Dirac
string singularities which have to be taken into account if one uses
Stokes' theorem. For instance, if
$A^{1}=\frac{p^{1}}{4\pi}(\cos{\theta}+1)d\varphi$,\footnote{Here we are using
  a different normalization for the charge $p^{1}$, for the sake of
  simplicity.} the string lies at $\theta=0$

\begin{equation}
  \int_{S^{2}_{r}}F^{1}
=  
  \int_{S^{2}_{r}}dA^{1}
  =
  \lim_{\epsilon\to 0}\int_{\theta=\epsilon}A^{1}
  = \lim_{\epsilon\to 0} \frac{p^{1}}{2}(\cos{\epsilon}+1)
  =
  p^{1}\,,
\end{equation}

\noindent
independently of the radius of the sphere we have integrated on.

Now, let us consider the integral of the closed form $d(\chi A^{1})$ over the
same $\Sigma^{3}$

\begin{equation}
  \begin{aligned}
  0
   & =
  \int_{\Sigma^{3}}d d(\chi A^{1})
  =
  \int_{S^{2}_{r_{+}}}d(\chi A^{1})
  -\int_{S^{2}_{r_{-}}}d(\chi A^{1})
  =
  \int_{S^{2}_{r_{+}}}\chi(r) F^{1}
     -\int_{S^{2}_{r_{-}}}\chi(r) F^{1}
    \\
    & \\
  & =
\left[\chi(r_{+})-\chi(r_{-})\right]p^{1}\,,
  \end{aligned}
\end{equation}

\noindent
which does not vanish unless $\chi$ is constant. This contradiction arises
because $d(\chi A^{1})$ has string singularities that need to be taken into
account when we use Stokes' theorem in the first step by adding a tubular
boundary around the string. If $A^{1}$ has the Dirac string at $\theta=0$ we
must add a boundary term\footnote{The sign is due to the orientation.}  
 
\begin{equation}
  \lim_{\epsilon\to 0} -\int_{\theta=\epsilon\, r\in[r_{+},r_{-}]}d(\chi A^{1})
  =
  \lim_{\epsilon\to 0} -\int_{\theta=\epsilon\, r\in[r_{+},r_{-}]}
  \frac{p^{1}}{4\pi}\chi'(r)(\cos{\theta}+1)dr\wedge d\varphi\,,
\end{equation}

\noindent
that cancels the value obtained in the right-hand side of the previous
equation.

The conclusion we draw from this example is that, for magnetic $A^{1}$, a
non-constant $\chi$ is associated to string singularities that must be taken
into account whenever we apply Stokes' theorem to derive a Gauss law for the
5-dimensional electric charge: integrating over $S^{2}_{r}\times S^{1}$, only
the term that contains the $\mathfrak{h}^{(1)}$ factor contributes and value
of the 5-dimensional charge will depend on $r$. If we are interested in
topologically simple situations in which the 5-dimensional electric charge
satisfies a Gauss law and we get the same value integrating at spatial
infinity or at the horizon, we must demand $\chi$ to be constant
$\chi=\chi_{\infty}$. Then, we find the following relation 5- and
4-dimensional charges 

\begin{equation}
  \label{eq:q1versusQ}
Q= -\tfrac{1}{\sqrt{3}} k_{\infty}^{-1/2} \left(q_{1}-3\chi_{\infty}p^{1}\right)\,.  
\end{equation}

The condition $\chi=\chi_{\infty}$ that allows us to relate the charges and
potentials of 5- and 4-dimensional black holes also follows from the on-shell
closure of the generalized Komar charge for a Killing vector $l$: as we have
already mentioned,\footnote{See Eq.~(\ref{eq:h2versusQl-}) and the preceding
  discussion.} the obstruction for that closure is the 2-form charge
$\mathbf{Q}_{l\, -}$ that occurs in the magnetic momentum map
Eq.~(\ref{eq:magneticmomentummapequationCS2}). When $\mathbf{Q}_{l\, -}\neq 0$
we are dealing with 5-dimensional solitons (\textit{fuzzballs}) rather than
with black holes and we cannot prove the generalized 0\textsuperscript{th} law
for the magnetic potential either. $\mathbf{Q}_{l\, -}= 0$ implies the
vanishing of the axion charge, which is consistent with the condition
$\chi=\chi_{\infty}$. In what follows, we will use this condition to relate
the values of charges and potentials of 5- and 4-dimensional black holes,
only.

\subsubsection{U$(1)$ magnetic charge}

In the KK setting we can define 5-dimensional magnetic (``dipole'') 3-form
charges associated to $\hat{G}$ with Eq.~(\ref{eq:magneticcharged-2form})
($d=5,p=0$) using $\mathfrak{h}^{(1)}$

\begin{equation}
  \hat{\mathbf{P}}[\mathfrak{h}^{(1)}]
  \equiv
   \frac{1}{16\pi G_{N}^{(5)}} \mathfrak{h}^{(1)} \wedge \hat{G}\,.
\end{equation}

\noindent
Using the decomposition Eq.~(\ref{eq:formreductionformulae}), we find that
this charge is related to the magnetic 2-form charge of the 4-dimensional
matter field $A^{1}$ by

\begin{equation}
  \begin{aligned}
    \hat{\mathbf{P}}[\mathfrak{h}^{(1)}]
    =
    \sqrt{3}k_{\infty}^{-1/2}
    \left[\mathbf{P}^{1}-\frac{1}{16\pi G_{N}^{(4)}}d\left(\chi A^{0}\right) \right]
    \wedge  \frac{\mathfrak{h}^{(1)}}{2\pi \ell}\,.
  \end{aligned}
\end{equation}

Again, the total derivative involves a product of $\chi$ with a gauge field,
$A^{0}$ in this case and the same discussion we had in the previous section
applies here. Integrating over $S^{2}_{r}\times S^{1}$ and setting
$\chi=\chi_{\infty}$ we find the following relation between 5- and
4-dimensional charges

\begin{equation}
  \label{eq:p1versusP}
P =\sqrt{3} k_{\infty}^{-1/2}\left(p^{1}+\chi_{\infty}p^{0}\right)\,.  
\end{equation}

\subsubsection{Local symmetries: GCTs and momentum in the internal direction}

Here we are only going to consider the charges associated to Killing vectors,
that can be computed by integrating the corresponding generalized Komar charge
at infinity.

In the KK context we always have the Killing vector
$\hat{k}=\partial_{\underline{z}}$, that generates translations along the
compact direction. The associated conserved charge is the momentum along that
direction $\hat{p}_{z}$. Let us consider each of the three terms in the
generalized Komar charge Eq.~(\ref{eq:generalizedKomar5d}) separately,
starting with the second, which is the 5-dimensional electric 3-form charge
multiplied by the momentum map associated to $\hat{k}$.  Taking into account
Eq.~(\ref{eq:PkinKKtheory3}), we find that the magnetic momentum map equation
(\ref{eq:magneticmomentummapequationCS2}) is solved by

\begin{equation}
  \tilde{\hat{P}}_{\hat{k}}
  =
  \sqrt{3}e^{-\phi_{\infty}}\left[\tfrac{1}{3}A_{1}
    +\chi^{2}(\mathfrak{h}^{(1)}e^{\phi_{\infty}}+A^{0})\right]\,, 
\end{equation}

\noindent
with vanishing $\mathfrak{h}^{(2)}$ and

\begin{equation}
  \label{eq:momentumcompactdirection}
    \hat{\mathbf{K}}[\hat{k}]
     = d\left[ -e^{\phi_{\infty}}
      (A_{0}-\tfrac{1}{3}\chi A_{1})\right]\wedge \mathfrak{h}^{(1)}
      +e^{-2\phi_{\infty}}\left\{\tfrac{2}{3}\mathbf{J}_{1}
      +d\left[\tfrac{1}{3}A_{1}\wedge (A^{1}+\chi A^{0})\right] \right\}\,,
\end{equation}

\noindent
which, under the conditions mentioned in the previous cases that force
$\chi$ to be constant and upon integration over $S^{2}\times S^{1}$, leads to
the well-known relation between momentum in the compact direction $P_{z}$ and
electric charge of the KK vector field $A^{0}$ in 4 dimensions, $q_{0}$

\begin{equation}
  \label{eq:momentumzversuscharge}
  P_{z}
  =
  e^{\phi_{\infty}}(q_{0}-\tfrac{1}{3}\chi_{\infty}q_{1})\,.
\end{equation}

\subsubsection{Topological charge in the internal direction}

In 4 dimensions there is a topologically-conserved 2-form charge whose
integral gives the NUT charge of Lorentzian Taub-NUT spaces
\cite{Bossard:2008sw}

\begin{equation}
  \label{eq:NUTchargedef}
  N
  =
  -\frac{1}{8\pi G_{N}^{(4)}} \int_{S^{2}_{\infty}} d\hat{\partial_{t}}\,,
\end{equation}

\noindent
where $\hat{\partial_{t}}$ is the 1-form dual to the Killing vector
$\partial_{t}$.  This charge occurs in a very natural way in the generalized
Komar charge of General Relativity when the Einstein--Hilbert term is
supplemented by the Holst term (see Ref.~\cite{Cerdeira:2025agq} and
references therein). There is no analogue of the Holst term in 5 dimensions,
but, in the KK setting, it is not difficult to define a 5-dimensional
generalization of the above definition which is also topologically closed:

\begin{equation}
  \label{eq:KKcharge}
  \mathbf{KK}
  \equiv
  \frac{1}{16\pi G_{N}^{(5)}} \mathfrak{h}^{(1)}\wedge d \hat{\hat{k}}\,,
\end{equation}

\noindent
where, as before, $\hat{\hat{k}}$ is the 1-form dual to the 5-dimensional
Killing vector $\hat{k}=\partial_{\underline{z}}$.
It is not difficult to see that

\begin{equation}
  \label{eq:KKchargeversusmagneticcharge}
  \mathbf{KK}
  =
  e^{\phi_{\infty}}\left[e^{-4\phi}\mathbf{P}^{0}+4 e^{-4\phi}d\phi\wedge
  A^{0}\right] \wedge \frac{ \mathfrak{h}^{(1)}}{2\pi \ell}
\end{equation}

Integrating over $S^{2}_{\infty}\times S^{1}$ and calling $KK$ the result for
lack of a better name, we find

\begin{equation}
  \label{eq:KKchargeversusmagneticcharge2}
  KK
  =
  e^{-3\phi_{\infty}}p^{0}\,,
\end{equation}

\noindent
where $p^{0}$ is the magnetic charge of the KK vector field.

\subsubsection{Global symmetries}
\label{sec-KKreductionofglobalcurrentsandcharges}

The two global symmetries of the 5-dimensional theory with KK boundary
conditions $\delta_{\eta}$ and $\delta_{\epsilon}$ defined in
Eqs.~(\ref{eq:deltaeta5d}) and (\ref{eq:deltaepsilon5d}) respectively, give
rise to the two on-shell closed 4-form currents $\mathbf{J}_{\eta}$ and
$\mathbf{J}_{\epsilon}$ in Eqs.~(\ref{eq:Jeta}) and
(\ref{eq:Jepsilon}).\footnote{Here we want to use the ``original'' form of
  these currents. If we remove total derivatives as we did in
  Eq.~(\ref{eq:Jepsilon2}), we get much more complicated expressions.}

We find 

\begin{subequations}
  \begin{align}
    \hat{\mathbf{J}}_{\eta}
    & =
      \tfrac{1}{\sqrt{3}}\mathbf{J}_{-} \wedge \frac{\mathfrak{h}^{(1)}}{2\pi\ell}
      +\mathbf{J}^{(4)}_{\eta}\,,
      \\
    & \nonumber \\
    \hat{\mathbf{J}}_{\epsilon}
    & =
      \tfrac{2}{3}\mathbf{J}_{1}\wedge \frac{\mathfrak{h}^{(1)}}{2\pi \ell}
       +\mathbf{J}^{(4)}_{\epsilon}\,,
  \end{align}
\end{subequations}

\noindent
where $\mathbf{J}^{(4)}_{\eta}$ and $\mathbf{J}^{(4)}_{\epsilon}$ are two
4-forms (\textit{i.e.}~volume forms in 4 dimensions), whose quite complicated
explicit form is irrelevant. This is the result we expected given the
relations between these 5-dimensional higher-form symmetries and the
4-dimensional global symmetries of the T$^{3}$ model found in
Section~\ref{sec-KKreductionsymmetries}.

The decomposition of the 5-dimensional 3-form charges
$\hat{\mathbf{Q}}_{\eta\, l}$ and $\hat{\mathbf{Q}}_{\epsilon\, l}$ given in
Eqs.~(\ref{eq:Qetal}) and (\ref{eq:Qepsilonl}), respectively, is

\begin{subequations}
  \begin{align}
    \label{eq:Qetalreduction}
    \hat{\mathbf{Q}}_{\eta\, l}
    & =
      \tfrac{1}{\sqrt{3}}\mathbf{Q}_{l\, -}\wedge \frac{\mathfrak{h}^{(1)}}{2\pi \ell}\,,
    \\
    & \nonumber \\
    \label{eq:Qepsilonreduction}
    \hat{\mathbf{Q}}_{\epsilon\, l}
  & =
  \tfrac{2}{3}\mathbf{Q}_{l\, 1}\wedge \frac{\mathfrak{h}^{(1)}}{2\pi \ell}\,,
  \end{align}
\end{subequations}

\noindent
plus uninteresting, trivially conserved 3-form currents coming from the volume
4-forms.

\subsubsection{Local symmetries: GCTs and the mass and angular momentum}
\label{sec-massandangularmomentum}

Let us now consider the conserved charge associated to the 5-dimensional
Killing vector field $\hat{l}$ which is related to a 4-dimensional Killing
vector field $l$ by

\begin{equation}
  \hat{l}
  =
  l -\chi_{l}\hat{k}\,,  
  \hspace{1cm}
  \hat{k}
  =
  \partial_{\underline{z}}\,,
\end{equation}

\noindent
where $\chi_{l}$ is given by Eq.~(\ref{eq:chil3}) which we reproduce here for
the sake of convenience

\begin{equation}
  \chi_{l}
  =
  \imath_{l}A -\bar{P}_{l}\,,
  \hspace{1cm}
  \bar{P}_{l}
  \equiv
  P_{l}-P_{l}(\mathcal{H})\,,
\end{equation}

\noindent
where we have chosen $P_{l}$ to vanish at infinity, so that $\bar{P}_{l}$
vanishes on the horizon. This particular choice is meant to deal with the case
in which $\hat{l}$ and $l$ generate, respectively, the 5- and 4-dimensional
Killing horizons but can be used in other cases setting $P_{l}(\mathcal{H})=0$
and relaxing the boundary conditions on $P_{l}$ which, on the other hand, will
only be used in the computation of the generalized Komar integrals.

From the 5- and 4-dimensional electric momentum map equations
(\ref{eq:pformmomentummap}) we find

\begin{equation}
  \hat{P}_{\hat{l}}
  =
  -\sqrt{3}e^{-\phi_{\infty}}\left(P_{l}{}^{1}+\chi \bar{P}_{l}{}^{0} \right)\,.
\end{equation}

In the black-hole case $\chi=\chi_{\infty}$ and with the boundary conditions we
have assumed this implies the following relation between 5- and 4-dimensional
electric potentials at infinity and on the horizon:

\begin{subequations}
  \begin{align}
    \label{eq:5dPhiinfinity}
    \hat{\Phi}_{\infty}
    & =
      \sqrt{3}e^{-\phi_{\infty}}\chi_{\infty}\Phi_{\mathcal{H}}{}^{0}\,,
    \\
    & \nonumber \\
    \label{eq:5dPhiH}
    \hat{\Phi}_{\mathcal{H}}
    & =
      -\sqrt{3}e^{-\phi_{\infty}}\Phi_{\mathcal{H}}{}^{1}\,,
  \end{align}
\end{subequations}

The 5-dimensional magnetic momentum map equation
(\ref{eq:magneticmomentummapequationCS2}) is satisfied with

\begin{subequations}
    \label{eq:dimensionalreductionmomentummap}
  \begin{align}
    \tilde{\hat{P}}_{\hat{l}}
    & =
      -\sqrt{3}e^{-\phi_{\infty}}\left[
      \left(\tfrac{1}{3} P_{l\, 1}
      -2\chi P_{l}{}^{1}-\chi^{2}\bar{P}_{l}{}^{0}\right)\wedge(\mathfrak{h}^{(1)}
      +e^{-\phi_{\infty}} A^{0}) +\tfrac{1}{3}P_{l}{}^{0}(\mathcal{H})A_{1} \right]\,,
    \\
    & \nonumber \\
      \label{eq:h2versusQl-}
    \mathfrak{h}^{(2)}
    & =
      \frac{e^{-2\phi_{\infty}}(16\pi G_{N}^{(4)})}{\sqrt{3}} \mathbf{Q}_{l\, -}\,.
  \end{align}
\end{subequations}

This result implies that, in general, the 5-dimensional generalized Komar
charge is not closed as long as $\mathbf{Q}_{l\, -}\neq 0$, as we have
advanced.

Combining the above results we get

\begin{equation}
  \begin{aligned}
    \tilde{\hat{P}}_{\hat{l}}
    -\tfrac{2}{\sqrt{3}}\hat{P}_{\hat{l}}\hat{V}
    & =
    -\sqrt{3}e^{-\phi_{\infty}} \left( \tfrac{1}{3}P_{l\, 1}
      +\chi^{2}\bar{P}_{l}{}^{0}\right)
      \wedge(\mathfrak{h}^{(1)}+e^{-\phi_{\infty}} A^{0})
    \\
    & \\
    & \hspace{.5cm}
    -\sqrt{3}e^{-\phi_{\infty}}
    \left[P_{l}{}^{0}(\mathcal{H})A_{1} +2(P_{l}{}^{1}+\chi\bar{P}_{l}{}^{0})A^{1} \right]\,,
  \end{aligned}
\end{equation}

\noindent
which implies the following relations

\begin{subequations}
  \begin{align}
    \label{eq:5dtildePhiinfinity}
    \tilde{\hat{\Phi}}_{\infty}
    & =
      \sqrt{3}e^{-\phi_{\infty}} \chi_{\infty}^{2}\Phi_{\mathcal{H}}{}^{0}\,,
    \\
    & \nonumber \\
    \label{eq:5dtildePhiH}
    \tilde{\hat{\Phi}}_{\mathcal{H}}
    & =
      -\frac{e^{-\phi_{\infty}}}{\sqrt{3}}\Phi_{\mathcal{H}\, 1}\,.
  \end{align}
\end{subequations}

Finally, the standard 5-dimensional Komar charge term decomposes as
follows:\footnote{Here we use boldface for the 1-forms dual to the Killing
  vectors: $\hat{\mathbf{l}}$ is the 5-dimensional 1-form dual to the
  5-dimensional Killing vector $\hat{l}$ and $\mathbf{l}_{E}$ is the
  4-dimensional 1-form dual to the 4-dimensional vector $l$ using the
  Einstein-frame metric.}

\begin{equation}
  \label{eq:standardKomarreduction1}
  \begin{aligned}
    \hat{\star} d\hat{\mathbf{l}}
    & =
      e^{-2\phi_{\infty}}
      \left[-\star_{E}d\mathbf{l}_{E} +2\imath_{l}\star_{E}d\phi
      -\bar{P}_{l}{}^{0}\left(F_{0} -\chi F_{1}+3\chi^{2}F^{1} +\chi^{3}F^{0}\right)
      \right]\wedge(\mathfrak{h}^{(1)}+e^{-\phi_{\infty}} A^{0})
    \\
    & \\
    & \hspace{.5cm}
      +e^{-3\phi_{\infty}} \left[\star_{E}\imath_{l}F^{0} +4 \bar{P}_{l}{}^{0}
      \star_{E}d\phi\right]\,.
  \end{aligned}
\end{equation}

This expression can be rewritten in terms of 4-dimensional charges and
currents:

\begin{equation}
  \label{eq:standardKomarreduction2}
  \begin{aligned}
    \frac{1}{16\pi G_{N}^{(5)}} \hat{\star} d\hat{\mathbf{l}}
    & =
      \frac{1}{2\pi \ell}\left\{ \mathbf{K}[l]
      -\bar{P}_{l}{}^{0} \left(\tfrac{2}{3}\chi\mathbf{Q}_{1}
      -3\chi^{2}\mathbf{P}^{1}-\chi^{3}\mathbf{P}^{0} \right)
      \right.
    \\
    & \\
    & \hspace{.5cm}
      -P_{l}{}^{0}(\mathcal{H}) \left(\mathbf{Q}_{0}-\tfrac{1}{3}\chi
      \mathbf{Q}_{1} \right)
      +\tfrac{1}{3}\chi P_{l\, 1}\mathbf{P}^{0} +2\chi P_{l}{}^{1}\mathbf{P}^{1}
      -\tfrac{2}{3} P_{l}{}^{1}\mathbf{Q}_{1} +\tfrac{1}{3}P_{l\, 1}\mathbf{P}^{1}
    \\
    & \\
    & \hspace{.5cm}
      \left.
      -\tfrac{1}{3}\left(\mathbf{Q}_{l\, 1}-\chi \mathbf{Q}_{l\, -}
      \right)
      \right\}\wedge(\mathfrak{h}^{(1)}+e^{-\phi_{\infty}} A^{0})
    \\
    & \\
    & \hspace{.5cm}
      +\frac{e^{-\phi_{\infty}}}{2\pi \ell} \left\{ e^{6\phi}\mathbf{l}_{E} \wedge
      \left(\mathbf{Q}_{0}-\chi\mathbf{Q}_{1}
      +3\chi^{2}\mathbf{P}^{1}+\chi^{3}\mathbf{P}^{0} \right)
      \right.
    \\
    & \\
    & \hspace{.5cm}
      \left.
      +\bar{P}_{l}{}^{0}
      \left[\tfrac{2}{3}\left(\mathbf{J}_{1}-\chi\mathbf{J}_{-}\right)
      +F_{0}\wedge A^{0}+\tfrac{1}{3}F_{1}\wedge A^{1} -\tfrac{2}{3}\chi
      F_{1}\wedge A^{0} -2\chi F^{1}\wedge A^{1}\right]
      \right\}\,.
  \end{aligned}
\end{equation}

Integrating this term at infinity in the form of
Eq.~(\ref{eq:standardKomarreduction1}) with the chosen boundary conditions
($\chi=\chi_{\infty},\mathbf{Q}_{l\, -}=0$ ) and the normalization factor
$(16\pi G_{N}^{(5)})^{-1}$, we get

\begin{equation}
  \label{eq:standardKomarintegralatinfinity}
    \frac{1}{16\pi G_{N}^{(5)}} \int_{S^{2}_{\infty}\times S^{1}}\hat{\star} d\hat{\mathbf{l}}
    =
    \tfrac{1}{2}\left(M+\Sigma^{\phi}\right) -\Omega J
    -\Phi_{\mathcal{H}}{}^{0}\left(q_{0} -\chi_{\infty} q_{1}
      +3\chi_{\infty}^{2}p^{1} +\chi_{\infty}^{3}p^{0}\right)\,,
\end{equation}

\noindent
where $M$ is the 4-dimensional ADM mass and $\Sigma^{\phi}$ is the dilaton
scalar charge defined conventionally in
Eq.~(\ref{eq:conventionaldefinitionscalarcharges}) and in a covariant way in
Eq.~(\ref{eq:covariantdefinitionscalarcharges}) and expressed in terms of
electric and magnetic potentials and charges in
Eq.~(\ref{eq:valuesscalarcharges}).

Integrating instead over the bifurcation sphere $\mathcal{BH}$ at which $l=0$,
we get

\begin{equation}
  \label{eq:standardKomarintegralatbifurcationsphere}
  \frac{1}{16\pi G_{N}^{(5)}} \int_{\mathcal{BH}}\hat{\star} d\hat{\mathbf{l}}
  =
  ST\,.
\end{equation}

Combining  all the partial results, we arrive at the following decomposition
of the 5-dimensional generalized Komar 3-form charge: 

\begin{equation}
  \label{eq:reductionKomarcharge}
  \hat{\mathbf{K}}[\hat{l}]
  =
  \tilde{\mathbf{K}}[l] \wedge
  \frac{\mathfrak{h}^{(1)}}{2\pi \ell}
  +\mathbf{J}_{l}\,,
\end{equation}

\noindent
where we have defined the following 4-dimensional 2-form charge

\begin{equation}
  \tilde{\mathbf{K}}[l]
  =
  \mathbf{K}[l]
  -\tfrac{1}{3}\left(\mathbf{Q}_{l\, 1}-\chi\mathbf{Q}_{l\, -}\right)
+\frac{P_{l}{}^{0}(\mathcal{H})}{16\pi G_{N}^{(4)}}
d\left(A_{0}-\tfrac{1}{3}\chi A_{1}\right)\,,
\end{equation}

\noindent
and 3-form current

\begin{equation}
  \label{eq:newcurrent}
  \begin{aligned}
\mathbf{J}_{l}
    & \equiv 
      \frac{k_{\infty}^{1/2}}{2\pi \ell} \left\{
      \tilde{\mathbf{K}}[l] \wedge A^{0}
      +e^{6\phi}\mathbf{l}_{E}\wedge \left(\mathbf{Q}_{0}-\chi \mathbf{Q}_{1}
      +\chi^{3}\mathbf{P}^{0}+3\chi^{2}\mathbf{P}^{1} \right)
      \right.
    \\
    & \\
    & \hspace{.5cm}
    +\bar{P}_{l}{}^{0}  \left(\tfrac{2}{3}\mathbf{J}_{1}
      +F_{0}\wedge A^{0} +\tfrac{1}{3}F_{1}\wedge A^{1}\right)
    \\
      & \\
    & \hspace{.5cm}
+P_{l}{}^{1}\left(\tfrac{2}{3}\mathbf{J}_{-}+\tfrac{2}{3}F_{1}\wedge
      A^{0}+2F^{1}\wedge A^{1}\right)
    \\
    & \\
    & \hspace{.5cm}
      \left.
      +\tfrac{1}{3}P_{l}{}^{0}(\mathcal{H})(F^{1}+\chi F^{0})\wedge A_{1}
      \right\}\,.
  \end{aligned}
\end{equation}

In the above expressions, $\mathbf{K}[l]$ is the 4-dimensional, on-shell
closed, generalized Komar 2-form charge in Eq.~(\ref{eq:4dGKC}),
$\mathbf{Q}_{l\, 1}$ and $\mathbf{Q}_{l\, -}$ are the on-shell closed 2-form
charges in Eqs.~(\ref{eq:Ql1}) and (\ref{eq:Ql-})\footnote{The combination
  $\left(\mathbf{Q}_{l\, 1}-\chi\mathbf{Q}_{l\, -}\right)\sim
  -6\imath_{l}\star_{E}d\phi +\cdots$.} and $P_{l}{}^{0}(\mathcal{H})$ is the
momentum map $P_{l}{}^{0}$ evaluated over the horizon (a constant),
$\mathbf{J}_{1}$ and $\mathbf{J}_{-}$ are the on-shell closed 3-form currents
in Eqs.~(\ref{eq:J1}) and (\ref{eq:J-}).

The 2-form $\tilde{\mathbf{K}}[l]$ is not closed if $\chi$ is not
constant. This is consistent with the fact that, in general,
$\hat{\mathbf{K}}[\hat{l}]$ is not closed on-shell and satisfies, instead,

\begin{equation}
  d\hat{\mathbf{K}}[\hat{l}]
  \doteq
  -\frac{1}{3\cdot 16\pi G_{N}^{(5)}}\hat{G}\wedge \mathfrak{h}^{(2)}\,,
\end{equation}

\noindent
where, as we have shown in Eq.~(\ref{eq:h2versusQl-})

\begin{equation}
  \mathfrak{h}^{(2)}
  =
  \tfrac{e^{-2\phi_{\infty}}(16\pi G_{N}^{(4)})}{\sqrt{3}}\mathbf{Q}_{l\, -}\,.  
\end{equation}

Then,

\begin{subequations}
  \begin{align}
  d\tilde{\mathbf{K}}[l]
  & \doteq
  \frac{1}{3}d\chi\wedge \mathbf{Q}_{l\, -} \wedge
    \frac{\mathfrak{h}^{(1)}}{2\pi \ell}\,,
    \\
    & \nonumber \\
    d\mathbf{J}_{l}
    & \doteq
      \frac{e^{-\phi_{\infty}}}{3\cdot 2\pi \ell} d(A^{1}+\chi A^{0})\wedge \mathbf{Q}_{l\, -}\,.
  \end{align}
\end{subequations}

Thus, again, we arrive to the conclusion that 4-dimensional regular black
holes are related to 5-dimensional regular black holes if $\chi=\chi_{\infty}$
which implies $\mathbf{Q}_{l\,-}=0$. 

Under those conditions the 3-form current $\mathbf{J}_{l}$ is also closed
on-shell.  While this is not too difficult to prove, it is somewhat surprising
because it means that there is a new, $l$-dependent, on-shell closed 3-form
current in the 4-dimensional theory with no obvious relation to any known
global symmetry.\footnote{Since, by assumption, there is an isometry generated
  by $l$, there is a global symmetry but the Noether 3-form current is
  trivially (off-shell) conserved because it is a particular case of the
  Noether 3-form current associated to the invariance of the theory under all
  GCTs. This off-shell conservation implies the existence of the Noether--Wald
  2-form charge, which, as we have explained is not closed on-shell even for
  isometries. The conserved charge is the generalized Komar 2-form charge
  associated to the isometry, which is closed on-shell by construction. The
  above 3-form current does not coincide with any of these objects and,
  therefore, it is not the Noether 3-form current of the global symmetry
  associated to the assumed isometry although it must be related to it.}

$\tilde{\mathbf{K}}[l]$ contains two terms in addition to $\mathbf{K}[l]$.
The second term is related to the momentum in the internal direction, $P_{z}$,
given in Eq.~(\ref{eq:momentumzversuscharge}) and to the associated potential
and it originates in the term $-\chi_{l}\hat{k}$ that one has to add to the
Killing vector $l$ that generates the 4-dimensional Killing horizon to get the
5-dimensional Killing vector $\hat{l}$ that generates the 5-dimensional
Killing horizon.

The first term is identical to the one found in Ref.~\cite{Barbagallo:2025fkg}
and leads to the same problem: when $l=\partial_{t}$, the integral at spatial
infinity of the 5-dimensional Komar charge with KK boundary conditions gives
the 4-dimensional ADM mass $M$ plus a term proportional to the scalar charge
associated to the constant shifts of the KK scalar
($-\tfrac{1}{3}\mathcal{Q}_{1}$ in this case, once we have set
$\chi=\chi_{\infty}$ and $\mathbf{Q}_{l\,-}=0$).  If we want a 5-dimensional
Komar charge whose integral at spatial infinity gives directly the
4-dimensional ADM mass (which is the only meaningful definition of
5-dimensional mass with KK boundary conditions) we have to remove the term
$\mathbf{Q}_{l\, 1}$ in Eq.~(\ref{eq:reductionKomarcharge}). This can be done
directly in 5-dimensional language using the charges associated to the
higher-form symmetries as explained in Ref.~\cite{Barbagallo:2025fkg} and we
will not do it explicitly here. The addition of an on-shell conserved charge
will not modify the Smarr formula.

\subsection{5- versus 4-dimensional black-object thermodynamics}
\label{sec-5vs4dthermodynamics}

In the KK case there is a compact isometric direction (S$^{1}$) with an
associated harmonic 1-form $\mathfrak{h}^{(1)}$ defined in the whole space
(and not just on the horizon, as in the asymptotically-flat black ring
case). Spatial infinity is now S$^{2}_{\infty}\times$S$^{1}$.

Which kind of black-object solutions can we expect in minimal 5-dimensional
supergravity with these boundary conditions? They are essentially those whose
dimensional reduction leads to asymptotically-flat 4-dimensional black holes,
namely non-extremal versions of pp-waves, Gross--Perry--Sorkin KK monopoles,
Reissner--Nordstr\"om--Tangherlini (RNT) black holes smeared in the compact
direction and charged strings wrapping the compact direction. These elementary
solutions are reviewed in Appendix~\ref{app-KKBHs}. Each of them carries the
5-dimensional version of one of the 4 kinds of electric and magnetic charges
associated to the two 4-dimensional gauge fields that 4-dimensional black
holes can carry:

\begin{enumerate}
\item The non-extremal versions of pp-waves (Appendix~\ref{app-electricKKBH})
  carry momentum in the compact direction which can be computed using the
  generalized Komar charge for the Killing vector that generates translations
  in the compact direction Eq.~(\ref{eq:momentumcompactdirection}). This
  momentum becomes electric charge of the 4-dimensional vector $A^{0}$
  according to Eq.~(\ref{eq:momentumzversuscharge}) and, therefore, the
  5-dimensional pp-waves give rise to 4-dimensional electrically-charged black
  holes.
  
\item The non-extremal versions of the KK monopole
  (Appendix~\ref{app-magneticKKBH}) carry a topological charge that measures
  how the compact direction is fibered over 2-spheres. This charge can be
  computed using the topologically-closed 3-form charge defined in
  Eq.~(\ref{eq:KKcharge}) and it becomes magnetic charge of the vector $A^{0}$
  according to Eq.~(\ref{eq:KKchargeversusmagneticcharge2}). Thus, these
  solutions give rise to 4-dimensional magnetically-charged black holes.

\item The smeared non-extremal RNT black hole
  (Appendix~\ref{app-electricblackhole}) carries 5-dimensional electric charge
  which becomes electric charge of the 4-dimensional vector field $A^{1}$
  according to Eq.~(\ref{eq:q1versusQ}). Thus, it gives rise to 4-dimensional
  electrically-charged black holes.

\item The non-extremal string wrapped around the compact direction
  (Appendix~\ref{app-magneticblackstring}) is electrically charged with
  respect to the 2-form dual to the 5-dimensional vector field $\hat{V}$ and
  it carries ``magnetic dipole charge'' with respect to it which becomes
  magnetic charge of the 4-dimensional vector field $A^{1}$. Thus, it gives
  rise to 4-dimensional magnetically-charge black holes.

\end{enumerate}

Generic solutions carry combinations of these 4 kinds of charge. The dyonic KK
black hole (Appendix~\ref{app-dyonicKKBH}) carries the first two kinds of
charge mentioned above.

Apart from these, the 5-dimensional black objects can have mass and only one
angular momentum, whose only meaningful definition is that they are those the
dimensionally-reduced object.

It is interesting to see in detail how the 5-dimensional Smarr formula
Eq.~(\ref{eq:5dSmarrforumula}) gives the 4-dimensional
Eq.~(\ref{eq:4dSmarrformula}) one once the boundary conditions and the
relations between all the charges and thermodynamical potentials have been
taken into account.

First of all, let us rewrite the 5-dimensional Smarr formula
Eq.~(\ref{eq:5dSmarrforumula}) in the way in which it has been obtained, with
the result of the integral at infinity on the left-hand side and the result of
the integral over the bifurcation surface on the right-hand side:

\begin{equation}
  \tfrac{2}{3} M
  -\Omega^{1} J_{1}
  -\Omega^{2} J_{2}
  =
  ST
  +\tfrac{2}{3}\Phi_{\mathcal{H}} Q +\tfrac{1}{3}\tilde{\Phi}_{\mathcal{H}}P\,. 
\end{equation}

First of all, with KK boundary conditions only one independent angular
momentum can be carried by black objects and, therefore, we have to replace
$ -\Omega^{1} J_{1} -\Omega^{2} J_{2}$ by $-\Omega J$.  Furthermore, in this
expression, $M$ is the 5-dimensional mass with asymptotically-flat boundary
conditions. Finally, we were forced to use boundary conditions that give
additional contributions at infinity coming from the electric and magnetic
terms.

The left-hand slide must be replaced by a combination of
Eqs.~(\ref{eq:standardKomarintegralatinfinity}), (\ref{eq:5dPhiinfinity}),
(\ref{eq:q1versusQ}), (\ref{eq:5dtildePhiinfinity}) and (\ref{eq:p1versusP})

\begin{equation}
  \begin{aligned}
  \int_{S^{2}_{\infty}\times S^{1}}\hat{\mathbf{K}}[\hat{l}]
  & = 
    \int_{S^{2}_{\infty}}\tilde{\mathbf{K}}[l]
    =
      \tfrac{1}{2}\left(M+\Sigma^{\phi}\right) -\Omega J
    -\Phi_{\mathcal{H}}{}^{0}\left(q_{0} -\chi_{\infty} q_{1}
    +3\chi_{\infty}^{2}p^{1} +\chi_{\infty}^{3}p^{0}\right)
    \\
    & \\
    & \hspace{.5cm}
      +\tfrac{2}{3}\left(\sqrt{3}e^{-\phi_{\infty}}\chi_{\infty}\Phi_{\mathcal{H}}{}^{0} \right)
      \left[-\tfrac{1}{\sqrt{3}} e^{\phi_{\infty}} \left(q_{1}-3\chi_{\infty}p^{1}\right) \right]
    \\
    & \\
    & \hspace{.5cm}
    +\tfrac{1}{3}\left(\sqrt{3}e^{-\phi_{\infty}}
      \chi_{\infty}^{2}\Phi_{\mathcal{H}}{}^{0} \right)
      \left[\sqrt{3} e^{\phi_{\infty}}\left(p^{1}+\chi_{\infty}p^{0}\right) \right]
    \\
    & \\
    & =
      \tfrac{1}{2}M+\tfrac{1}{2}\Sigma^{\phi} -\Omega J
    -\Phi_{\mathcal{H}}{}^{0}\left(q_{0} -\tfrac{1}{3}\chi_{\infty} q_{1}\right)\,,
  \end{aligned}
\end{equation}

\noindent
which, apart from the contribution of the scalar charge, that can be removed
by the procedure proposed in Ref.~\cite{Barbagallo:2025fkg} (being replaced by
an identical term with opposite sign from the integral over the bifurcation
surface) the result is, as expected, the 4-dimensional ADM mass, the angular
momentum multiplied by the angular velocity of the horizon and the momentum in
the internal direction multiplied by the electrostatic potential evaluated at
the horizon. If we use the value of the scalar charge in
Eq.~(\ref{eq:valuesscalarcharges}), we arrive at

\begin{equation}
\begin{aligned}
  \int_{S^{2}_{\infty}\times S^{1}}\hat{\mathbf{K}}[\hat{l}]
  & =
    \tfrac{1}{2}M -\Omega J
    -\tfrac{1}{2}\left(\Phi_{\mathcal{H}}{}^{0}q_{0} +\Phi_{\mathcal{H}\,0}p^{0} \right)
    +\tfrac{1}{6}\left(\Phi_{\mathcal{H}}{}^{1}q_{1}+\Phi_{\mathcal{H}\,1}p^{1} \right)
        \\
    & \\
    & \hspace{.5cm}
-\chi_{\infty}\left(2\Phi_{\mathcal{H}}{}^{1}p^{1}
      +\tfrac{1}{3}\Phi_{\mathcal{H}\, 1}p^{0}\right)\,.
  \end{aligned}
\end{equation}

At the right-hand side, we get a combination of the results in
Eqs.~(\ref{eq:standardKomarintegralatbifurcationsphere}) (\ref{eq:5dPhiH}),
(\ref{eq:q1versusQ}), (\ref{eq:5dtildePhiH}) and  (\ref{eq:p1versusP})

\begin{equation}
  \begin{aligned}
  \int_{\mathcal{BH}_{4}\times S^{1}}\hat{\mathbf{K}}[\hat{l}]
  & = 
    \int_{\mathcal{BH}_{4}}\tilde{\mathbf{K}}[l]
    =
ST +\tfrac{2}{3}\Phi_{\mathcal{H}}{}^{1}q_{1}-\tfrac{1}{3}\Phi_{\mathcal{H}\, 1}p^{1}
      -\chi_{\infty}\left(2\Phi_{\mathcal{H}}{}^{1}p^{1}+\tfrac{1}{3}\chi_{\infty}\Phi_{\mathcal{H}\, 1}p^{0}\right)\,,
  \end{aligned}
\end{equation}

\noindent
and equating the two results we get the 4-dimensional Smarr formula.

\section{Conclusions}
\label{sec-discussion}

In this paper we have studied the local and global symmetries and conserved
currents and charges of minimal 5-dimensional supergravity (and some
higher-rank generalizations that include 11-dimensional supergravity) with
asymptotically-flat and asymptotically-KK boundary conditions, relating them
in the second case to the global and local symmetries and the conserved
currents and charges of the T$^{3}$ model of $\mathcal{N}=2,d=4$ supergravity.

Apart from the general expressions we have found a number of specially
interesting or intriguing results:

\begin{itemize}

\item We have found that a subgroup of the global symmetry group of the
  T$^{3}$ model (that does not include electric-magnetic duality rotations)
  corresponds to a group of higher-form symmetries that act on the metric and
  the vector field of the 5-dimensional theory.

\item We have found that, with KK boundary conditions, it is possible to find
  the obstruction to the on-shell closure of the 5-dimensional theory that
  justifies the possible existence of globally regular, horizonless,
  asymptotically-KK solutions. In principle, these solutions may give rise to
  globally-regular, horizonless, asymptotically-flat  4-dimensional solutions
  such as those constructed in Ref.~ \cite{Saxena:2005uk}, but reaching a
  definite conclusion requires further study.

\item We have found a new conserved 3-form current, Eq.~(\ref{eq:newcurrent}),
  which should be related to a new global symmetry which we are currently
  trying to identify \cite{kn:BCMO}.

\item We have shown that the harmonic terms in the momentum map equations are
  related to the generalized symmetric ansatz in which fields are invariant
  under the action of a GCT up to the action of a global symmetry that, in
  the cases considered, is a higher-form symmetry.

  In the magnetic case, since the field involved does not occur explicitly in
  the action, there is no explicit dependence on the coordinate adapted to the
  GCT. These harmonic terms support solitonic solutions (the fuzzballs of
  5-dimensional supergravity).
  
  In the electric case, the fields involved do occur explicitly in the action
  and they do depend on the coordinate adapted to the GCT. Interestingly, in
  this case one can also construct solitonic solutions (and hairy black holes,
  see, for instance, Refs.~\cite{Herdeiro:2014goa,Herdeiro:2015gia}).

  All this suggests that, both in the electric as well as in the magnetic
  case, the harmonic terms could be responsible for the existence of the
  solitonic solutions. In the magnetic case, this is the main result of
  Ref.~\cite{Gibbons:2013tqa} which we have rewritten in terms of an
  obstruction to the closure of the generalized Komar charge. The electric
  case for scalars was considered in \cite{Ballesteros:2024prz}, albeit with a
  slightly different perspective in which the generalized Komar charges are
  on-shell closed and it could be helpful to relate the obstruction
  to the closure of the generalized Komar charge  directly to the generalized symmetric
  ansatz.

  It could be interesting to obtain a generic relation between the obstruction
  to the closure of the generalized Komar charge and the existence of
  solitonic (massive, static, globally regular, horizonless,
  asymptotically-flat) solutions. For instance, in
  Ref.~\cite{Ballesteros:2025wvs} the obstruction to the closure of the
  generalized Komar charge in Einstein--scalar--Gauss--Bonnet theories has
  been identified and it would be interesting to find such solitonic solutions
  in that theory.

\end{itemize}

Extending the results of this paper to more general, higher-dimensional
theories should be the next step in our long-term program. Work in this
direction is well underway.

\section*{Acknowledgments}

To would like to thank E.~Bergshoeff, J.J.~Fern\'andez-Melgarejo and
J.A.~Rosabal for many useful conversations and the Van Swinderen Institute of
the University of Groningen, the Physics Department of the University of
Murcia and CERN's Theory group for their hospitality and financial support.
The work of GB, CG-F, TO and JLVC has been supported in part by the MCI, AEI,
FEDER (UE) grants PID2021-125700NB-C21 and PID2024-155685NB-C21 (``Gravity,
Supergravity and Superstrings'' (GRASS)) and IFT Centro de Excelencia Severo
Ochoa CEX2020-001007-S. The work of PM has been supported by the MCI, AEI,
FEDER (UE) grant PID2021-123021NB-I00 and by FICYT through the Asturian grant
SV-PA-21-AYUD/2021/52177.  The work of GB has been supported by the fellowship
CEX2020-001007-S-20-5. The work of CG-F was supported by the MU grant
FPU21/02222. The work of JLVC has been supported by the CSIC JAE-INTRO grant
JAEINT-24-02806. TO wishes to thank M.M.~Fern\'andez for her permanent
support.

\appendix

\section{The $p$-form momentum map and the
  generalized symmetric ansatz}
\label{app-generalized}

Let us consider the theory of $(p+1)$-forms $V$ given in Eq.~(\ref{eq:CStheoryaction}),
which is invariant under $p$-form gauge transformations
Eq.~(\ref{eq:pformgaugetrans}). If the manifolds on which we are working
admit a harmonic $(p+1)$-form $\mathfrak{h}^{(p+1)}$ (just one, for the sake
of simplicity), it is also invariant under a global transformation of the
form Eq.~(\ref{eq:globalhigherformp}).

As discussed in Section~\ref{sec-gctsCS}, the action of a GCT generated by a
vector field $\xi$ on $V$ includes an induced gauge transformation and has the
general form Eq.~(\ref{eq:deltaxigeneral}). Thus, if the vector $k$ generates
a symmetry of a given $V$, we write

\begin{equation}
  \label{eq:deltakV}
  \delta_{k}V
  =
  -\mathcal{L}_{k}V +\delta_{\Lambda_{k}}V
  =
  -\imath_{k}G +d\left(\Lambda_{k}-\imath_{k}V\right)
  =
  0\,.
\end{equation}

A symmetry of $V$ should also be a symmetry of its gauge-invariant field
strength $G$, that is

\begin{equation}
  \delta_{k}G
  =
  -\mathcal{L}_{k}G 
  =
  -d\imath_{k}G 
  =
  0\,,
\end{equation}

\noindent
after use of the Bianchi identity $dG=0$.

Under the assumptions we have made, the above equation is solved by

\begin{equation}
  \label{eq:extendedpformmomentummapequation}
  \imath_{k}G
  =
  c\ \mathfrak{h}^{(p+1)} -dP_{k}\, , 
\end{equation}

\noindent
for some constant $c$. The derivative of the momentum map $P_{k}$ is (minus)
the exact part of $\imath_{k}G$ and we can take this property as its
definition.

If we substitute the above solution into Eq.~(\ref{eq:deltakV}), we get

\begin{equation}
  \label{eq:deltakV2}
  \delta_{k}V
  =
  -\mathcal{L}_{k}V +\delta_{\Lambda_{k}}V
  =
  -c\mathfrak{h}^{(p+1)} +d\left(\Lambda_{k}+P_{k}-\imath_{k}V\right)
  =
  0\,,
\end{equation}

\noindent
which is only consistent if $c=0$.

The solution (\ref{eq:extendedpformmomentummapequation}) is consistent with the following generalization of
Eq.~(\ref{eq:deltakV}), though:

\begin{equation}
  \label{eq:deltakVgeneralized}
  \delta_{k}V
  =
  -\mathcal{L}_{k}V +\delta_{\Lambda_{k}}V
  =
  -\imath_{k}G +d\left(\Lambda_{k}-\imath_{k}V\right)
  =
  \delta_{-c}V
  \equiv 
  -c\ \mathfrak{h}^{(p+1)}\,,
\end{equation}

\noindent
which states that $V$ is only invariant under the GCT generated by $k$, up to a
global symmetry transformation. In other words: $V$ satisfies a
\textit{generalized symmetry ansatz} (see Ref.~\cite{Ballesteros:2024prz} and
references therein).

On the other hand, for any value of $c$, the momentum map equation
(\ref{eq:extendedpformmomentummapequation}) defines the $p$-form momentum map
$P_{k}$ up to a closed $p$-form. Thus, we expect all expressions involving
physical quantities, such as the Smarr formula, to be invariant under

\begin{equation}
  \label{eq:Ctransformations}
  \delta_{C}P_{k}=C\,,
  \hspace{1cm}
  dC=0\,.
\end{equation}

\section{The magnetic momentum map in higher-rank-form
  theories with Chern--Simons terms}
\label{app:magneticmomentummap}

$\star G$ is a gauge-invariant tensor and
$\delta_{k}\star G = -\mathcal{L}_{k}\star G$. Assuming that the
transformation $\delta_{k}$ leaves invariant all the fields, it follows that

\begin{equation}
  \imath_{k}d\star G
  =
  -d\imath_{k}\star G\,.
\end{equation}

If we take the interior product of $k$ with the equation of motion of the
$(p+1)$-form Eq.~(\ref{eq:EVCS}) and use the above result, the $p$-form
momentum map Eq.~(\ref{eq:pformmomentummap}) and the Bianchi identity $dG=0$,
we find 

\begin{equation}
  \begin{aligned}
    \imath_{k}\mathbf{E}
    & =
      d\left\{ \imath_{k}\star G
      -N(N+1)\gamma P_{k}\wedge
      \underbrace{G\wedge \cdots \wedge G}_\text{N-1 times}
      \right\}\,.
  \end{aligned}
\end{equation}

Then, on-shell and locally, we can define the magnetic $\tilde{p}$-form
momentum map $\tilde{P}_{k}$, with $\tilde{p}\equiv d-p-4$, through the
\textit{magnetic momentum map equation}

\begin{equation}
  \label{eq:magneticmomentummapequationCS}
  \imath_{k}\star G
  -N(N+1)\gamma P_{k}\wedge
  \underbrace{G\wedge \cdots \wedge G}_\text{N-1 times} +d\tilde{P}_{k}
  \doteq
  0\,.
\end{equation}

As noticed by Gibbons and Warner in Ref.~\cite{Gibbons:2013tqa} in the context
of $\mathcal{N}=1,d=5$ supergravity coupled to vector multiplets, if the
manifold admits harmonic $(\tilde{p}+1)$-forms $\mathfrak{h}^{(\tilde{p}+1)}$,
the above equation should include them

\begin{equation}
  \label{eq:magneticmomentummapequationCS2}
  \imath_{k}\star G
  -N(N+1)\gamma P_{k}\wedge
  \underbrace{G\wedge \cdots \wedge G}_\text{N-1
    times}
  +d\tilde{P}_{k}
  +\mathfrak{h}^{(\tilde{p}+1)}
  \doteq
  0\,.
\end{equation}

Actually, if the field theory has on-shell closed but not exact
$(\tilde{p}+1)$-forms (charges or currents), they can play exactly the same
role as $\mathfrak{h}^{\tilde{p}+1}$ in the above equation. In minimal
5-dimensional supergravity $\tilde{p}=1$ and the field strength $G$ can play
this role but so can other closed 2-form charges, as we have shown in
Section~\ref{sec-massandangularmomentum}.

As in the electric case, this equation, with
$\mathfrak{h}^{(\tilde{p}+1)}\neq 0$ can also be interpreted in terms of a
generalized symmetric ansatz formulated in terms of the dual
$(\tilde{p}+1)$-form potential defined in
Section~\ref{sec-magneticchargesCS}..

$\tilde{P}_{k}$ is defined by this equation up to a closed 
$\tilde{p}$-form and we expect all expressions involving
physical quantities, such as the Smarr formula, to be invariant under

\begin{equation}
  \label{eq:Dtransformations}
  \delta_{D}\tilde{P}_{k}
  =
  D\,,
  \hspace{1cm}
  dD
  =
  0\,.
\end{equation}

Furthermore, since this equation depends on $P_{k}$, the transformations
Eq.~(\ref{eq:Ctransformations}) will act on $\tilde{P}_{k}$ on it as well:

\begin{equation}
  \label{eq:CtransformationsontildeP}
  \delta_{C}\tilde{P}_{k}
  =
  N(N+1)\gamma C \wedge
  \underbrace{G\wedge \cdots \wedge G}_\text{N-2 times}\wedge V\,.
\end{equation}

\section{Relation between the 4-and 5-dimensional fields}
\label{app-4-5fields}

The fields of minimal 5-dimensional supergravity have the following
decomposition in terms of the (Einstein-frame) fields of the T$^{3}$ model:

\begin{subequations}
  \begin{align}
    \hat{e}^{a}
    & =
      \left(k/k_{\infty}\right)^{-1/2}e^{a}_{E}\,,
    \\
    & \nonumber \\
    \hat{e}^{z}
    & =
      k(dz+k_{\infty}^{1/2} A^{0})\,,
    \\
    & \nonumber \\
      ds_{(5)}^{2}
   & =
   \left(k/k_{\infty}\right)^{-1}ds_{(4)\, E}^{2} -k^{2}\left(dz+k_{\infty}^{1/2}A^{0}\right)^{2}\,,
    \\
    & \nonumber \\
    \hat{V}
    & =
      -\sqrt{3}\left[k_{\infty}^{1/2}A^{1} +\chi (dz+k_{\infty}^{1/2} A^{0})\right]\,.
  \end{align}
\end{subequations}

The inverse relations are

\begin{subequations}
  \begin{align}
    e^{a}_{E\, \mu}
    & =
      \left(k/k_{\infty}\right)^{1/2}
      e^{a}_{\mu}
      =
    k_{\infty}^{-1/2}  (-\hat{g}_{\underline{z}\underline{z}})^{1/4}\hat{e}^{a}{}_{\mu}\,,
    \\
    & \nonumber \\
    g_{E\,\mu\nu}
    & =
      \left(k/k_{\infty}\right) g_{\mu\nu}
      =
      k_{\infty}^{-1}
      (-\hat{g}_{\underline{z}\underline{z}})^{1/2}\left(\hat{g}_{\mu\nu}
      -\hat{g}_{\mu\underline{z}}\hat{g}_{\nu\underline{z}}/\hat{g}_{\underline{z}\underline{z}}
      \right)\,,
    \\
    & \nonumber \\
    A^{0}{}_{\mu}
    & =
      A_{E\, \mu}
      =
      k_{\infty}^{-1/2}A_{\mu}
      =
      k_{\infty}^{-1/2}\hat{g}_{\mu\underline{z}}/\hat{g}_{\underline{z}\underline{z}}\,,
    \\
    & \nonumber \\
    A^{1}{}_{\mu}
    & =
      -\tfrac{1}{\sqrt{3}}V_{E\, \mu}
      =
      -\tfrac{1}{\sqrt{3}}k_{\infty}^{-1/2}V_{\mu}
      =
      -\tfrac{1}{\sqrt{3}}k_{\infty}^{-1/2}
      \left(\hat{V}_{\mu}-\hat{V}_{\underline{z}}\hat{g}_{\mu\underline{z}}/\hat{g}_{\underline{z}\underline{z}}\right)\,,
    \\
    & \nonumber \\
    e^{-2\phi}
    & =
      k
      =
      (-\hat{g}_{\underline{z}\underline{z}})^{1/2}\,,
    \\
    & \nonumber \\
    \chi
    & =
      -\tfrac{1}{\sqrt{3}}l
      =
      -\tfrac{1}{\sqrt{3}}\hat{V}_{\underline{z}}\,.
  \end{align}
\end{subequations}

\section{Relation between the 4-and 5-dimensional charges and potentials}
\label{app-4-5chargesandpotentials}

In this appendix we summarize the relations between the 4- and 5-dimensional
charges assuming KK boundary conditions plus $\chi=\chi_{\infty}$. These
relations are valid to relate the charges and potentials of 4- and
5-dimensional black holes.

First of all, with KK boundary conditions we have adopted the convention that
the only meaningful definitions of 5-dimensional mass and angular momentum are
the 4-dimensional ones in Einstein frame, $M$ and $J$. (The KK boundary
conditions eliminate the possiblity of having another angular momentum.)

The 5-dimensional electric and magnetic charges $Q$ and $P$, the momentum in
the internal direction $P_{z}$ and the topological charge of KK monopoles,
$KK$, decompose into the 4-dimensional ones $q_{0},q_{1},p^{0},p^{1}$ as

\begin{subequations}
  \begin{align}
    Q
    & =
      -\tfrac{1}{\sqrt{3}} e^{\phi_{\infty}}
      \left(q_{1}-3\chi_{\infty}p^{1}\right)\,,
    \\
    & \nonumber \\
    P & =
        \sqrt{3} e^{\phi_{\infty}}\left(p^{1}+\chi_{\infty}p^{0}\right)\,,
    \\
    & \nonumber\\
    P_{z}
    & =
      e^{\phi_{\infty}}(q_{0}-\tfrac{1}{3}\chi_{\infty}q_{1})\,,
    \\
    KK
    & =
      e^{-3\phi_{\infty}}p^{0}\,.
  \end{align}
\end{subequations}

The 5-dimensional electric and magnetic potentials $\hat{\Phi}$ and
$\tilde{\hat{\Phi}}$ at infinity and on the horizon are related to the
4-dimensional ones on the horizon as

\begin{subequations}
  \begin{align}
    \hat{\Phi}_{\infty}
    & =
      \sqrt{3}e^{-\phi_{\infty}}\chi_{\infty}\Phi_{\mathcal{H}}{}^{0}\,,
    \\
    & \nonumber \\
    \hat{\Phi}_{\mathcal{H}}
    & =
      -\sqrt{3}e^{-\phi_{\infty}}\Phi_{\mathcal{H}}{}^{1}\,,
    \\
    & \nonumber \\
    \tilde{\hat{\Phi}}_{\infty}
    & =
      \sqrt{3}e^{-\phi_{\infty}} \chi_{\infty}^{2}\Phi_{\mathcal{H}}{}^{0}\,,
    \\
    & \nonumber \\
    \tilde{\hat{\Phi}}_{\mathcal{H}}
    & =
      -\frac{e^{-\phi_{\infty}}}{\sqrt{3}}\Phi_{\mathcal{H}\, 1}\,.
  \end{align}
\end{subequations}

\section{Black-hole solutions of the T$^{3}$ model and their 5-dimensional origin}
\label{app-KKBHs}

For the sake of completeness, we describe in this appendix several simple
solutions of minimal 5-dimensional supergravity that give rise to
asymptotically-flat, static, charged, non-extremal 4-dimensional black-hole
solutions of the T$^{3}$ model and can be used to test the main results of the
paper. The solutions we are going to consider fall into two different
categories: those which correspond to 5-dimensional, purely gravitational
solutions and are solely described by a Ricci-flat metric (usually known as
Kaluza--Klein (KK) black holes) and those which correspond to 5-dimensional
solutions with a non-trivial gauge field $V$. The 4-dimensional electric and
magnetic charges of KK black holes have purely gravitational origin while
those of the 4-dimensional black holes in the latter category originate in the
charges of the 5-dimensional gauge field $V$.

We are going to start with 3 KK black-hole solutions:

\begin{enumerate}
\item A purely electric KK black-hole solution whose electric charge is
  5-dimensional momentum along the compact direction $z$. This solution was
  studied in more detail in Ref.~\cite{Gomez-Fayren:2023wxk}.
\item A purely magnetic KK black-hole solution whose magnetic charge
  originates in a Gross--Perry--Sorkin \cite{Gross:1983hb,Sorkin:1983ns} KK
  monopole (Euclidean Taub--NUT) in which the compact dimension is fibered
  over a 2-sphere.
\item A dyonic KK black-hole solution \cite{Gibbons:1985ac} that reduces to
  the two previous cases when one of the charges vanishes and to a dyonic
  Reissner--Nordstr\"om black hole \cite{kn:Reiss,kn:No8} when the two charges
  (with appropriate moduli factors) are identical. The 5-dimensional solution
  has momentum along the compact direction which is non-trivially fibered over
  2-spheres.
\end{enumerate}

All these solutions are also solutions of the Einstein--Maxwell--Dilaton (EMD)
model with\footnote{
  The static, asymptotically-flat, spherically
  symmetric solutions of the EMD model where found in
  Refs.~\cite{Gibbons:1982ih,Gibbons:1984kp,Holzhey:1991bx} for all values of
  $a$. Some of the solutions we are going to discuss are merely stationary in
  5 dimensions \cite{Gomez-Fayren:2023wxk}.
  }
$a=\sqrt{3}$, but we must take into account
that in the T$^{3}$ model the dilaton has a different normalization and one
has to do the rescaling $\phi \to -\frac{1}{2\sqrt{3}}\phi$
($\phi_{\infty} \to -\frac{1}{2\sqrt{3}}\phi_{\infty}$) to bring the T$^{3}$
solution to the standard form of the EMD solutions. The electric and magnetic
solutions are related by a discrete electric-magnetic duality transformation

\begin{equation}
  q\to p\,,
  \hspace{1cm}
  p\to -q\,,
  \hspace{1cm}
  \phi_{\infty}\to -\phi_{\infty}\,,
  \hspace{1cm}
  \Sigma\to -\Sigma\,,
\end{equation}

\noindent
that leaves invariant the dyonic one.

Then, we are going to review two solutions in the second category:

\begin{enumerate}
\item A purely electric black hole obtained by smearing and compactifying a
  5-dimensional Reissner--Nordstr\"om--Tangherlini black hole
  \cite{kn:Reiss,kn:No8,Tangherlini:1963bw} along the compact direction $z$.
\item A purely magnetic black hole that corresponds to a 5-dimensional string
  wrapped around the compact direction and is electrically charged with
  respect to the 2-form dual to the gauge field $V$.
\end{enumerate}

These two solutions are also solutions of the EMD model with $a=-1/\sqrt{3}$
but, now, we must also take into account the different normalization of the
gauge field $A^{1}$ in the T$^{3}$ model and (apart from the dilaton) one has
to rescale $A^{1}\to \tfrac{1}{\sqrt{3}}A^{1}$ to bring the T$^{3}$ solution
to the standard form of the EMD solutions.

These two solutions are related by electric-magnetic duality in 4
dimensions. In this case, no dyonic solution is known.

There are more general black-hole solutions of the T$^{3}$, with more than one
active electric or magnetic charge and more than one active scalar, but the
ones we have chosen can be seen as their building blocks and are rich enough
for our purposes.

In all the solutions that we are going to describe, as usual, the compact
coordinate $z$ takes values $z\in[0,2\pi \ell]$ and the proper length of the
compact dimension is $2\pi R_{z}$. Furthermore, the asymptotic value of the
scalar $\phi$ is related to $R_{z}$ and $\ell$ by

\begin{equation}
e^{\phi_{\infty}}=(R_{z}/\ell)^{-1/2}\,.
\end{equation}

We are going to describe most solutions using two functions $H,W$ of the form

\begin{equation}
  \label{eq:HWfunctions}
  H = 1+\frac{h}{r}\,,
  \hspace{1cm}
  W = 1+\frac{w}{r}\,,
\end{equation}

\noindent
chosen in such a way that the event horizon is placed at $r=-w$. Sometimes we
will use the more standard non-extremality parameter $r_{0}=-w/2$ which is
related to the Hawking temperature $T$ and Bekenstein-Hawking entropy $S$ by

\begin{equation}
  \label{eq:2STrelation}
  2ST
  =
  \frac{r_{0}}{G_{N}^{(4)}}\,.
\end{equation}

As the name indicates, the extremal limit is always reached when $r_{0}=0$
(which can be seen as a saturated Bogomol'nyi bound) so that $W=1$. Except for
the dyonic black hole, in this limit all the solutions have only one
independent function and parameter (apart from the modulus $\phi_{\infty}$)
which indicates that all their charges are related. In all cases except
the dyonic one, extremality is related to unbroken supersymmetry.

\subsection{The purely electric KK black hole}
\label{app-electricKKBH}

Upon use of the relations in Appendix~\ref{app-4-5fields}, the purely electric
4-dimensional KK black hole solution of the 4-dimensional T$^{3}$ model

\begin{equation}
  \label{eq:4delectricKKBH}
  \begin{aligned}
    ds^{2}_{E\, (4)}
    & =
          H^{-1/2}Wdt^{2}-H^{1/2}\left(W^{-1}dr^{2}+r^{2}d\Omega_{(2)}^{2}\right)\,,
    \\
    & \\
    A^{0}
    & =
    \alpha  e^{3\phi_{\infty}}\left(H^{-1}-1\right)dt\,,
    \\
    & \\
    e^{\phi}
    & =
    e^{\phi_{\infty}}H^{-1/4}\,,
  \end{aligned}
\end{equation}

\noindent
where the integration constants $h,w$ and $\alpha$ must satisfy the relation

\begin{equation}
  w
=
  h\left(1-\alpha^{2}\right)\,,
\end{equation}

\noindent
can be seen to correspond to the Ricci-flat 5-dimensional metric

\begin{equation}
  ds^{2}_{(5)}
  =
  H^{-1}Wdt^{2}
  -W^{-1}dr^{2} - r^{2}d\Omega^{2}_{(2)}
    -H\left[\frac{R_{z}}{\ell} dz +\alpha (H^{-1}-1)dt\right]^{2}\,.
\end{equation}

The integration constants $h,w,\alpha$ are functions of the 4-dimensional
independent physical parameters: the ADM mass $M$, the electric charge $q_{0}$
and $\phi_{\infty}$

\begin{subequations}
  \label{eq:whandphysicalparameterselectric}
  \begin{align}
    w
    & =
      -2r_{0}
      =
      -G_{N}^{(4)}\left[3M -\sqrt{M^{2}+8e^{6\phi_{\infty}}q^{2}}\right]\,,
    \\
    & \nonumber \\
    h
    & =
      -2G_{N}^{(4)}\left[M -\sqrt{M^{2}+8e^{6\phi_{\infty}}(q_{0})^{2}}\right]\,,
    \\
    & \nonumber \\
\alpha
    & =
      \frac{2e^{3\phi_{\infty}}q_{0}}{M -\sqrt{M^{2}+8e^{6\phi_{\infty}}(q_{0})^{2}} }\,,
  \end{align}
\end{subequations}

The 4-dimensional scalar charge $\Sigma$ is also a function of
them\footnote{Here we are using the scalar charge associated to the
  (canonically-normalized) dilaton of the EMD model because it gives rise to
  more standard expressions. The scalar charge of the dilaton in the T$^{3}$
  model is $2\sqrt{3}$ times this one. We will do the same in the following
  examples as well.}

\begin{equation}
  \label{eq:SigmaKKelectric}
  \Sigma
  =
  -\sqrt{3}\left[M-\sqrt{M^{2}+8e^{6\phi_{\infty}}(q_{0})^{2}}\right]\,,
\end{equation}

\noindent
in agreement with the no-hair theorem
\cite{Pacilio:2018gom,Ballesteros:2023iqb}.

The non-extremality parameter $r_{0}=-w/2$ satisfies the relation

\begin{equation}
    \left(r_{0}/G_{N}^{(4)}\right)^{2}
  =
  M^{2}+\tfrac{1}{4}\Sigma^{2}-4e^{6\phi_{\infty}}(q_{0})^{2}\,.
\end{equation}

The Hawking temperature and Bekenstein-Hawking temperature of the
4-dimensional black hole are

\begin{subequations}
  \begin{align}
    \label{eq:temperatureKK}
    T
    & =
    \frac{1}{4\pi[2r_{0}(2r_{0}+h)]^{1/2}}\,,
    \\
    & \nonumber \\
    \label{eq:entropyKK}
    S
    & =
    \frac{\pi}{G_{N}^{(4)}}\left(2r_{0}+h\right)^{1/2}(2r_{0})^{3/2}\,,    
  \end{align}
\end{subequations}

\noindent
and they satisfy the relation Eq.~(\ref{eq:2STrelation}). This leads to the
Smarr formula

\begin{equation}
  \label{eq:Smarrelectric}
  M
  =
  2ST +\Phi^{0} q_{0}\,,  
\end{equation}

\noindent
with

\begin{equation}
  \label{eq:PhiKK}
  \Phi^{0}
  =
  \frac{4 e^{6\phi_{\infty}}q_{0}}{M+\sqrt{M^{2}+8e^{6\phi_{\infty}}(q_{0})^{2}} }\,. 
\end{equation}

The first law takes the form

\begin{equation}
  \delta M
  =
  T\delta S +\Phi^{0}\delta q_{0} +\tfrac{1}{4}\Sigma \delta\phi_{\infty}\,.
\end{equation}

\subsection{The purely magnetic KK black hole}
\label{app-magneticKKBH}

The purely magnetic 4-dimensional KK black hole solution corresponds to the
following Ricci-flat metric

\begin{equation}
  ds^{2}_{(5)}
  =
  Wdt^{2}
  -H\left[W^{-1}dr^{2} +r^{2}d\Omega^{2}_{(2)}\right]
  -H^{-1}\left[\frac{R_{z}}{\ell} dz
    +\alpha h\cos{\theta}d\varphi\right]^{2}\,,
\end{equation}

\noindent
and to the 4-dimensional solution of the T$^{3}$ model

\begin{equation}
  \label{eq:4dmagneticKKBH}
  \begin{aligned}
    ds^{2}_{E\, (4)}
    & =
          H^{-1/2}Wdt^{2}-H^{1/2}\left(W^{-1}dr^{2}+r^{2}d\Omega_{(2)}^{2}\right)\,,
    \\
    & \\
    A^{0}
    & =
    e^{3\phi_{\infty}}\alpha h \cos{\theta}d\varphi\,,
    \\
    & \\
    e^{\phi}
    & =
    e^{\phi_{\infty}}H^{1/4}\,.
  \end{aligned}
\end{equation}

The constants $R_{z}$ and $\ell$ have the same meaning as in the purely
electric case and are related to the asymptotic value of the dilaton in the
same way. The integration constants $h,w,\alpha$ are now functions of
the physical parameters of the corresponding 4-dimensional solution, namely
the ADM mass $M$, the magnetic charge $p$ and $\phi_{\infty}$

\begin{subequations}
  \label{eq:whandphysicalparametersmagnetic}
  \begin{align}
    w
    & =
      -2r_{0}
      =
      -G_{N}^{(4)}\left[3M -\sqrt{M^{2}+8e^{-6\phi_{\infty}}(p^{0})^{2}}\right]\,,
    \\
    & \nonumber \\
    h
    & =
      -2G_{N}^{(4)}\left[M -\sqrt{M^{2}+8e^{-6\phi_{\infty}}(p^{0})^{2}}\right]\,,
    \\
    & \nonumber \\
\alpha
    & =
      -\frac{2e^{-3\phi_{\infty}}p^{0}}{M -\sqrt{M^{2}+8e^{-6\phi_{\infty}}(p^{0})^{2}} }\,.
  \end{align}
\end{subequations}

The 4-dimensional scalar charge $\Sigma$ is also a function of them

\begin{equation}
  \label{eq:SigmaKKmagnetic}
  \Sigma
  =
  \sqrt{3}\left[M-\sqrt{M^{2}+8e^{-6\phi_{\infty}}(p^{0})^{2}}\right]\,,
\end{equation}

\noindent
and the non-extremality parameter $r_{0}=-w/2$ satisfies the relation

\begin{equation}
    \left(r_{0}/G_{N}^{(4)}\right)^{2}
  =
  M^{2}+\tfrac{1}{4}\Sigma^{2}-4e^{-6\phi_{\infty}}(p^{0})^{2}\,.
\end{equation}

The Hawking temperature and Bekenstein-Hawking entropy of the 4-dimensional
black hole have the same form as in the electric case as functions of
$r_{0}=-w/2$ and $h$, Eqs.~(\ref{eq:temperatureKK}) and
(\ref{eq:entropyKK}). The Smarr formula now takes the form

\begin{equation}
M = 2ST -\Phi_{0}p^{0}\,,  
\end{equation}

\noindent
with

\begin{equation}
  \Phi_{0}
  =
  -\frac{4e^{-6\phi_{\infty}} p^{0}}{M+\sqrt{M^{2}+8e^{-6\phi_{\infty}}(p^{0})^{2}}}\,.
\end{equation}

The sign of $\Phi_{0}$ is unconventional for a thermodynamic potential but it
is required for the Smarr formula to be electric-magnetic duality invariant
\cite{Ortin:2022uxa}.

\subsection{The dyonic KK black hole}
\label{app-dyonicKKBH}

Shifting the radial coordinate of Ref.~\cite{Gibbons:1985ac}
$r\rightarrow r+M+w/2$ and introducing an arbitrary asymptotic value for the KK
scalar (set to zero in that reference), the dyonic KK black hole solution can
be written in the 5-dimensional form

\begin{equation}
  ds^{2}_{(5)}
   =
  H_{e}^{-1}Wdt^{2}
  -H_{m}\left[W^{-1}dr^{2} +r^{2}d\Omega^{2}_{(2)}\right]
  -H_{e}H_{m}^{-1}\left[\frac{R_{z}}{\ell} dz
    +A\right]^{2}\,,
\end{equation}

\noindent
where, now

\begin{subequations}
  \label{eq:HKW}
  \begin{align}
    H_{e} & =
        \left[1+\frac{G_{N}^{(4)}\left(M+\frac{w}{2}+\frac{\Sigma}{2\sqrt{3}}\right)}{r}\right]^{2}
        -\frac{4}{\sqrt{3}}\frac{(G_{N}^{(4)}e^{3\phi_{\infty}}q_{0})^{2}\Sigma}{\left(M+\frac{\Sigma}{2\sqrt{3}}\right)r^{2}}\,,
    \\
      & \nonumber \\
    H_{m} & =
        \left[1+\frac{G_{N}^{(4)}\left(M+\frac{w}{2}-\frac{\Sigma}{2\sqrt{3}}\right)}{r}\right]^{2}
        -\frac{4}{\sqrt{3}}\frac{(G_{N}^{(4)}e^{-3\phi_{\infty}}p^{0})^{2}\Sigma}{\left(M-\frac{\Sigma}{2\sqrt{3}}\right)r^{2}}\,,    
    \\
      & \nonumber \\
    W
      & =
        1+\frac{w}{r}\,,
    \\
      & \nonumber \\
    A
      & =
        \frac{4G_{N}^{(4)}e^{3\phi_{\infty}}q_{0}}{r H_{e}}
        \left[1+\frac{G_{N}^{(4)}
        \left(M+\frac{w}{2}-\frac{\Sigma}{2\sqrt{3}}\right)}{r}\right]dt
        +4e^{-3\phi_{\infty}}G_{N}^{(4)}p^{0}\cos{\theta}d\varphi\,,
  \end{align}
\end{subequations}

\noindent
where the non-extremality parameter $r_{0}=-w/2$ can be written in the form

\begin{equation}
  \left(r_{0}/G_{N}^{(4)}\right)^{2}
  =
  M^{2}+\tfrac{1}{4}\Sigma^{2} -4e^{6\phi_{\infty}}(q_{0})^{2}
  -4e^{-6\phi_{\infty}}(p^{0})^{2}\,,
\end{equation}

\noindent
and where the dependence of the scalar charge $\Sigma$ on the independent
physical parameters $M,p,q,\phi_{\infty}$ is implicit in the relation

\begin{equation}
  \label{eq:SigmaKKdyonic}
  \Sigma
  =
  \sqrt{3}\left[\frac{4e^{6\phi_{\infty}}(q_{0})^{2}}{M+\frac{\Sigma}{2\sqrt{3}}}
  -\frac{4e^{-6\phi_{\infty}}(p^{0})^{2}}{M-\frac{\Sigma}{2\sqrt{3}}}\right]\,.
\end{equation}

In the purely electric case ($p^{0}=0$) the above equation can be solved,
recovering Eq.~(\ref{eq:SigmaKKelectric}) and the relation

\begin{equation}
M+\frac{w}{2}  = \frac{\Sigma}{2\sqrt{3}}\,,
\end{equation}

\noindent
which can be used to recover the electric solution as described in
Appendix~\ref{app-electricKKBH}. The purely magnetic ($q=0$) solution is
recovered in the form given in Appendix~\ref{app-magneticKKBH} in a similar
fashion.

The physical parameters in the fields correspond to those of the following
4-dimensional solution of the T$^{3}$ model

\begin{equation}
  \label{eq:4ddyonicKKBH}
  \begin{aligned}
    ds^{2}_{E\, (4)}
    & =
      \left(H_{e}H_{m}\right)^{-1/2}Wdt^{2}
      -\left(H_{e}H_{m}\right)^{1/2}\left(W^{-1}dr^{2}+r^{2}d\Omega_{(2)}^{2}\right)\,,
    \\
    & \\
    A^{0}
    & =
        \frac{4G_{N}^{(4)}e^{6\phi_{\infty}}q_{0}}{r H_{e}}
        \left[1+\frac{G_{N}^{(4)}
        \left(M+\frac{w}{2}-\frac{\Sigma}{2\sqrt{3}}\right)}{r}\right]dt
        +4G_{N}^{(4)} p^{0}\cos{\theta}d\varphi\,,
    \\
    & \\
    e^{\phi}
    & =
    e^{\phi_{\infty}}\left(H_{e}/H_{m}\right)^{-1/4}\,.
  \end{aligned}
\end{equation}

When $e^{6\phi_{\infty}}(q_{0})^{2}=e^{-6\phi_{\infty}}(p^{0})^{2}$, $\Sigma=0$ and
$H_{e}=H_{m}$, which leads to a constant scalar. The metric and vector field
above solution are identical to those of the dyonic Reissner--Nordstr\"om
black hole with equal electric and magnetic charges and the corresponding
5-dimensional metric is a particular embedding of that solution in higher
dimensions and, being purely gravitational, in the string effective action
\cite{Cano:2019ycn}. The extremal limit of this particular embedding is not
supersymmetric \cite{Khuri:1995xq}.

The Hawking temperature and the Bekenstein--Hawking entropy have complicated
expressions that can be written in the compact form

\begin{subequations}
  \begin{align}
    \label{eq:temperatureKKdyonic}
    T
    & =
      \frac{1}{8\pi r_{0} \left[H(2r_{0})K(2r_{0})\right]^{1/2}}\,,
    \\
    & \nonumber \\
    \label{eq:entropyKKdyonic}
    S
    & = \frac{\pi}{G_{N}^{(4)}}
      \left[H_{e}(2r_{0})H_{m}(2r_{0})\right]^{1/2}(2r_{0})^{2}\,,    
  \end{align}
\end{subequations}

\noindent
where $H_{e}(2r_{0})$ and $H_{m}(2r_{0})$ are the functions of $r$ in
Eqs.~(\ref{eq:HKW}) evaluated on the horizon which lies at $r=2r_{0}$.

The Smarr formula must take the general form

\begin{equation}
M= 2ST +\Phi^{0}q_{0}-\Phi_{0}p^{0}\,,  
\end{equation}

\noindent
and it has been checked in Ref.~\cite{Gibbons:1985ac} using
$\Phi^{0}(M,q,p,\phi_{\infty})=A^{0}_{t}(-\omega)$ and defining
$\Phi_{0}(M,q,p,\phi_{\infty}) = \Phi^{0}(M,p,-q,-\phi_{\infty})$.

\subsection{Black holes from electrically charged black holes}
\label{app-electricblackhole}

The 5-dimensional Reissner-Nordstr\"om--Tangherlini (RNT) black hole
\cite{kn:Reiss,kn:No8,Tangherlini:1963bw} can be written in the form

\begin{equation}
  \begin{aligned}
  ds^{2}_{(5)}
   & =
    \hat{H}^{-2}\hat{W}dt^{2}
     -\hat{H}\left[\hat{W}^{-1}d\rho^{2}+\rho^{2}d\Omega_{(3)}^{2}\right]\,,
    \\
    & \\
    V
    & =
      -\alpha  \left(\hat{H}^{-1}-1\right)dt\,,
  \end{aligned}
\end{equation}

\noindent
where

\begin{equation}
  \hat{H} = 1 +\frac{\hat{h}}{\rho^{2}}\,,
  \hspace{1cm}
  \hat{W} = 1 +\frac{\hat{w}}{\rho^{2}}\,,
\end{equation}

\noindent
and where the integration constants satisfy the constraint

\begin{equation}
\hat{w}=\hat{h}\left(1 -\alpha^{2}/3\right)\,.  
\end{equation}

The RNT black hole is electrically charged, as black holes cannot carry
magnetic charges in 5 dimensions: the dual of the gauge field is a 2-form that
couples to strings. The electric-magnetic dual is, therefore, a string
electrically charged with respect to that form and magnetically charged
(``dipole magnetic charge'') with respect to the original gauge field. The
electric-magnetic duality transformation is straightforward in 4 dimensions
but, first, one has to reduce the RNT solution in one dimension along an
isometric direction. The 3-sphere has quite few, but the $\rho^{2}$ factor in
front of its metric leads to a KK scalar that blows up at infinity and to
other pathologies.

The standard procedure to carry out the dimensional reduction, in extremal
cases in which $\hat{W}=1$ and $\hat{H}$ is an arbitrary harmonic function in
$\mathbb{R}^{4}$, is to \textit{smear} the solution along one of the
$\mathbb{R}^{4}$ directions, constructing a multicenter solution and deriving
a leading-order approximation to it. The procedure\footnote{See
  Ref.~\cite{Ortin:2015hya} and references therein.} guarantees that one
obtains a solution of the equations of motion.

In the non-extremal case that we are considering the procedure cannot be
applied. However, there is a family of solutions that can be understood as the
result of smearing the above RNT solution along direction parametrized by the
coordinate $z$ ($\rho^{2}=r^{2}+z^{2}$ where $r$ is the radial coordinate in 3
space dimensions) \cite{Ortin:2015hya}. It is given by

\begin{equation}
  \begin{aligned}
  ds^{2}_{(5)}
   & =
    H^{-2}Wdt^{2}
     -H\left[W^{-1}dr^{2}+r^{2}d\Omega_{(2)}^{2}
     +\left(\frac{R_{z}}{\ell}dz\right)^{2}\right]\,,
    \\
    & \\
    V
    & =
      -\alpha  \left(H^{-1}-1\right)dt\,,
  \end{aligned}
\end{equation}

\noindent
where the functions $H$ and $W$ now have the usual form
Eq.~(\ref{eq:HWfunctions}) and where the integration constants $h,w$ and $\alpha$ must satisfy the constraint

\begin{equation}
  \label{eq:hwconstraintRNT}
w=h\left(1 -\alpha^{2}/3\right)\,.  
\end{equation}

The corresponding 4-dimensional solution of the T$^{3}$ model reads

\begin{equation}
  \label{eq:electricKKBH-2}
  \begin{aligned}
    ds_{E\, (4)}^{2}
    & =
      H^{-3/2}Wdt^{2}
      -H^{3/2}\left[W^{-1}dr^{2}+r^{2}d\Omega^{2}_{(2)}\right]\,.
    \\
    & \\
    A^{1}
    & =
      \tfrac{1}{\sqrt{3}}e^{\phi_{\infty}}\alpha \left(H^{-1}-1\right)dt\,,
    \\
    & \\
    e^{\phi}
    & =
      e^{\phi_{\infty}}H^{-1/4}\,.
  \end{aligned}
\end{equation}

The relation between the integration constants and the 4-dimensional
independent physical constants is

\begin{subequations}
  \begin{align}
    w
    & =
      G_{N}^{(4)}\left[M-3\sqrt{M^{2}
      -\tfrac{8}{9}e^{2\phi_{\infty}}(q_{1})^{2}}\right]\,,
    \\
    & \nonumber \\
    h
    & =
      2G_{N}^{(4)}\left[M-\sqrt{M^{2}
      -\tfrac{8}{9}e^{2\phi_{\infty}}(q_{1})^{2}}\right]\,,
    \\
    & \nonumber \\
    \alpha
    & =
      \frac{-2e^{\phi_{\infty}} q_{1}}{\sqrt{3}\left[M-\sqrt{M^{2}
      -\tfrac{8}{9}e^{2\phi_{\infty}}(q_{1})^{2}}\right]}\,,
  \end{align}
\end{subequations}

\noindent
and the scalar charge is given by

\begin{equation}
  \Sigma
  =
  \sqrt{3} \left[M-\sqrt{M^{2}
      -\tfrac{8}{9}e^{2\phi_{\infty}}(q_{1})^{2}}\right]\,.
\end{equation}

The non-extremality parameter satisfies the relation

\begin{equation}
  \left(r_{0}/G_{N}^{(4)}\right)^{2}
  =
  M^{2}+\tfrac{1}{4}\Sigma^{2}-\tfrac{4}{3}e^{2\phi_{\infty}}q_{1}^{2}\,,
\end{equation}

\noindent
with

\begin{equation}
  \Phi^{1}
  =
  \frac{4e^{2\phi_{\infty}}q_{1}}{3\left[M+\sqrt{M^{2}
      -\tfrac{8}{9}e^{2\phi_{\infty}}(q_{1})^{2}}\right]}\,,
\end{equation}

\noindent
and using Eq.~(\ref{eq:2STrelation}) we can check that the Smarr formula

\begin{equation}
  M
  =
  2ST +\Phi^{1}q_{1}\,,
\end{equation}

\noindent
is also satisfied.

In terms of the non-extremality parameter $r_{0}$ and $h$, the temperature and
entropy are given by

\begin{subequations}
  \begin{align}
    \label{eq:temperatureKKstring}
    T
    & =
    \frac{(2r_{0})^{1/2}}{4\pi(2r_{0}+h)^{3/2}}\,,
    \\
    & \nonumber \\
    \label{eq:entropyKKstring}
    S
    & =
    \frac{\pi}{G_{N}^{(4)}}\left(2r_{0}+h\right)^{3/2}(2r_{0})^{1/2}\,.    
  \end{align}
\end{subequations}

\subsection{Black holes from magnetically charged black strings}
\label{app-magneticblackstring}

As discussed before we can use 4-dimensional electric-magnetic duality to
generate a new solution of minimal 5-dimensional supergravity. The result is a
solution that describes a black string that is magnetically charged with respect to
the vector field and electrically charged with respect to the dual
2-form. They are given by the following metric and vector field (see,
\textit{e.g.}~Ref.~\cite{Ortin:2015hya}):

\begin{equation}
  \begin{aligned}
  ds^{2}_{(5)}
   & =
    H^{-1}\left[Wdt^{2}-\frac{R_{z}}{\ell}dz^{2}\right]
     -H^{2}\left[W^{-1}dr^{2}+r^{2}d\Omega_{(2)}^{2}\right]\,,
    \\
    & \\
    V
    & =
      \alpha h \cos{\theta}d\varphi\,,
  \end{aligned}
\end{equation}

\noindent

The non-vanishing fields of the corresponding 4-dimensional solution of the
T$^{3}$ model, dual of Eq.~(\ref{eq:electricKKBH-2}), are

\begin{equation}
  \label{eq:magneticKKBH-2}
  \begin{aligned}
    ds_{E\, (4)}^{2}
    & =
      H^{-3/2}Wdt^{2}
      -H^{3/2}\left[W^{-1}dr^{2}+r^{2}d\Omega^{2}_{(2)}\right]\,.
    \\
    & \\
    A^{1}
    & =
      -\frac{1}{\sqrt{3}}e^{\phi_{\infty}}\alpha h \cos{\theta}d\varphi\,,
    \\
    & \\
    e^{\phi}
    & =
      e^{\phi_{\infty}}H^{1/4}\,.
  \end{aligned}
\end{equation}

\noindent
which is a solution to the EMD system with $a=1/\sqrt{3}$.

The integration constants $\alpha,w$ and $h$ have the following expressions in
terms of the independent physical constants:\footnote{Notice that, due to the
  normalization of the kinetic term of $A^{1}$ and the definitions of electric
  and magnetic charges we have made, the action of electric-magnetic duality
  on the electric and magnetic charges is $q_{1}\to 3 p^{1}$,
  $p^{1}\to -\tfrac{1}{3}q_{1}$. As usual, $\phi_{\infty}\to -\phi_{\infty}$.}

\begin{subequations}
  \begin{align}
    w
    & =
      G_{N}^{(4)}\left\{M -3\sqrt{M^{2}-8e^{-2\phi_{\infty}}(p^{1})^{2}}\right\}\,,
    \\
    & \nonumber \\
    h
    & =
    2 G_{N}^{(4)}\left\{  M -\sqrt{M^{2}-8e^{-2\phi_{\infty}}(p^{1})^{2}}\right\}\,,
    \\
    & \nonumber \\
    \alpha
    & =
      -\frac{2\sqrt{3}e^{-\phi_{\infty}}p^{1}}{M -\sqrt{M^{2}-8e^{-2\phi_{\infty}}(p^{1})^{2}}}\,.
  \end{align}
\end{subequations}

The scalar charge is given by 

\begin{equation}
  \label{eq:SigmaKKstringmegnetic}
  \Sigma
  =
  -\sqrt{3}\left[M-\sqrt{M^{2}-8e^{-2\phi_{\infty}}(p^{1})^{2}}\right]\,.
\end{equation}

\noindent
and the non-extremality parameter can be written in the form

\begin{equation}
    \left(r_{0}/G_{N}^{(4)}\right)^{2}
  =
  M^{2}+\tfrac{1}{4}\Sigma^{2}-12e^{-2\phi_{\infty}}(p^{1})^{2}\,.
\end{equation}

The Smarr formula has the form

\begin{equation}
  M =
  2ST -\Phi_{1}p^{1}\,,
\end{equation}

\noindent
with

\begin{equation}
  \Phi_{1}
  =
  -\frac{12e^{-2\phi_{\infty}}(p^{1})^{2}}{M+\sqrt{M^{2}-8e^{-2\phi_{\infty}}p^{1}}}\,.
\end{equation}

In terms of $r_{0}$ and $h$, the temperature and entropy have the form
Eqs.~(\ref{eq:temperatureKKstring}) and (\ref{eq:entropyKKstring}).


\end{document}